\documentclass[cernpreprint,coverpage=false,atlasdraft=false,texlive=2016,UKenglish,texmf]{atlasdoc}
 
\usepackage{atlaspackage}
\usepackage{atlasbiblatex}

\usepackage{atlascontribute}
 
\RequirePackage[version=4]{mhchem} 
\usepackage{ANA-HION-2018-27-PAPER-defs} 
 
\usepackage{atlasphysics}
\usepackage{atlasunit}
\usepackage{atlasxref}
\usepackage{siunitx}
\usepackage{hyperref}
 
\graphicspath{{logos/}{figures/}}

\AtlasTitle{Two-particle azimuthal correlations in photonuclear ultraperipheral Pb+Pb collisions at 5.02~TeV with ATLAS}
 
\AtlasAbstract{
Two-particle long-range azimuthal correlations are measured in photonuclear collisions using 1.7~nb$^{-1}$ of 5.02~TeV Pb+Pb collision data collected by the ATLAS experiment at the LHC.
Candidate events are selected using a dedicated high-multiplicity photonuclear event trigger, a combination of information from the zero-degree calorimeters and forward calorimeters, and from pseudorapidity gaps constructed using calorimeter energy clusters and charged-particle tracks.
Distributions of event properties are compared between data and Monte Carlo simulations of photonuclear processes.
Two-particle correlation functions are formed using charged-particle tracks in the selected events, and a template-fitting method is employed to subtract the non-flow contribution to the correlation. Significant nonzero values of the second- and third-order flow coefficients are observed and presented as a function of charged-particle multiplicity and transverse momentum.
The results are compared with flow coefficients obtained in proton--proton and proton--lead collisions in similar multiplicity ranges, and with theoretical expectations.
The unique initial conditions present in this measurement provide a new way to probe the origin of the collective signatures previously observed only in hadronic collisions.
}
 
\author{The ATLAS Collaboration}
 
\AtlasRefCode{ATLAS-HION-2018-027}
 
\PreprintIdNumber{CERN-EP-2020-246}

\AtlasJournal{Phys.\ Rev.\ C. 104 (2021) 014903}
\AtlasDOI{10.1103/PhysRevC.104.014903}
 
\hypersetup{pdftitle={ATLAS document},pdfauthor={The ATLAS Collaboration}}
 
\begin{document}
 
\maketitle
 
\tableofcontents

\section{Introduction}
 
In ultrarelativistic collisions of lead nuclei at the Large Hadron Collider (LHC), the typical processes studied are those for which the nuclei have an impact parameter less than
twice their radius ($b \lesssim 2R_A$).
Such lead--lead (\PbPb) collisions are understood to create a quark--gluon plasma and result in a large number of particles in the final state which participate in collective motion as a consequence of the plasma evolution~\cite{Busza:2018rrf,Romatschke:2017ejr,Heinz:2013th}.
In addition to the particles produced in \PbPb collisions, those produced in high-energy proton--proton (\pp) and proton--lead (\pPb) collisions also exhibit a collective behavior which manifests as an event-wide azimuthal variation persisting broadly in pseudorapidity, initially observed as a `ridge'~\cite{HION-2015-09,HION-2016-01,CMS-HIN-14-006,HION-2017-11}. This behavior is characterized in terms of nonzero single-particle azimuthal anisotropies, given by $n^\mathrm{th}$-order Fourier coefficients, and here referred to as flow coefficients $v_n$. Nonzero $v_n$ values have also been observed in significantly lower-energy ($19.6$--$200$~\GeV) $d$+Au collisions at the Relativistic Heavy Ion Collider~\cite{Aidala:2017ajz}. A natural question is whether such signatures persist in even smaller collision systems~\cite{Nagle:2017sjv} and, if so, how this may influence the interpretation of these signatures in $pp$ or $p$+Pb collisions.
 
In the prevailing paradigm, these observed anisotropies arise from the creation of a miniature region of quark--gluon plasma~\cite{Nagle:2018nvi,PHENIX:2018lia}, in which a hydrodynamic-like expansion of the system converts spatial nonuniformities in the initial state of the system into momentum-space anisotropies of the final-state particles. However, momentum correlations already present in the initial state of the collision may also persist into the final state~\cite{Greif:2017bnr}. The relative importance of these two explanations can be tested in collision systems where one or both of the `beams' has a significantly simpler initial state. Recently, studies were performed in archived $e^+e^-$ collision data at $\sqrt{s} = 91$~\GeV\ from the ALEPH detector~\cite{Badea:2019vey} and in archived $ep$ collision data at $\sqrt{s} = 316$~\GeV\ from the ZEUS detector~\cite{ZEUS:2019jya} with a deep-inelastic-scattering selection $Q^2 > 1$~\GeV$^2$. A ridge signature was not observed, and experimental upper limits were set on the possible magnitude of $v_n$ coefficients.
 
In addition to the hadronic \PbPb interactions described above, the strong electromagnetic~(EM) fields of the fully ionized nuclei can induce interactions even when the nuclei have significantly larger impact parameters such that no hadronic interaction occurs ($b \gtrsim 2R_A$). In the equivalent photon approximation~\cite{Fermi:1924tc,vonWeizsacker:1934nji,Williams:1934ad}, these strong EM fields correspond to a flux of quasi-real, high-energy photons. These photons can be emitted coherently from the entire nucleus, producing a flux enhanced by a factor of $Z^2$ ($Z = 82$ for Pb) for photons up to 80~\GeV\ at the LHC.
These ultraperipheral collisions (UPC)~\cite{Bertulani:2005ru,Klein:2020fmr} have appreciable rates and include photon--photon ($\gamma\gamma$) and photonuclear ($\gamma+\mathrm{A}$) interactions.
 
\begin{figure}[t]
\centering
\includegraphics[width=0.49\textwidth]{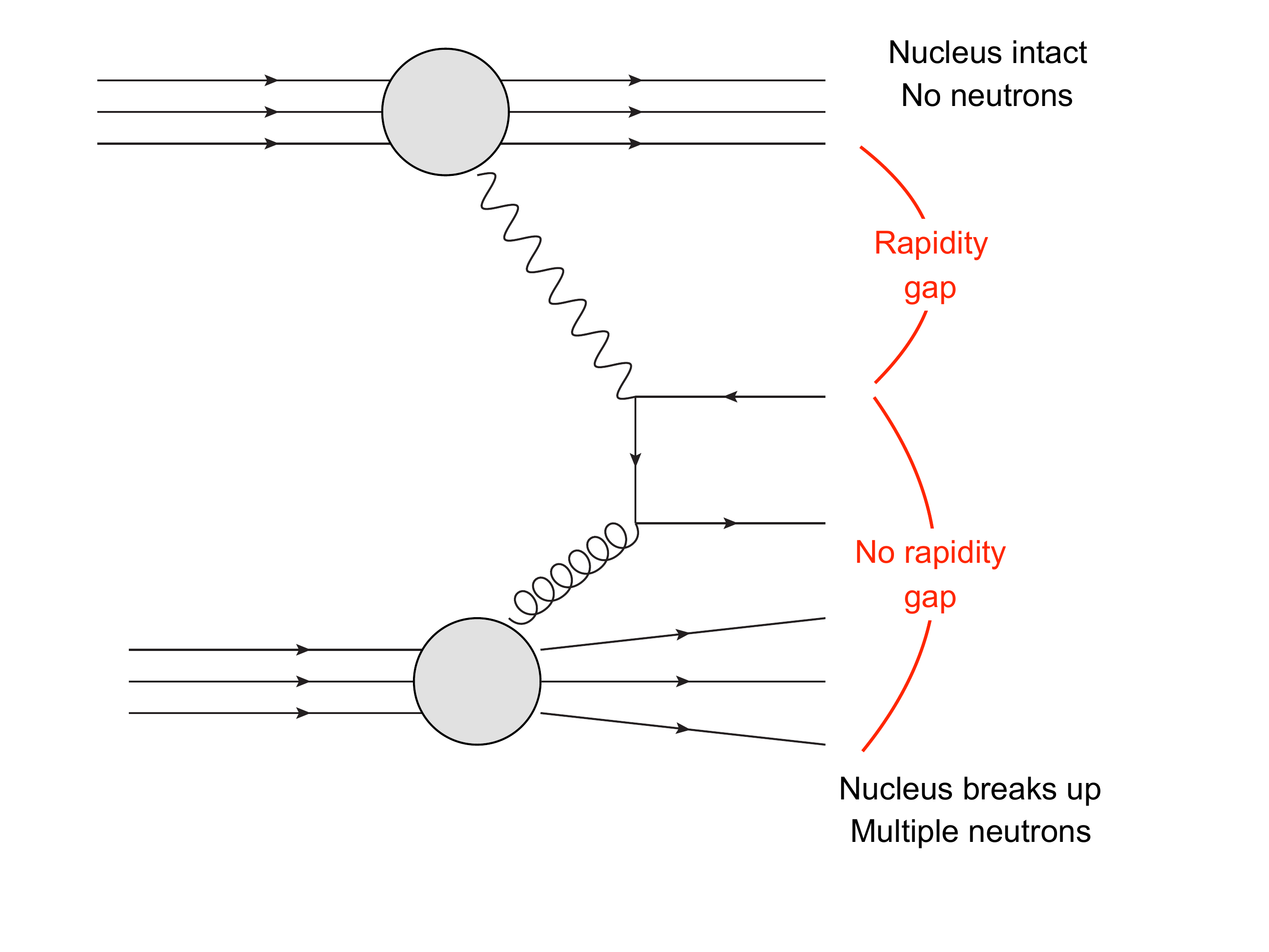}
\includegraphics[width=0.49\textwidth]{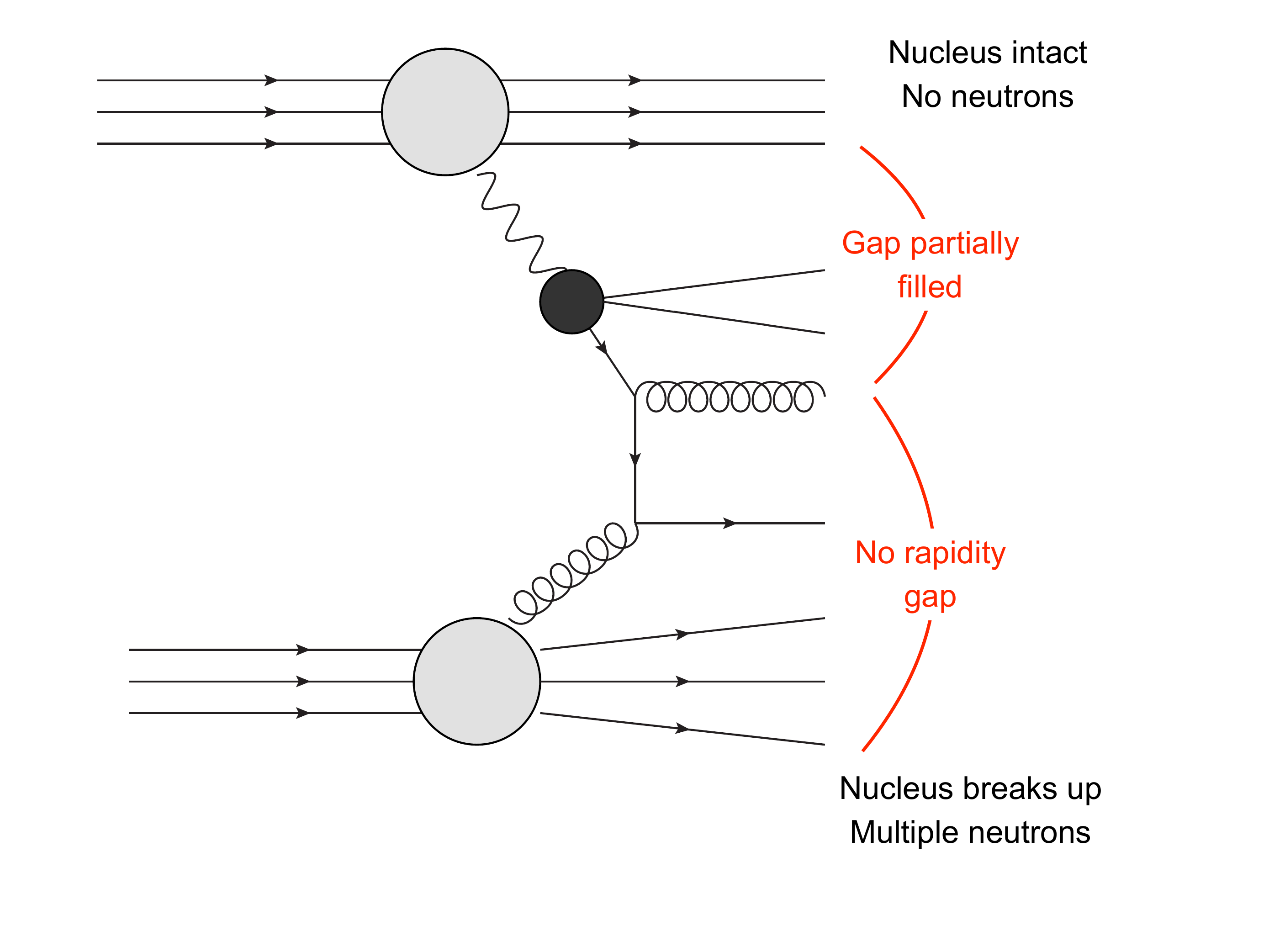}
\caption{Diagrams representing different types of photonuclear collisions and the general features of their event topologies. {\em Left:} the direct process, in which the photon itself interacts with the nucleus. {\em Right:} the resolved process, in which the photon fluctuates into a hadronic state. }
\label{fig:diagrams}
\end{figure}

At the LHC, previous measurements of ultraperipheral processes in Pb+Pb collisions include light-by-light scattering ($\gamma\gamma\to\gamma\gamma$)~\cite{HION-2016-05,HION-2018-19,Aad:2020cje,CMS-FSQ-16-012}, exclusive dilepton production ($\gamma\gamma\to{e}^+{e}^-$ and $\gamma\gamma\to\mu^+\mu^-$)~\cite{HION-2016-02,Sirunyan:2020vvm,Abbas:2013oua}, and the photonuclear production of various meson states ($\gamma+\mathrm{A}\to{h}+X$)~\cite{Abbas:2013oua,CMS-HIN-12-009,Acharya:2019vlb}.
In the photonuclear case, the photon may act as a point-like particle interacting with a parton in the nucleus (the `direct' case). However, the vector-meson dominance picture~\cite{Sakurai:1960ju,Bertulani:2005ru} suggests that the photon often fluctuates into a vector-meson state such as a $\rho$ or $\omega$ (the `resolved' case). In this case, the interaction proceeds as a meson--nucleus collision at an energy lower than that of the associated nucleon--nucleon collision. \Fig{\ref{fig:diagrams}} illustrates the direct and resolved photonuclear interactions. The photon--nucleon collision energy and the boost of the center-of-mass relative to the nucleus--nucleus rest frame depends on the photon energy and thus varies event to event. For photons with energies at the upper boundary of the coherence region, $E = 80$~\GeV, the resulting photon--nucleon center-of-mass energy is approximately $900$~\GeV. Thus photonuclear collisions may be used to probe the dynamics of a system with a novel energy and geometry compared to $pp$ or $p$+A collisions at the LHC, and to $e^+e^-$ or $ep$ collisions at LEP and HERA. Since photonuclear events are the photoproduction limit of deep inelastic scattering on nuclei, these measurements may also shed light on possible collective signatures at the future Electron Ion Collider~\cite{Accardi:2012qut,Shi:2020djm}.
 
This paper presents a measurement of azimuthal anisotropies obtained via two-particle correlations in photonuclear collisions, where such analyses have not previously been undertaken. The data were recorded using a trigger designed to select minimum-bias and high-multiplicity photonuclear events in 1.7~nb$^{-1}$ of Pb+Pb collisions at 5.02~\TeV\ per nucleon pair delivered by the LHC in 2018. Photonuclear event candidates are selected, and are distinguished from peripheral hadronic Pb+Pb events and other background events, using the topology of the distribution of particles in the event as measured in the zero-degree, forward, and  barrel calorimeters, as well as the tracking systems through the reconstruction of pseudorapidity gaps~\cite{STDM-2011-01,CMS-FSQ-12-005}. Properties of the selected events are compared with the expectations from Monte Carlo (MC) simulations of photonuclear processes.
 
\newcommand{\AtlasCoordFootnote}{
ATLAS uses a right-handed coordinate system with its origin at the nominal interaction point (IP)
in the center of the detector and the $z$-axis along the beam pipe.
The $x$-axis points from the IP to the center of the LHC ring,
and the $y$-axis points upwards.
Cylindrical coordinates $(r,\phi)$ are used in the transverse plane,
$\phi$ being the azimuthal angle around the $z$-axis.
The pseudorapidity is defined in terms of the polar angle $\theta$ as $\eta = -\ln \tan(\theta/2)$.
}

Two-particle correlations as a function of relative separation in azimuth ($\Delta\phi$) and pseudorapidity\footnote{\AtlasCoordFootnote} ($\Delta\eta$) are constructed for different selections of event charged-particle multiplicity and charged-particle kinematics. The non-flow contributions to the two-particle correlations (for example, jet correlations which give rise to a $\Delta\phi$ correlation structure) are suppressed by studying the correlations at large $\Delta\eta$, and the residual non-flow contribution is subtracted via the template-fitting method used extensively in prior ATLAS measurements~\cite{HION-2015-09, HION-2016-01,
HION-2018-23}. In the template method, the correlation in high-multiplicity events is described as a combination of the correlation in lower-multiplicity events plus a component modulated by $\cos(2\Delta\phi)$ (and higher order) Fourier terms.
The template method makes particular assumptions about how the non-flow component evolves with multiplicity. Although there are differences between the system explored in this measurement and those in previous two-particle correlation measurements, the sensitivity to the assumptions of the template method can be tested within the standard approach.
A test of the template method in simulated photonuclear events which do not include flow or initial-state correlation mechanisms is performed in Section~\ref{sec:result}.
The resulting magnitudes of the two-particle correlations are interpreted as arising from the product of global $v_{2}$ and $v_{3}$ values for individual particles, and are reported as a function of the reconstructed charged-particle multiplicity ($N_\mathrm{ch}^\mathrm{rec}$) and transverse momentum ($p_\mathrm{T}$). The results are compared with other small collision systems at the LHC, and theoretical expectations from initial- and final-state physics mechanisms are discussed.

\section{ATLAS detector and data sample}
\label{sec:detector}
 
The ATLAS detector~\cite{PERF-2007-01}
covers nearly the entire solid angle around the collision point.
The detector systems used for the measurements presented in this paper include the inner tracking detector, the electromagnetic (EM) and hadronic calorimeters, the zero-degree calorimeter (ZDC), and the trigger and data acquisition systems. The detector halves at positive and negative $z$-values are referred to as the A and C sides, respectively.
 
The inner-detector system is immersed in a \SI{2}{\tesla} axial magnetic field
and provides charged-particle tracking in the range $|\eta| < 2.5$. The high-granularity silicon pixel detector covers the vertex region and typically provides four measurements per track.
The innermost layer, the insertable B-layer~\cite{ATLAS-TDR-19,ATLAS-TDR-19-addm,PIX-2018-001}, has been operating as a part of the silicon pixel detector since 2015.
 
The pixel detector is followed by the silicon microstrip tracker~(SCT) which usually provides four measurement points per track.
These silicon detectors are complemented by the transition radiation tracker,
which enables radially extended track reconstruction up to $|\eta| = 2.0$.
 
The calorimeter system covers the pseudorapidity range $|\eta| < 4.9$.
Within the region $|\eta|< 3.2$, electromagnetic calorimetry is provided by barrel and
endcap high-granularity lead/liquid-argon (LAr) calorimeters,
with an additional thin LAr presampler covering $|\eta| < 1.8$.
Hadronic calorimetry is provided by a steel/scintillator-tile calorimeter,
segmented into three barrel structures within $|\eta| < 1.7$, and two copper/LAr hadronic endcap calorimeters.
The solid angle coverage is extended with the forward calorimeter (FCal), composed of copper/LAr and tungsten/LAr modules
optimized for EM and hadronic measurements, respectively. The FCal detectors cover the regions $3.2 < \left|\eta\right| < 4.9$.
 
The minimum-bias trigger scintillator
(MBTS) detects charged particles over
$2.07<|\eta|<3.86$ using two hodoscopes of 12 counters positioned at
$z=\pm$3.6~m.
The ZDC consists of EM and hadronic sections and plays a key role in identifying UPC events in heavy-ion collisions by primarily detecting neutrons resulting from the breakup of one or both nuclei. The ZDC modules are located at $z=\pm\si{140}{~\mathrm{m}}$
from the IP. They measure neutral particles at pseudorapidities $|\eta| \geq 8.3$ and consist of layers of alternating
quartz rods and tungsten plates.
 
A two-level trigger system~\cite{TRIG-2016-01} is used to select events.
The first-level trigger (L1) is implemented in hardware and uses a subset of the detector information
to restrict the accepted rate to at most \SI{100}{\kilo\hertz}.
This is followed by a software-based high-level trigger (HLT) stage that
reduces the accepted event rate to 1--\SI{4}{\kilo\hertz}
depending on the data-taking conditions during 2018 Pb+Pb operations.

The measurements presented in this paper were performed using the $\sqrt{s_\mathrm{NN}} = 5.02$~\TeV\ Pb+Pb dataset collected with a variety of triggers in 2018, with a total integrated luminosity of 1.7~nb$^{-1}$. Photonuclear candidate events were first selected by the trigger by requiring one ZDC side (referred to as the Pb-going side) to have a minimum amount of energy at L1, $E > 1$~\TeV, consistent with the presence of one or more neutrons. The other side (referred to as the photon-going side) was required to have an energy below a maximum-energy cutoff, $E < 1$~\TeV, consistent with no neutrons. The cut value of $E = 1$~\TeV\ is several multiples of the energy resolution away from the single-neutron peak at $E = 2.5$~\TeV~\cite{HION-2016-02}. Thus, the selected topology is referred to as ``0nXn'' in the figures in this paper. Events were also required to satisfy an upper bound of $200$~\GeV\ on the total transverse energy deposited in the calorimeter at L1, for further rejection of hadronic Pb+Pb events.
 
After these requirements at L1, events had to pass either a minimum-bias (MB) trigger or one of several high-multiplicity triggers (HMT) with further requirements at the HLT stage. The MB trigger was defined by requiring at least one reconstructed online track with $p_\mathrm{T} > 0.4$~\GeV. The HMTs were defined by requiring a larger number of tracks originating from the same vertex.
The primary vertex, defined as the one with the largest scalar sum of $p_\mathrm{T}^2$ values of associated tracks, was used.
An HMT with a threshold of 15 tracks was defined by also requiring a hit in the MBTS at L1. Two HMTs with thresholds of 25 and 35 tracks were defined by requiring a minimum amount of energy in the calorimeter at L1. Finally, all HMTs had an additional HLT requirement of less than 5~\GeV\ of transverse energy in the photon-going FCal, which rejected a large fraction of the 0nXn peripheral Pb+Pb background events. The 35-track HMT sampled the full luminosity of 1.7~nb$^{-1}$ during data-taking. The 25-track HMT, 15-track HMT, and the MB trigger were configured with increasingly larger prescale factors, sampling $1.6$~nb$^{-1}$, $0.13$~nb$^{-1}$, and $1.0~\mu$b$^{-1}$ of data respectively.

\section{Monte Carlo simulation}
\label{sec:MC}
 
Monte Carlo (MC) simulations of the relevant physics processes were used to understand the performance of the detector, and provide distributions to be compared with the data. For all the generators and configurations listed below, the simulated events were passed through a $\GEANT4$ simulation~\cite{Agostinelli:2002hh,SOFT-2010-01} of the detector and reconstructed under the same conditions as the data.
 
A sample of one million peripheral Pb+Pb events was generated using \textsc{Hijing} v1.383~\cite{Wang:1991hta} with impact parameters in the range $10 < b <20$~fm. These events were used to determine the charged-particle tracking efficiencies for the measurement, and to model the event properties of low-multiplicity, hadronic Pb+Pb events.
 
Several generators were used to simulate photonuclear interactions.
For each of the cases below, events were generated with different minimum requirements on $N_\mathrm{ch}^\mathrm{rec}$ to provide good statistical coverage over the $N_\mathrm{ch}^\mathrm{rec}$ range accessed in data.
First, the distribution of photon flux for $^{208}$Pb beams at the LHC was calculated using \textsc{STARlight}~\cite{Klein:2016yzr}. The flux distribution was passed to a multipurpose generator based on the Dual Parton Model (DPM) and referred to as \textsc{DPMJET}-III~\cite{Engel:1994vs,Roesler:2000he}, which simulated direct and resolved photon--lead ($\gamma$+Pb) interactions at the generator level, followed by a full $\GEANT4$ simulation of the ATLAS detector. Thirteen million $\gamma$+Pb events were simulated in this way. Additionally, the flux from \textsc{STARlight} was used to simulate two million $\gamma$+$p$ events using \textsc{DPMJET}-III, where the energy of the proton was set equal to the energy per nucleon in the Pb beams. Finally, for an alternative description of these processes, \PYTHIAV{8.240}~\cite{Sjostrand:2014zea} configured with the \textsc{NNPDF23LO} parton distribution functions~\cite{Ball:2012cx} and \textsc{A14} set of tuned parameters~\cite{ATL-PHYS-PUB-2014-021} was used to generate twelve million $\gamma$+$p$ events. The photon flux in \PYTHIA was reweighted at the event level to match that calculated by \textsc{STARlight}, and the simulation was configured to include both the direct and resolved photon interactions.
 
\section{Photonuclear event selection}
 
\subsection{Reconstruction and event selection}

Given the low particle multiplicities of UPC events, the charged-particle track and calorimeter energy-cluster reconstruction procedures follow those optimized for \pp data-taking~\cite{STDM-2010-06,Salzburger:2018442}.  Reconstructed charged-particle tracks are used in the analysis if they satisfy quality criteria as outlined in Ref.~\cite{STDM-2015-02}. These include criteria for the minimum number of hits in the pixel and SCT detectors. Tracks are further required to have $\pt>0.4~\GeV$, $|\eta|<2.5$, and distances of closest approach to the primary reconstructed vertex in the longitudinal and transverse directions of less than $1.5$~mm each.  Clusters in the range $|\eta|<4.9$ are constructed from topologically connected groups of calorimeter cells~\cite{PERF-2014-07}. They are required to have $\pt>0.2~\GeV$ and to meet the significance criteria for the measured energy as outlined in Ref.~\cite{STDM-2011-01} to suppress the contribution from electronic noise fluctuations.
 
Events are required to have a reconstructed primary vertex within $|z| < 90$~mm. Events compatible with multiple in-time interactions are tagged by the presence of multiple reconstructed vertices. These comprise less than 1\% of the sample and are rejected. Events must pass a version of the ZDC-based energy selection described in Section~\ref{sec:detector} which uses the same threshold but an improved calibration for the ZDC energies determined after data-taking.
Events are further characterized by their charged-particle multiplicity, $N_\mathrm{ch}^\mathrm{rec}$, which is defined as in previous ATLAS measurements of correlations in small systems~\cite{HION-2015-09,HION-2016-01,HION-2017-11} as the total number of reconstructed tracks, without efficiency correction, with $p_\mathrm{T} > 0.4$~\GeV\ and $\left|\eta\right| < 2.5$.
For the correlation analysis described in Section~\ref{sec:2PC}, events with a given $N_\mathrm{ch}^\mathrm{rec}$ were included in the analysis if they were selected by any of the HMTs which was more than 80\% efficient for events with this $N_\mathrm{ch}^\mathrm{rec}$ value.
For the study of event properties in Section~\ref{ssec:event_prop}, each $N_\mathrm{ch}^\mathrm{rec}$ range was only populated by the highest-sampled-luminosity HMT which was more than 99\% efficient over the entire $N_\mathrm{ch}^\mathrm{rec}$ range. In either case, events with $N_\mathrm{ch}^\mathrm{rec} < 15$ were populated only by the MB trigger.

Reconstructed pseudorapidity gap quantities, constructed using the tracks and clusters in each event, are used to distinguish between different physics processes such as photonuclear collisions, low-activity (peripheral) hadronic Pb+Pb collisions, and $\gamma\gamma\rightarrow X$ processes. The requirement of a rapidity gap above a minimum value in the photon-going direction can efficiently remove peripheral Pb+Pb events. Rather than the traditional pseudorapidity gap quantity~\cite{STDM-2011-01}, which determines the pseudorapidity difference between the edge of the detector and closest particle, an alternative `sum-of-gaps' definition is used, which adds together contiguous gaps separated by particle production concentrated in a narrow pseudorapidity regions. This alternative definition is used to retain a large selection efficiency for resolved photon events which may break up a large gap with a hadronic fragment localized in pseudorapidity (see Figure~\ref{fig:diagrams}).
The quantities \gapDef and \gapDefA correspond to the sum-of-gaps calculated in the photon-going and nucleus-going halves of the detector, respectively.
They are calculated by first sorting the tracks and clusters in $\eta$. The differences in $\eta$ between adjacent particles, $\Delta\eta$, are included in the sum if they are larger than $0.5$. The value of $0.5$ was observed in simulation to retain good efficiency for resolved photon events. The position at $\eta=0$ is treated as if it were a particle which ends the sum-of-gaps calculation. Thus, $\gapDef$ and $\gapDefA$ range from $0$ to $4.9$.
 
In addition to the sum-of-gaps quantities defined above, the traditional edge gap quantity, constructed using clusters, $\Delta\eta_\mathrm{edge}^\mathrm{cluster}$, is defined as the pseudorapidity difference between the edge of the detector on the photon-going side, at $\eta = +4.9$, and the first reconstructed cluster. In the mixed-event construction discussed in Section~\ref{sec:2PC}, events are classified by their $\Delta\eta_\mathrm{edge}^\mathrm{cluster}$ as a way to characterize their overall topology without using charged-particle tracks.
 
\subsection{Event properties}
\label{ssec:event_prop}
 
\begin{figure}[t]
\centering
\includegraphics[width=0.49\textwidth]{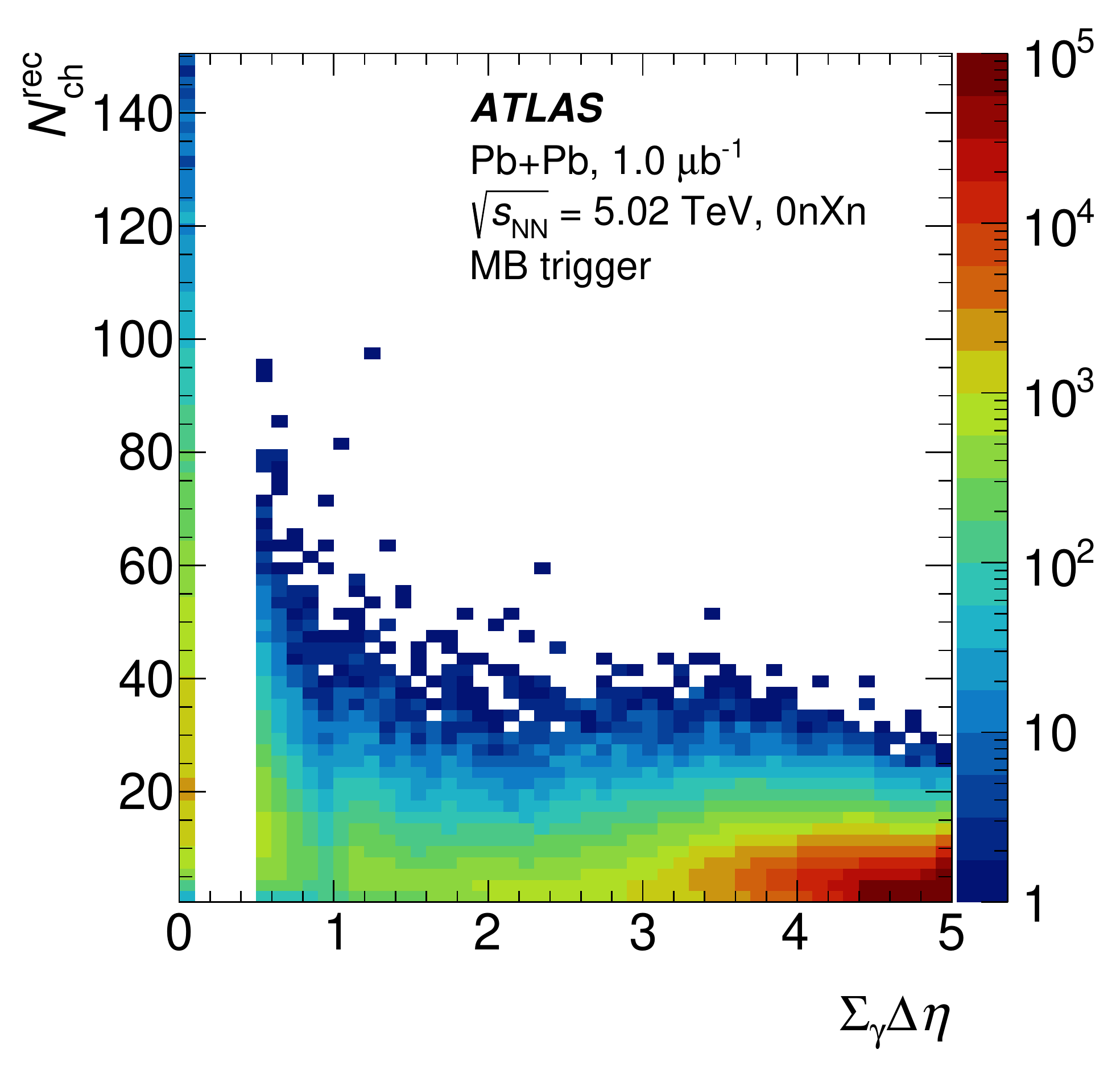}
\includegraphics[width=0.49\textwidth]{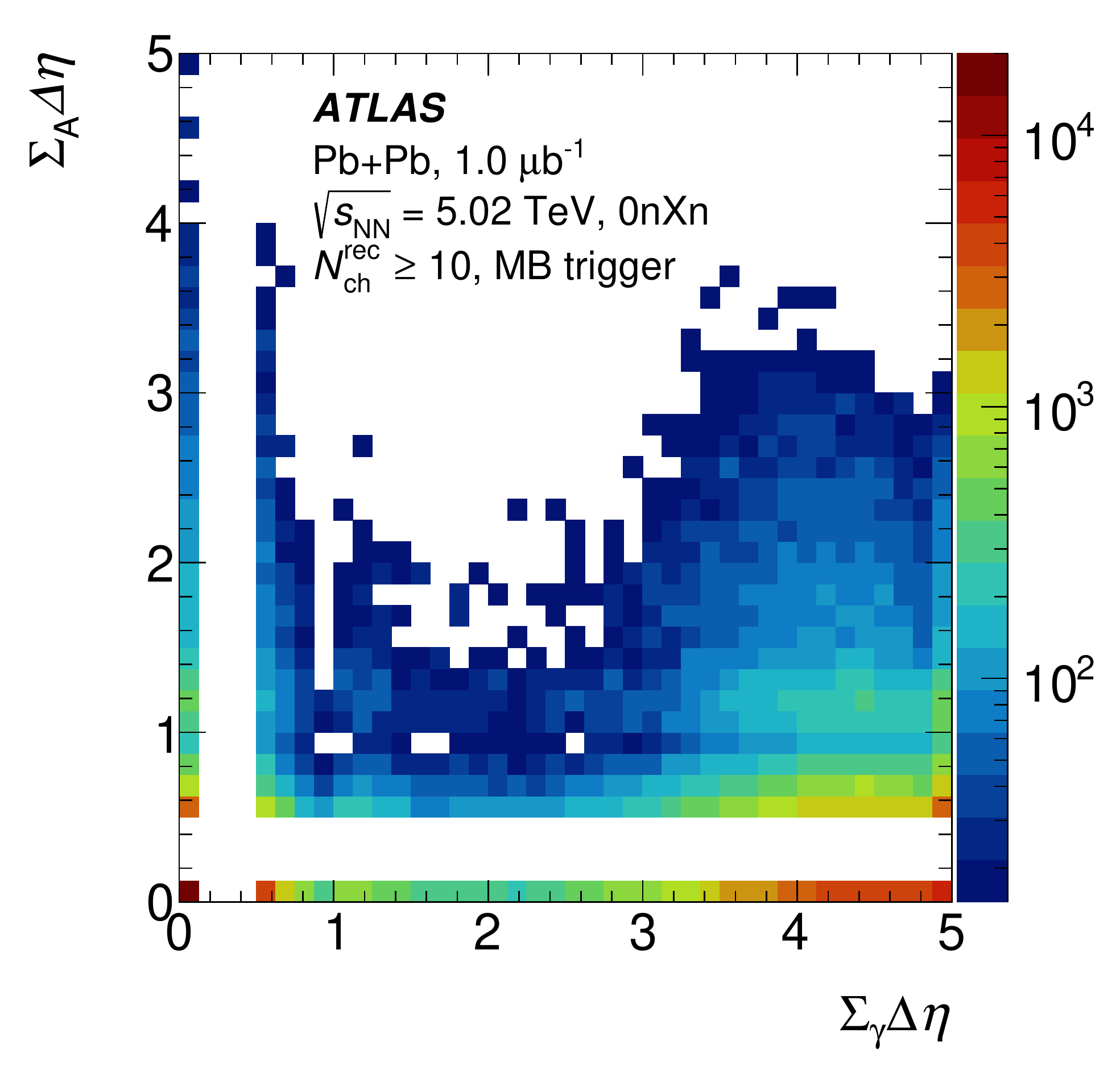}
\caption{Left: Correlation of $N_\mathrm{ch}^\mathrm{rec}$ and \gapDef for events selected by the MB trigger before application of gap-based event selection. Right: Correlation of \gapDef and \gapDefA for events selected by the MB trigger, with $N_\mathrm{ch}^\mathrm{rec} > 10$. }
\label{fig:gapCorr}
\end{figure}
 
\Fig{\ref{fig:gapCorr}} (left) shows the distribution of \gapDef and $N_\mathrm{ch}^\mathrm{rec}$ values for events recorded with the MB trigger.
The \gapDef can have a value of zero or fall into the range $0.5 \le \gapDef \le 4.9$, with the lower value determined by the minimum gap size included in the sum.
Events resulting from hadronic Pb+Pb collisions are at small \gapDef and have a broad multiplicity distribution, whereas events from photon-induced processes are at large \gapDef and have an $N_\mathrm{ch}^\mathrm{rec}$ distribution that falls steeply.  \Fig{\ref{fig:gapCorr}} (right) shows the distribution of \gapDef and \gapDefA values.  Most hadronic Pb+Pb collisions have a broad pseudorapidity distribution of particles and hence have both small \gapDef and small \gapDefA. In contrast, the photonuclear events have large \gapDef but smaller values of \gapDefA. There is no significant yield of events with a large sum-of-gaps on both sides, which could signal the presence of photon--photon hadronic process backgrounds.  The signal events for photonuclear collisions are defined by \gapDef > 2.5.  No event selection is made on \gapDefA.
 
\begin{figure}[t]
\centering
\includegraphics[width=0.49\textwidth]{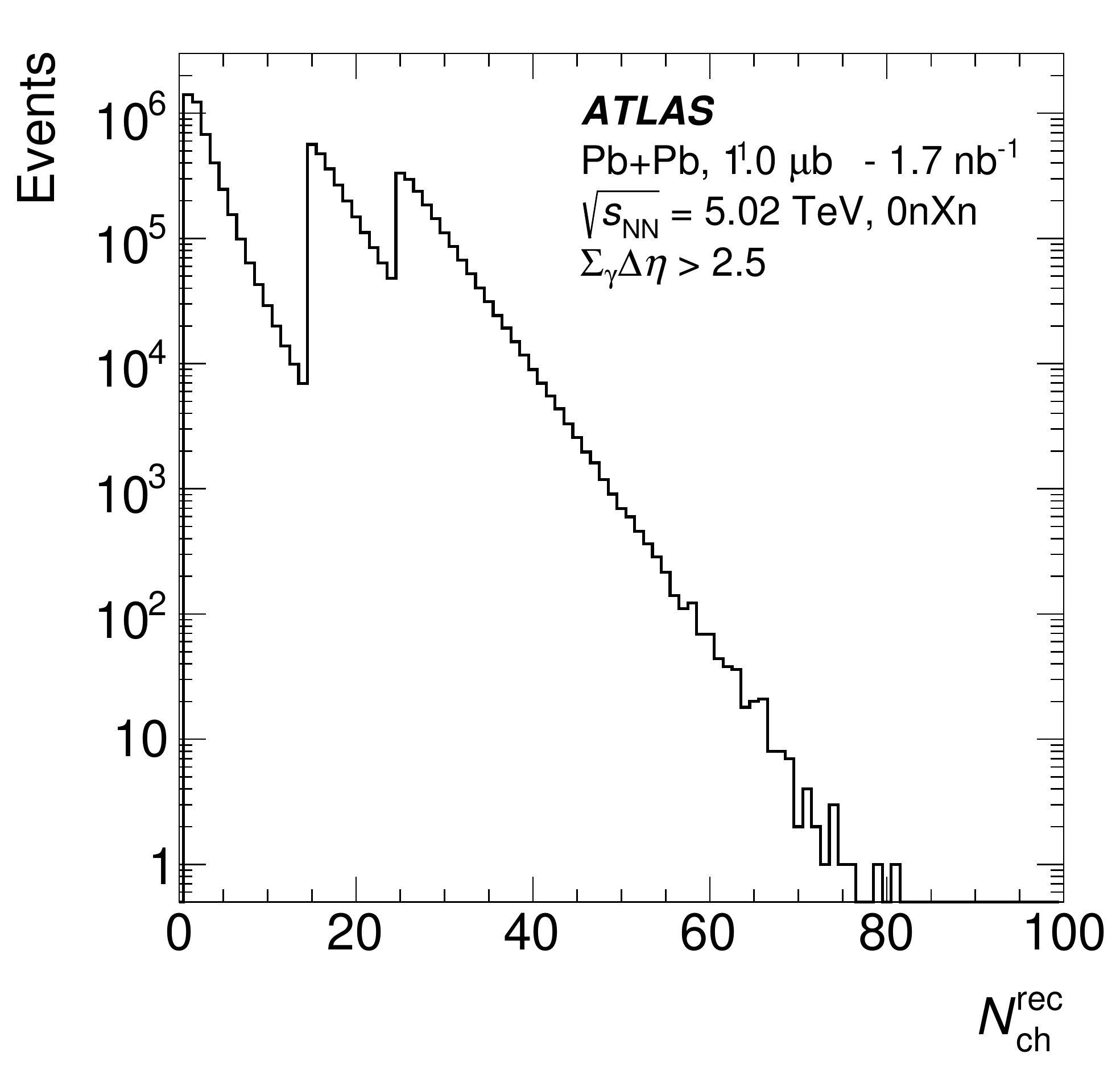}
\includegraphics[width=0.49\textwidth]{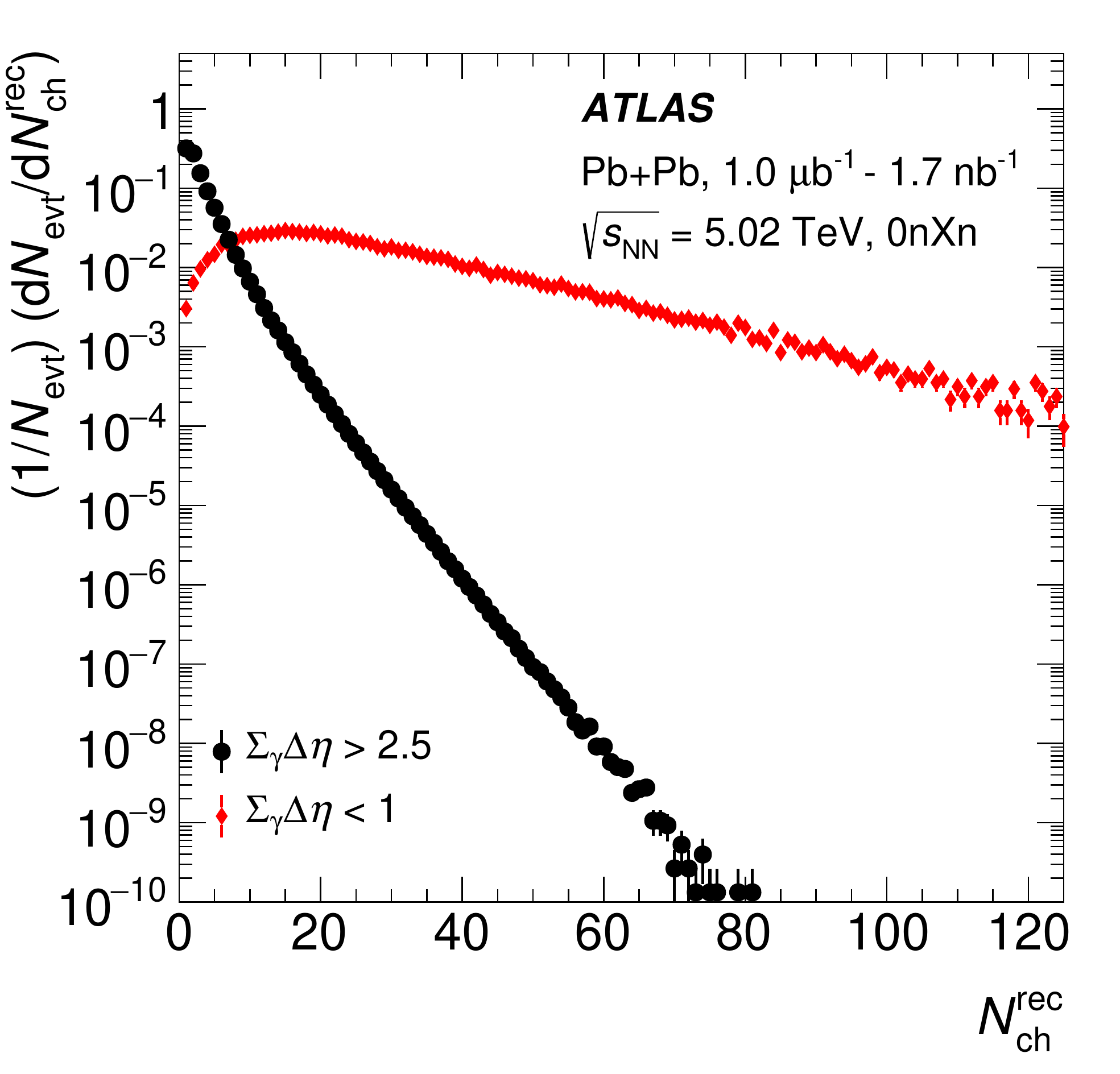}
\caption{ Left: Distribution of event charged-particle multiplicity, $N_\mathrm{ch}^\mathrm{rec}$, for the photonuclear event selection from multiple triggers, without prescale correction. Right: Distribution of event charged-particle multiplicity, $N_\mathrm{ch}^\mathrm{rec}$, for the photonuclear event selection (black) with corrections for trigger efficiency and prescale factors, and an alternative selection intended to select hadronic Pb+Pb events (red, see text). }
\label{fig:multUPC}
\end{figure}
 
\Fig{\ref{fig:multUPC}} summarizes the multiplicity distribution observed in data after the event selection described above.
The $N_\mathrm{ch}^\mathrm{rec}$ distribution for events passing the photonuclear event selection is shown in the left panel of Figure~\ref{fig:multUPC}. The sawtooth pattern arises from the inclusion of the HMTs at different $N_\mathrm{ch}^\mathrm{rec}$ thresholds as described above. In each $N_\mathrm{ch}^\mathrm{rec}$ range, the selected photonuclear events have a steeply falling multiplicity distribution.
The $N_\mathrm{ch}^\mathrm{rec}$ distribution for photonuclear events, fully corrected for the different luminosities sampled by the triggers and the $N_\mathrm{ch}^\mathrm{rec}$-dependent trigger efficiency, is shown using black circles in the right panel of Figure~\ref{fig:multUPC}. It is compared with the distribution from events with $\gapDef < 1$, which mostly selects Pb+Pb hadronic events, which are backgrounds in this analysis.  The selected photonuclear events have a significantly more steeply falling multiplicity distribution.
 
\begin{figure}[t]
\centering
\includegraphics[width=0.49\textwidth]{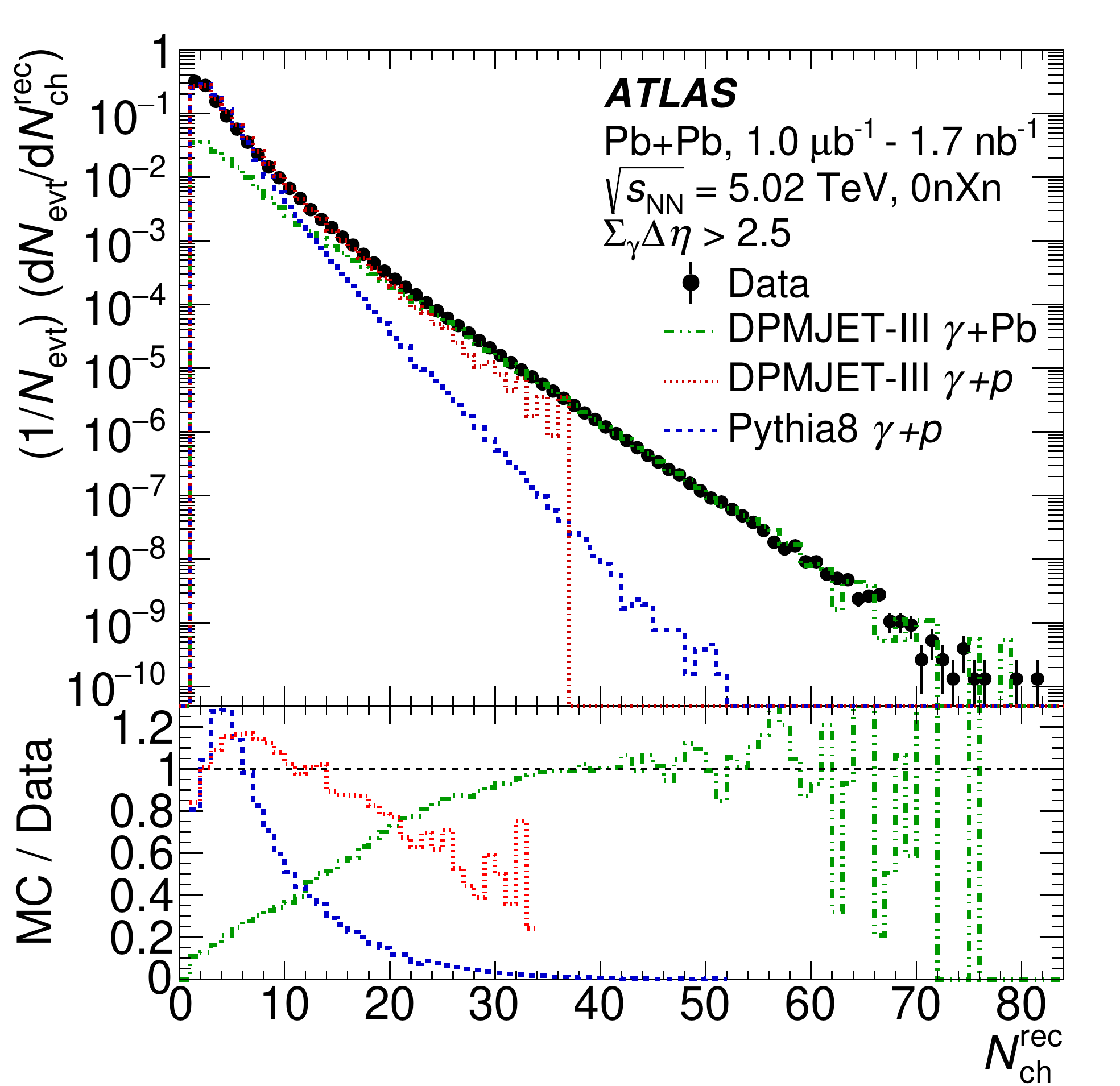}
\includegraphics[width=0.49\textwidth]{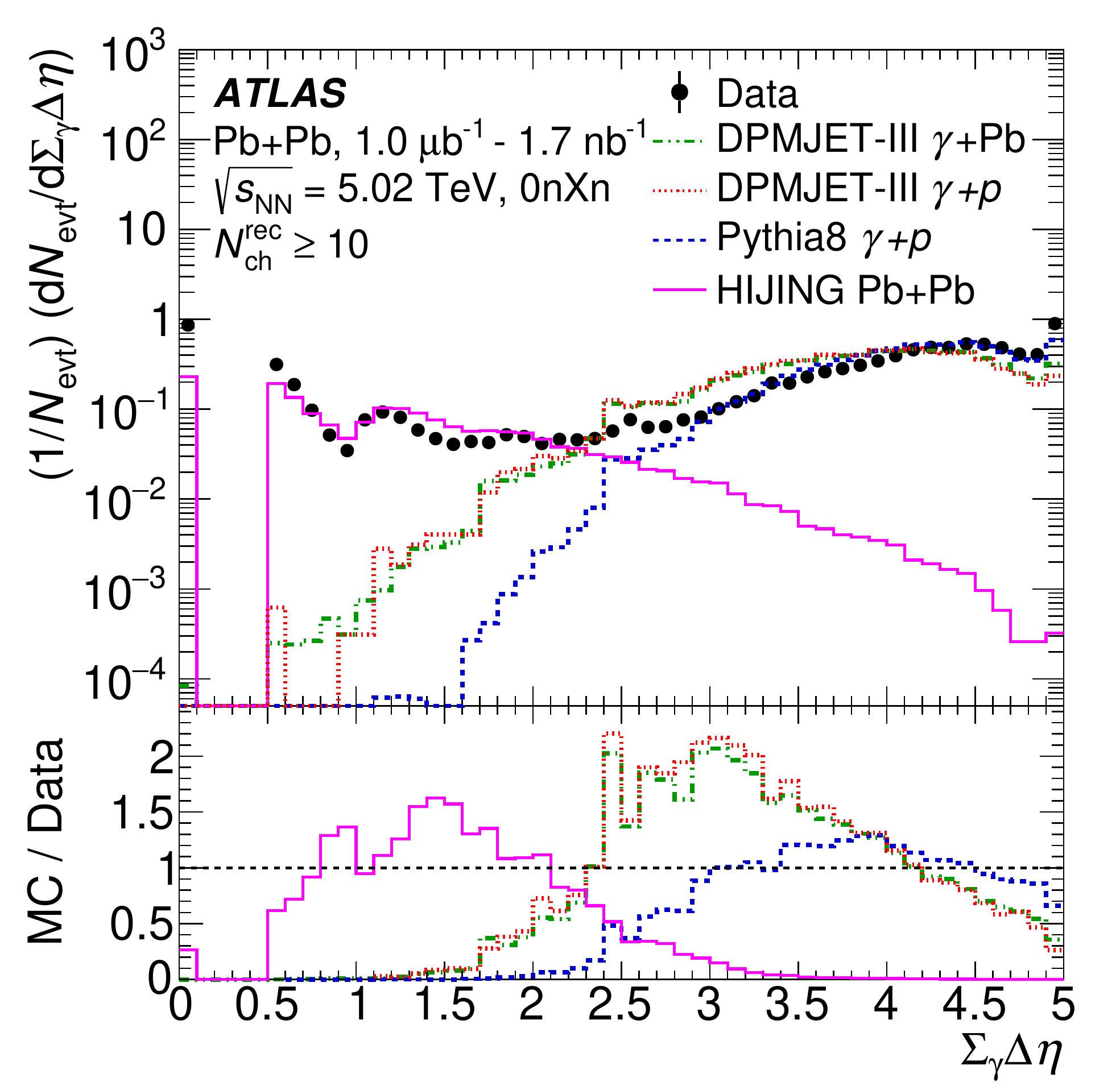}
\caption{Left: $N_\mathrm{ch}^\mathrm{rec}$ distribution in data, corrected for trigger and reconstruction efficiency and normalized per event (black points), compared with that in \textsc{DPMJET}-III $\gamma$+Pb (dot-dashed green histogram), \textsc{DPMJET}-III $\gamma$+$p$ (dotted red histogram), and \PYTHIA $\gamma$+$p$ (dashed blue histogram). The bottom panel shows the ratios of the MC distributions to the data distributions.
Right: \gapDef distribution in data for $N_\mathrm{ch}^\mathrm{rec} \ge 10$ (black points), normalized per event, and compared with that in \textsc{DPMJET}-III $\gamma$+Pb (dot-dashed green histogram), \PYTHIA $\gamma$+$p$ (dashed blue histogram), peripheral \textsc{Hijing} Pb+Pb (solid magenta histogram), and \textsc{DPMJET}-III $\gamma$+$p$ (dotted red histogram).}
\label{fig:DnDeta_Ntrk}
\end{figure}

\Fig{\ref{fig:DnDeta_Ntrk}} compares the multiplicity and sum-of-gap distributions in data and simulation.
The left panel of Figure~\ref{fig:DnDeta_Ntrk} shows the total $N_\mathrm{ch}^\mathrm{rec}$ distribution, corrected for the trigger efficiency and the different integrated luminosities sampled by the HMTs. This is compared with three distributions from MC generators, with the same gap-based selection requirements as the data. The ZDC selection is applied to data but not to the MC samples as the generators do not model the nuclear-breakup processes relevant for forward neutron spectators.
Including a ZDC requirement in MC events may thus impact the distributions if particle production in the detector is correlated with the nuclear fragmentation in the forward region.
Despite the limitations in the modeling of forward neutrons, a generator-level check requiring neutrons in the ZDC acceptance was performed and found not to impact the level of agreement between data and \textsc{DPMJET}-III.
The distributions in \PYTHIA and \textsc{DPMJET}-III $\gamma$+$p$ are normalized to have the same integral as the data over the full $N_\mathrm{ch}^\mathrm{rec}$ range. The models show good agreement with the data at low $N_\mathrm{ch}^\mathrm{rec}$, but systematically predict too low a relative yield at higher $N_\mathrm{ch}^\mathrm{rec}$. \textsc{DPMJET}-III $\gamma$+Pb does not describe the full distribution, but has been normalized to have the same integral as the data in $N_\mathrm{ch}^\mathrm{rec} > 35$ to highlight its good agreement in this region over many orders of magnitude. With this normalization, \textsc{DPMJET}-III $\gamma$+Pb systematically predicts too low a relative yield for $N_\mathrm{ch}^\mathrm{rec} < 15$. This comparison either suggests the presence of other, non-photonuclear, processes in data at such low $N_\mathrm{ch}^\mathrm{rec} < 10$ values, or points to the need for improved modeling of this region in the simulation.
 
The right panel of Figure~\ref{fig:DnDeta_Ntrk} shows the reconstructed \gapDef distributions in data and simulation for selected events with $N_\mathrm{ch}^\mathrm{rec} > 10$, without the $\gapDef > 2.5$ requirement. Structures in the distributions correspond to transitions between detector subsystems and the change in the detector response as a function of $\eta$. At large \gapDef values $\geq 2.5$, the shape of the distribution in data is qualitatively similar to that in \textsc{DPMJET}-III $\gamma$+Pb and Pythia $\gamma$+$p$ simulation. However, the distributions in the simulated photonuclear events decrease at smaller \gapDef values, while the distribution in data rises. At low \gapDef, the shape in data is qualitatively similar to that in peripheral \textsc{Hijing} Pb+Pb events. This comparison suggests that the trigger-selected events contain a mixture of peripheral Pb+Pb events and genuine photonuclear events, with the latter dominant at $\gapDef > 2.5$. The possible impact of residual peripheral Pb+Pb events in the set of selected events is discussed in Section~\ref{sec:syst}.
 
\begin{figure}[t]
\centering
\includegraphics[width=0.49\textwidth]{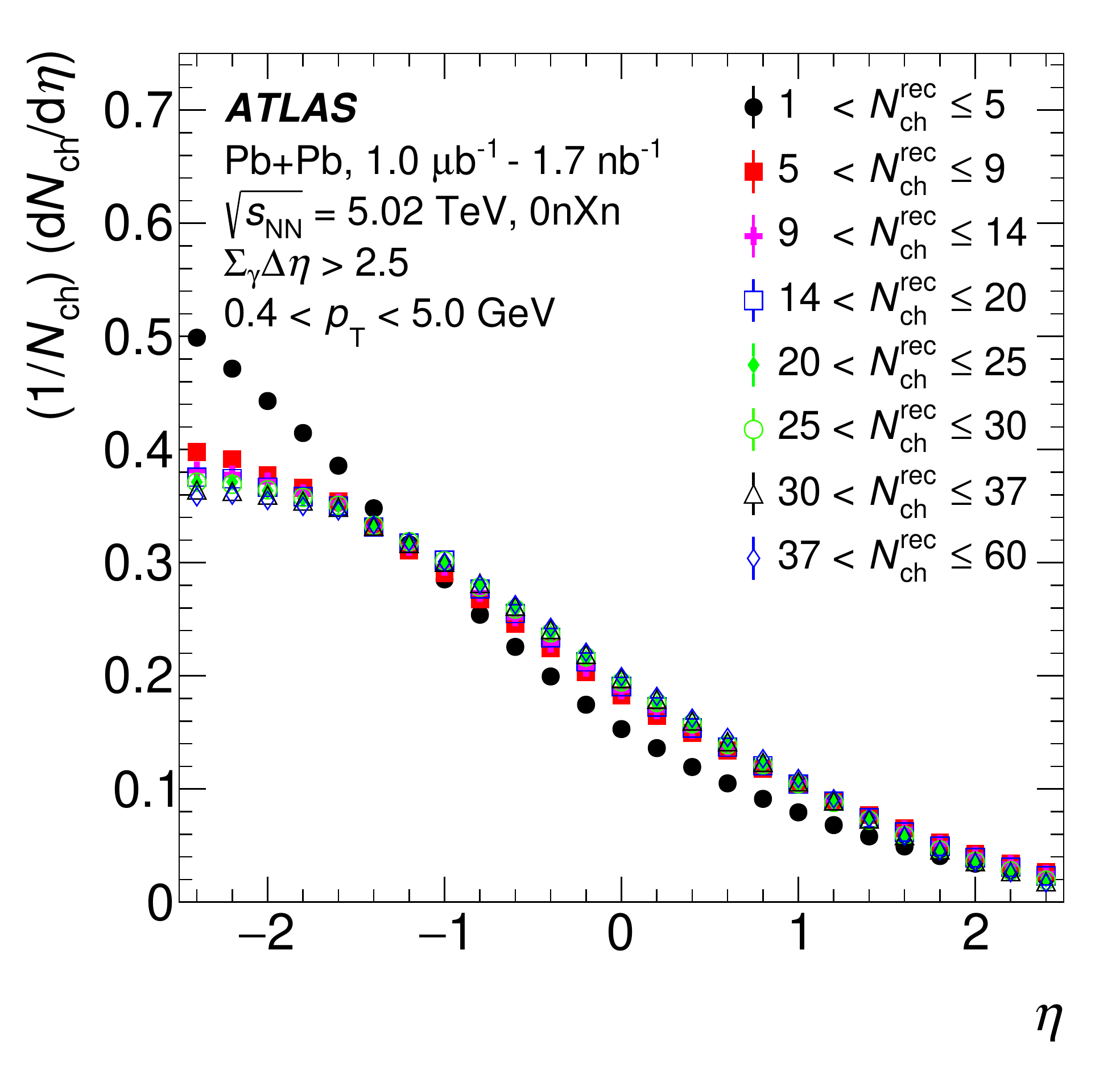}
\includegraphics[width=0.49\textwidth]{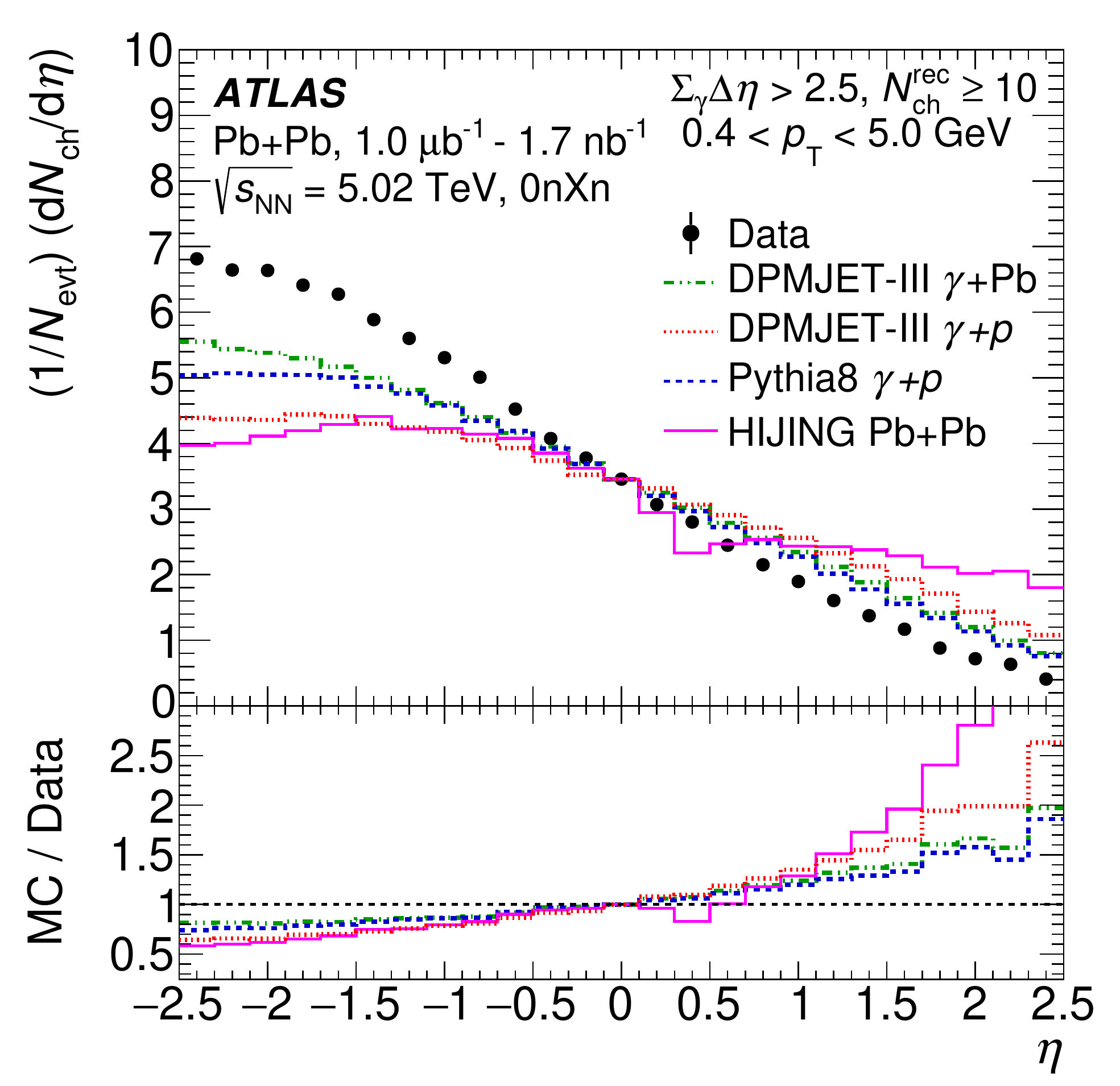}
\caption{ Left: Charged-particle pseudorapidity distribution, $dN_\mathrm{ch}/d\eta$, in selected $N_\mathrm{ch}^\mathrm{rec}$ ranges.   The distributions are normalized to the same integral and are shown in arbitrary units. Here, positive and negative $\eta$ denote the photon-going and nucleus-going directions, respectively.
Right: $dN_\mathrm{ch}/d\eta$ distribution in data for $N_\mathrm{ch}^\mathrm{rec} > 10$ (black points), normalized per event, and compared with that in \textsc{DPMJET}-III $\gamma$+Pb (dot-dashed green histogram), \PYTHIA $\gamma$+$p$ (dashed blue histogram), peripheral \textsc{Hijing} Pb+Pb (solid magenta histogram), and \textsc{DPMJET}-III $\gamma$+$p$ (dotted red histogram) with the same reconstruction-level selection as the data. All distributions have been normalized to have the same value as \textsc{DPMJET}-III $\gamma$+Pb at $\eta = 0$.}
\label{fig:MC_comp}
\end{figure}
 
\Fig{\ref{fig:MC_comp}} compares the charged-particle pseudorapidity distribution, $dN_\mathrm{ch}/d\eta$, in data and simulation. The left panel shows the $dN_\mathrm{ch}/d\eta$ in data, for charged particles with $0.4 < p_\mathrm{T} < 5$~\GeV, for multiple $N_\mathrm{ch}^\mathrm{rec}$ selections in photonuclear events. The distributions are corrected for tracking efficiency on a per-track basis, which ranges from $0.7$--$0.9$ depending on track $\eta$ and $p_\mathrm{T}$. To compare the relative shapes between $N_\mathrm{ch}^\mathrm{rec}$ selections, the distributions are each normalized to have an integral of one. In all cases, the pseudorapidity distributions are strongly asymmetric, peaking at $\eta = -2.5$ (the nucleus-going direction) and then monotonically decreasing until $\eta = +2.5$ (the photon-going direction). This overall shape is even more asymmetric than that observed in, for example, $p$+Pb collisions at the LHC~\cite{HION-2012-15}.
The $dN_\mathrm{ch}/d\eta$ shape remains similar over a wide range of multiplicities ($N_\mathrm{ch}^\mathrm{rec} > 10$).  The selection with $N_\mathrm{ch}^\mathrm{rec} < 5$ results in an even more strongly asymmetric  $dN_\mathrm{ch}/d\eta$ shape, suggesting that it may include photonuclear events at significantly lower energies or other processes. Additionally, as discussed above, the region $N_\mathrm{ch}^\mathrm{rec} < 10$ is not well described by simulation, and may contain a significant contribution from background events. On the basis of these observations, the two-particle correlation analysis is performed using events with $N_\mathrm{ch}^\mathrm{rec} \ge 10$.  Additionally, an upper bound of $N_\mathrm{ch}^\mathrm{rec} = 60$ is used to remove the region with very few events.
 
The right panel in Figure~\ref{fig:MC_comp} compares the pseudorapidity distribution of particles with $0.4 < p_\mathrm{T} < 5$~\GeV\ in data for $N_\mathrm{rec}^\mathrm{ch} \ge 10$ with the four sets of simulated events, which have the same reconstruction-level selections as the data. While all three generators considered here systematically fail to predict a steep enough slope for the $dN_\mathrm{ch}/d\eta$ distribution, the distribution for \textsc{DPMJET}-III $\gamma$+Pb events is the most similar to that observed in data.
 
\begin{figure}[t]
\centering
\includegraphics[width=0.55\textwidth]{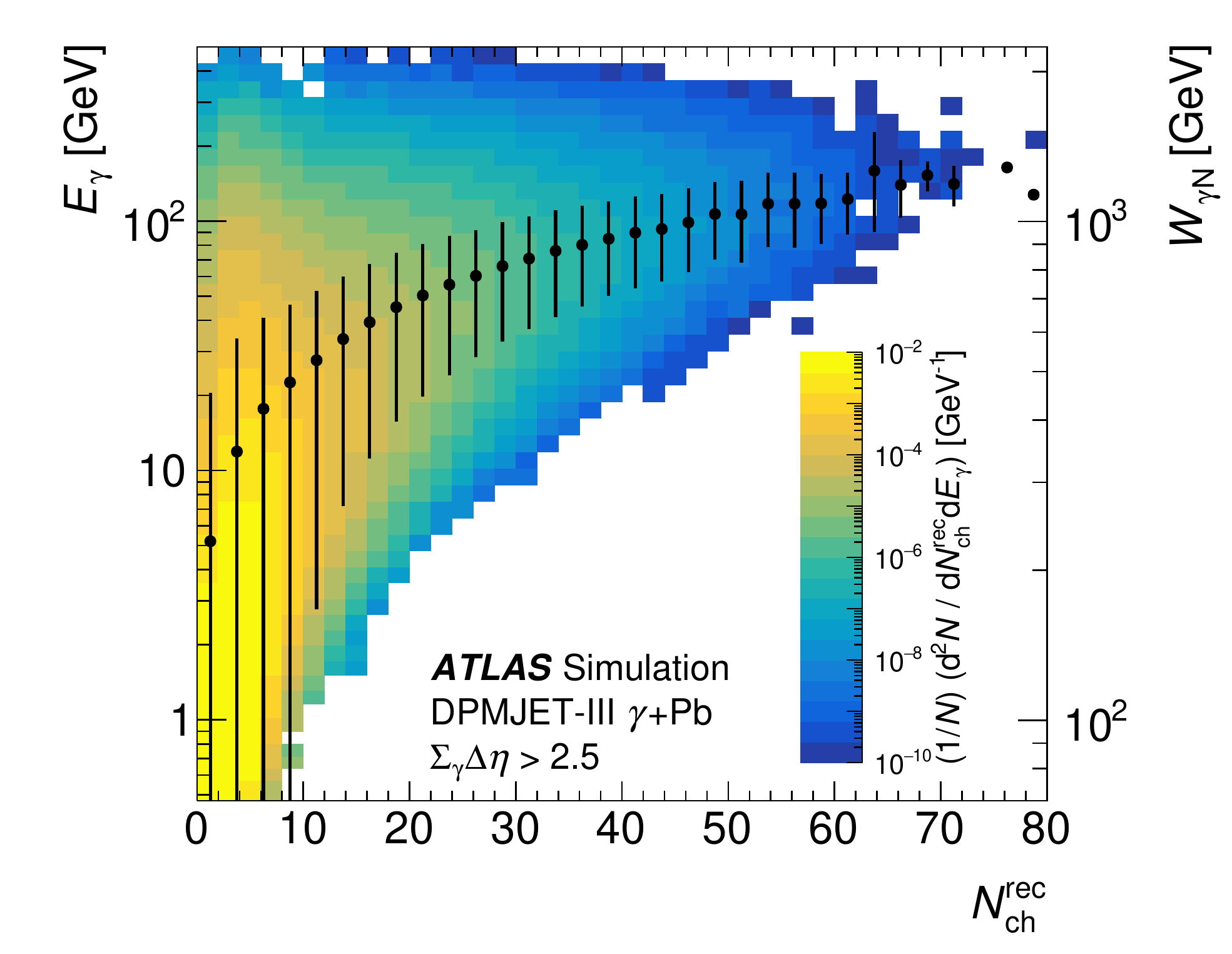}
\caption{Correlation between photon energy $E_{\gamma}$ and event multiplicity $N_\mathrm{ch}^\mathrm{rec}$ in simulated \textsc{DPMJET}-III $\gamma$+Pb events. The right-hand axis shows the center-of-mass energy $W$ for the photon--nucleon collision system. The markers and vertical errors bars show the mean and standard deviation, respectively, of $E_{\gamma}$ as a function of $N_\mathrm{ch}^\mathrm{rec}$.}
\label{fig:photonE_in_MC}
\end{figure}
 
In general, no single generator is able to quantitatively describe in detail all of the features observed in data.
The results here may be used to benchmark future developments in the modeling of photonuclear collisions.
The two-particle correlation analysis described in Section~\ref{sec:2PC}, due its data-driven nature, does not rely on a detailed description of photonuclear events in simulation and therefore its results are not sensitive to this mismodeling.
 
Since the \textsc{DPMJET}-III $\gamma$+Pb simulation and the data agree qualitatively in the $N_\mathrm{ch}^\mathrm{rec}$ and \gapDef distributions, the simulated events provide an estimate of what ranges of photon energy in the laboratory frame, $E_\gamma$, and photon--nucleon center-of-mass energy, $W_{\gamma\mathrm{N}}$, are selected for analysis by choosing various $N_\mathrm{ch}^\mathrm{rec}$ or \gapDef values.
These ranges are shown in Figure~\ref{fig:photonE_in_MC} as a function of $N_\mathrm{ch}^\mathrm{rec}$.
In \textsc{DPMJET}-III, photonuclear events with a larger $N_\mathrm{ch}^\mathrm{rec}$ have a larger average $E_\gamma$ and $W_{\gamma\mathrm{N}}$. At $N_\mathrm{ch}^\mathrm{rec} = 10$, the mean values are $\left<E_\gamma\right> = 26.8$~\GeV\ and $\left<W_{\gamma\mathrm{N}}\right> = 519$~\GeV, rising to $\left<E_\gamma\right> = 123$~\GeV\ and $\left<W_{\gamma\mathrm{N}}\right> = 1.11$~\TeV\ at $N_\mathrm{ch}^\mathrm{rec} = 60$. However, even a narrow range of $N_\mathrm{ch}^\mathrm{rec}$ values selects events with a broad distribution of $E_\gamma$ and $W_{\gamma\mathrm{N}}$. At lower values of $N_\mathrm{ch}^\mathrm{rec} < 10$, the mean photon energy decreases rapidly and the range of photon energies increases, supporting the data-driven minimum value of $N_\mathrm{ch}^\mathrm{rec}$ determined above.
 
\section{Two-particle correlations}
\label{sec:2PC}
 
This section describes the two-particle correlation and template fit procedures used to extract the azimuthal anisotropies, $v_n$. These procedures have been used in previous analyses of two-particle correlations in small systems~\cite{HION-2012-13,HION-2013-04,HION-2015-09, HION-2016-01,HION-2018-23}, and are summarized here.
 
Correlations between pairs of charged particles are reported as a function of the pair's relative separation in pseudorapidity, $\Delta\eta = \eta^a - \eta^b$, and azimuth, $\Delta\phi = \phi^a - \phi^b$. The labels $a$ and $b$ denote kinematic selections on the first and second particle, which are generally different. If two particles each meet selections $a$ and $b$, the pair is tabulated twice, once for each possible assignment. By convention, the distributions in $\Delta\eta$ and $\Delta\phi$, $Y(\Delta\phi, \Delta\eta)$, are averaged over events by normalizing them by the total yield of particles in selection $a$, $N_a$,
 
\begin{equation}\nonumber
Y(\Delta\phi, \Delta\eta) =  \frac{1}{N_{a}} \frac{d^2 N_{\mathrm{pair}}} {d\Delta\phi  \cdot d\Delta\eta}.
\end{equation}
 
To correct for the charged-particle reconstruction efficiency, entries in the distribution are weighted by the product of the inverse efficiencies evaluated for the kinematics of the two particles. These efficiencies are derived from the \textsc{Hijing} simulation described in Section~\ref{sec:MC}.
In this analysis, all charged-particle tracks within $\left|\eta\right| < 2.5$ are considered, except the tracks which lie in the region $\eta > 4.9 - \gapDef$ (with positive $\eta$ defined as the photon-going direction), if the sum-of-gaps extends into the tracker.
There are two reasons to exclude these tracks. First, in simulations of photonuclear events, such particles are typically fragments of resolved photons, and not the bulk particles produced in the $\gamma$+Pb interaction which are being measured for possible collective signatures. Second, since these tracks are typically near the edge of the inner detector's acceptance, the $\left|\Delta\eta\right|$ restriction described below would result in a small number of tracks being paired many times, and thus making an disproportionately large contribution to the correlation function.
 
The one-dimensional $\Delta\phi$ normalized pair yields, $Y(\Delta\phi)$, are  constructed by integrating over a restricted $|\Delta\eta|$ range, typically 2.0 to 5.0, designed to suppress the contributions to the correlation from non-flow effects (such as jets, particle decays, and resonances), as:
 
\begin{equation}\nonumber
Y(\Delta\phi) =  \int_{|\Delta\eta| = 2.0}^{|\Delta\eta| = 5.0} Y(\Delta\phi, |\Delta\eta|)~d|\Delta\eta|.
\end{equation}
 
In principle, due to finite coverage and detection efficiency introduced by a  real detector, $Y$ may contain acceptance effects that do not reflect physical particle pair correlations. To account for this, a mixed-event technique may be applied~\cite{HION-2011-01}, in which particles $a$ and $b$ are taken from different events that have similar overall characteristics. The two events are required to have a ZDC signature on the same side (i.e.\ same directions of the photon and nucleus), and a similar charged-particle multiplicity, primary vertex position, and cluster edge gap $\Delta\eta_\mathrm{edge}^\mathrm{cluster}$. The mixed-event distribution $d^2 N_{\mathrm{mixed}}/ (d\Delta\phi d\Delta\eta)$ contains the effects of the pair acceptance but not the physical correlations.
 
The acceptance effects can be corrected for by dividing the same-event pair distribution by the mixed-event ones. The two-dimensional (2-D) correlation function corrected for acceptance effects is
 
\begin{equation}\nonumber
C(\Delta\phi, \Delta\eta) = \frac{1}{N_a} \left. \frac{d^2 N_{\mathrm{pair}}}{d\Delta\phi d\Delta\eta} \right/ \frac{1}{N_{\mathrm{pair}}^\mathrm{mixed}} \frac{d^2 N_{\mathrm{mixed}}}{d\Delta\phi d\Delta\eta},
\end{equation}
 
where $N_a$ is the total yield of particles with selection $a$, $N_{\mathrm{mixed}}$ is the yield in mixed events, and $N_\mathrm{pair}^\mathrm{mixed}$ is the total integral of the mixed-event distribution $d^2 N_{\mathrm{mixed}}/ (d\Delta\phi d\Delta\eta)$. Without the mixing correction, the 2-D $Y(\Delta\phi, \Delta\eta)$ distributions have an artificial triangular shape in $\Delta\eta$, reflecting the convolution of the single-particle acceptances $\left|\eta^{a}, \eta^{b}\right| < 2.5$. This effect is removed by the mixing correction, and the $\Delta \eta$ dependence of structures in the 2-D $C(\Delta\phi, \Delta\eta)$ distributions become more evident.
 
Correlation functions are constructed for various selections on $N_\mathrm{ch}^\mathrm{rec}$ and $p_\mathrm{T}^a$, with the $p_\mathrm{T}$ of particle $b$ always in the range $0.4 < p_\mathrm{T}^{b} < 2$~\GeV. All events are given an equal statistical weight, and no corrections for the trigger efficiency or differences in sampled luminosity (for $N_\mathrm{ch}^\mathrm{rec}$ ranges that span contributions from multiple triggers) are performed. Although the mixed-event correction is necessary to properly construct the two-dimensional correlation functions presented below, it was found to ultimately have only a minor effect on the projected, one-dimensional $Y(\Delta\phi)$ distributions, producing compatible results but with increased statistical uncertainty. Thus the mixing correction is not applied for the nominal $v_2$ and $v_3$ results. Instead, it is used in the determination of systematic uncertainties associated with the pair acceptance of the detectors as described in Section~\ref{sec:syst}.

\begin{figure}[t]
\begin{center}
\includegraphics[width=0.46\textwidth]{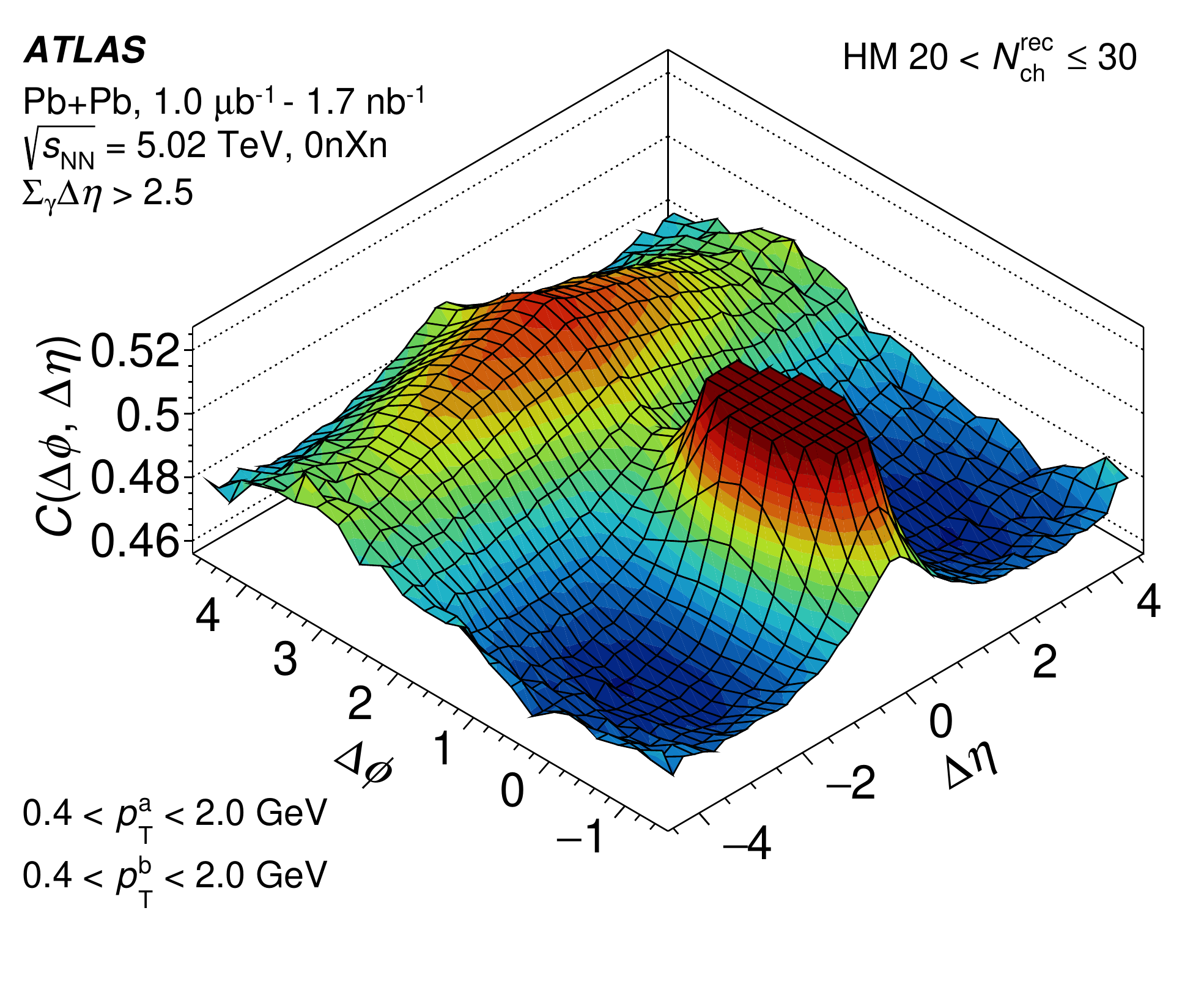}
\includegraphics[width=0.46\textwidth]{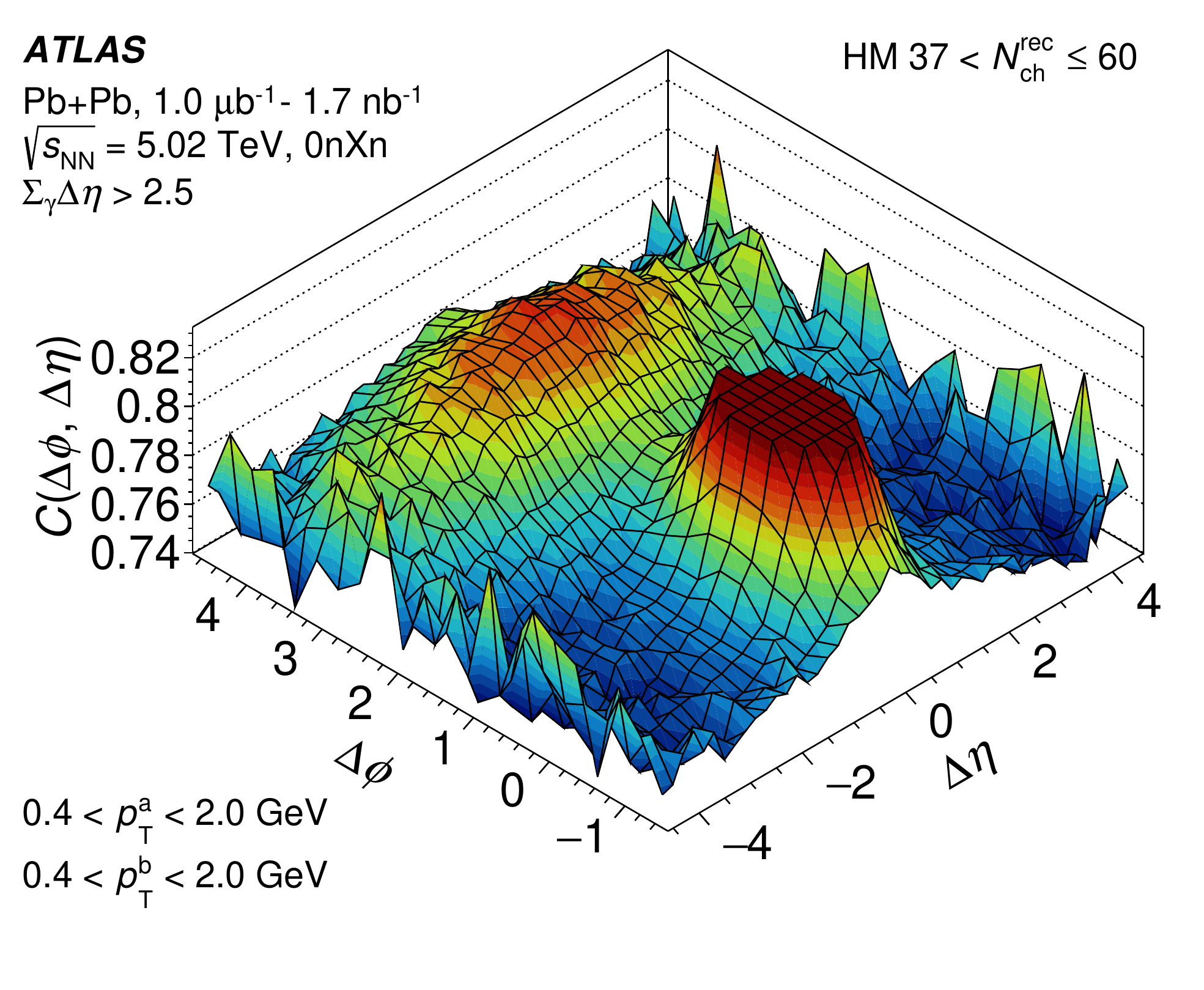}
\caption{Two-dimensional normalized particle pair distributions in photonuclear events, corrected for acceptance effects with the mixed-event distribution, and presented as a function of $\Delta\eta$ and $\Delta\phi$. The peak at $(\Delta\phi, \Delta\eta)=(0, 0)$ is truncated to better show the structure of the correlation function. Each panel represents a different charged-particle multiplicity $N_\mathrm{ch}^\mathrm{rec}$ range for the selection $0.4 < (p_\mathrm{T}^{a}, p_\mathrm{T}^{b}) < 2.0$~\GeV. }
\label{fig:PTY2D_mult}
\end{center}
\begin{center}
\includegraphics[width=0.46\textwidth]{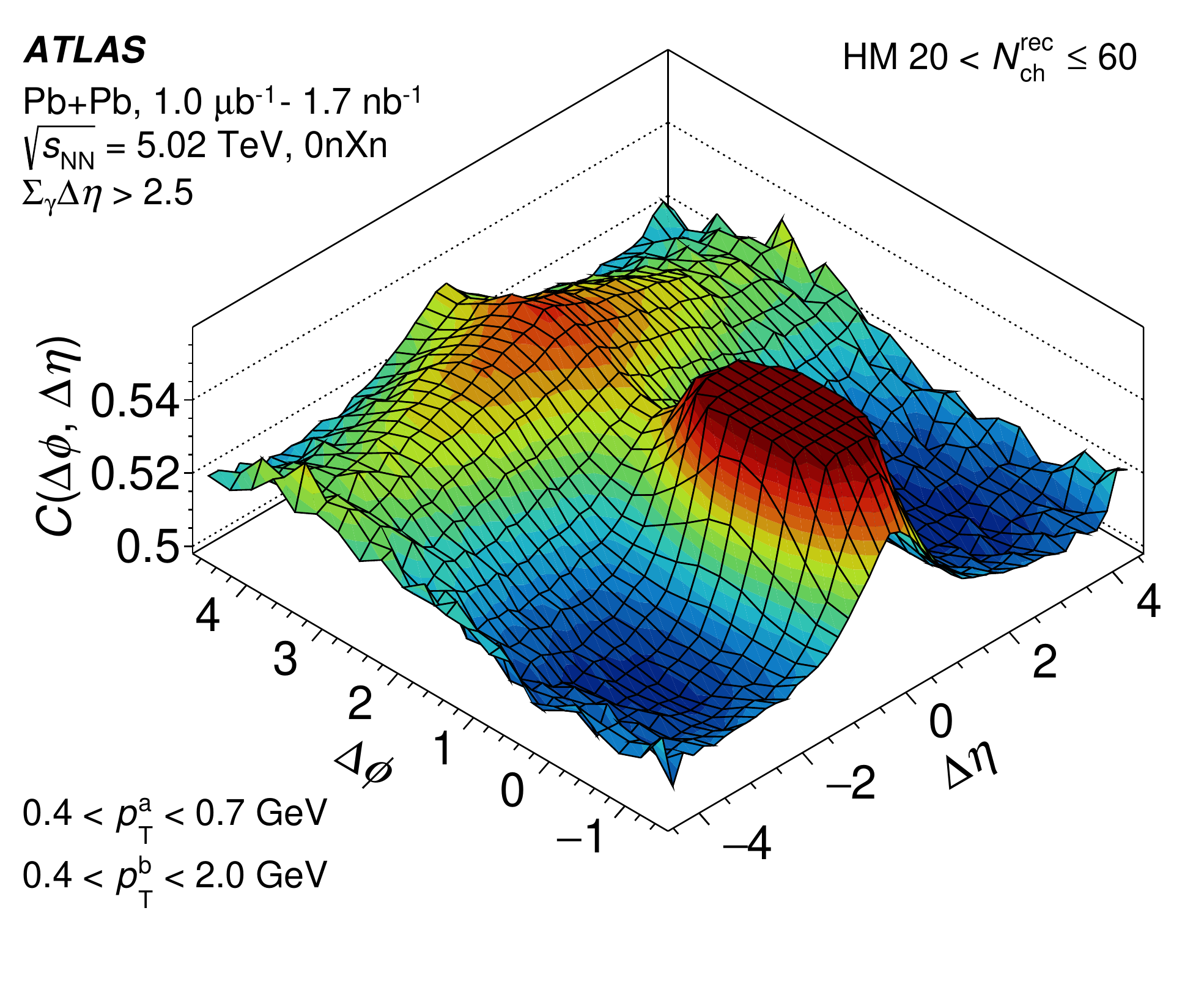}
\includegraphics[width=0.46\textwidth]{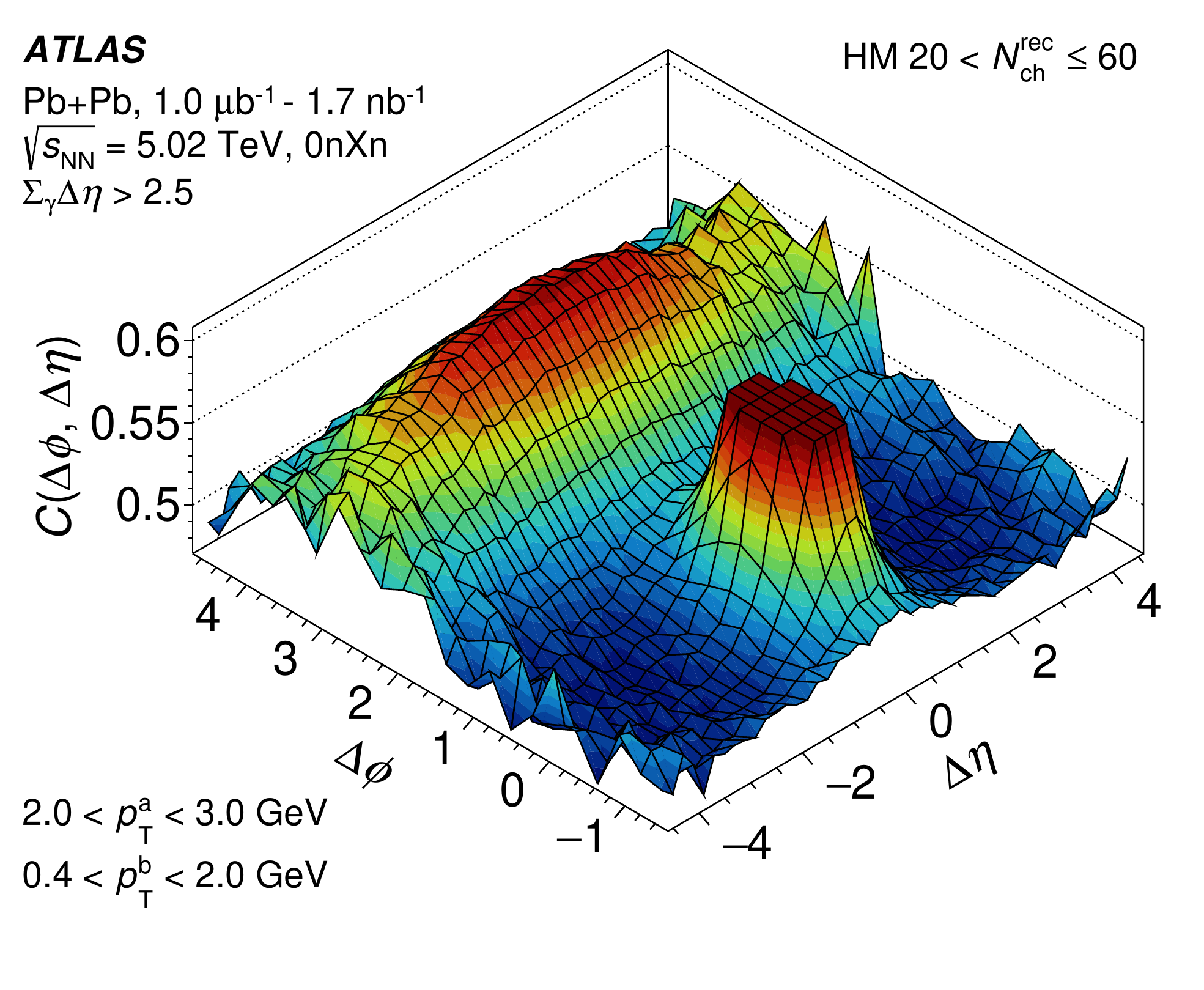}
\caption{Two-dimensional normalized particle pair distributions in photonuclear events, corrected for acceptance effects with the mixed-event distribution, and presented as a function of $\Delta\eta$ and $\Delta\phi$. The peak at $(\Delta\phi, \Delta\eta)=(0, 0)$ is truncated to better show the structure of the correlation function. Each panel represents a different $p_\mathrm{T}^{a}$ range for the selection $20 < N_\mathrm{ch}^\mathrm{rec} < 60$ and $0.4 < p_\mathrm{T}^{b} < 2.0$~\GeV. }
\label{fig:PTY2D_pTa}
\end{center}
\end{figure}

Examples of two-dimensional $C(\Delta\phi, \Delta\eta)$ correlation functions are presented in Figure~\ref{fig:PTY2D_mult} and Figure~\ref{fig:PTY2D_pTa}.
In Figure~\ref{fig:PTY2D_mult}, particles $a$ and $b$ are required to have $0.4 < p_\mathrm{T}^{a} < 2$~\GeV\ , and two example $N_\mathrm{ch}^\mathrm{rec}$ selections are shown. In Figure~\ref{fig:PTY2D_pTa}, correlation functions are presented for $20 < N_\mathrm{ch}^\mathrm{rec} < 60$,  with two example $p_\mathrm{T}$ selections for particle $a$ shown. The two-dimensional correlation functions have features which are broadly similar to those observed in $pp$ collisions. There is a localized `near-side' peak at $(\Delta\phi, \Delta\eta) \approx (0,0)$ from correlations between jet fragments, and an extended `away-side' ridge at $\Delta\phi \approx \pi$ which extends over a large $\Delta\eta$ range from correlations between fragments of azimuthally opposite jets.   A quantitative analysis and non-flow subtraction are necessary to discern if there are additional ridge structures, as detailed below.
 
\subsection{Non-flow subtraction}
 
To remove the contribution to the correlation function from non-flow effects that persist as long-range contributions on the away side ($\Delta\phi \sim \pi$), a template fit procedure is employed.
To perform the fit procedure, samples corresponding to events with low multiplicity (LM) and high multiplicity (HM) are selected.
The shape of the non-flow contribution is assumed in the template procedure to be the same in the LM and HM samples.
In this analysis, the LM events are chosen to have $15 \leq N_\text{ch}^\mathrm{rec} \leq 20$.
 
The $Y(\Delta\phi)$ in HM events is parameterized as the sum of an azimuthally modulated pedestal (which expresses the azimuthal anisotropy) and a non-flow component, as follows:
 
\begin{align}
Y^\mathrm{HM}(\Delta\phi)   &= FY^\mathrm{LM}(\Delta\phi) + G\bigg\{1+2\sum_{n=2}^{4}{v_{n,n}}\cos(n\Delta\phi)\bigg\} \nonumber \\
&= FY^\mathrm{LM}(\Delta\phi) + Y^\mathrm{ridge}(\Delta\phi).
\label{eq:tempfit_corr}
\end{align}
 
\begin{figure}[!t]
\centering
\includegraphics[width=0.49\textwidth]{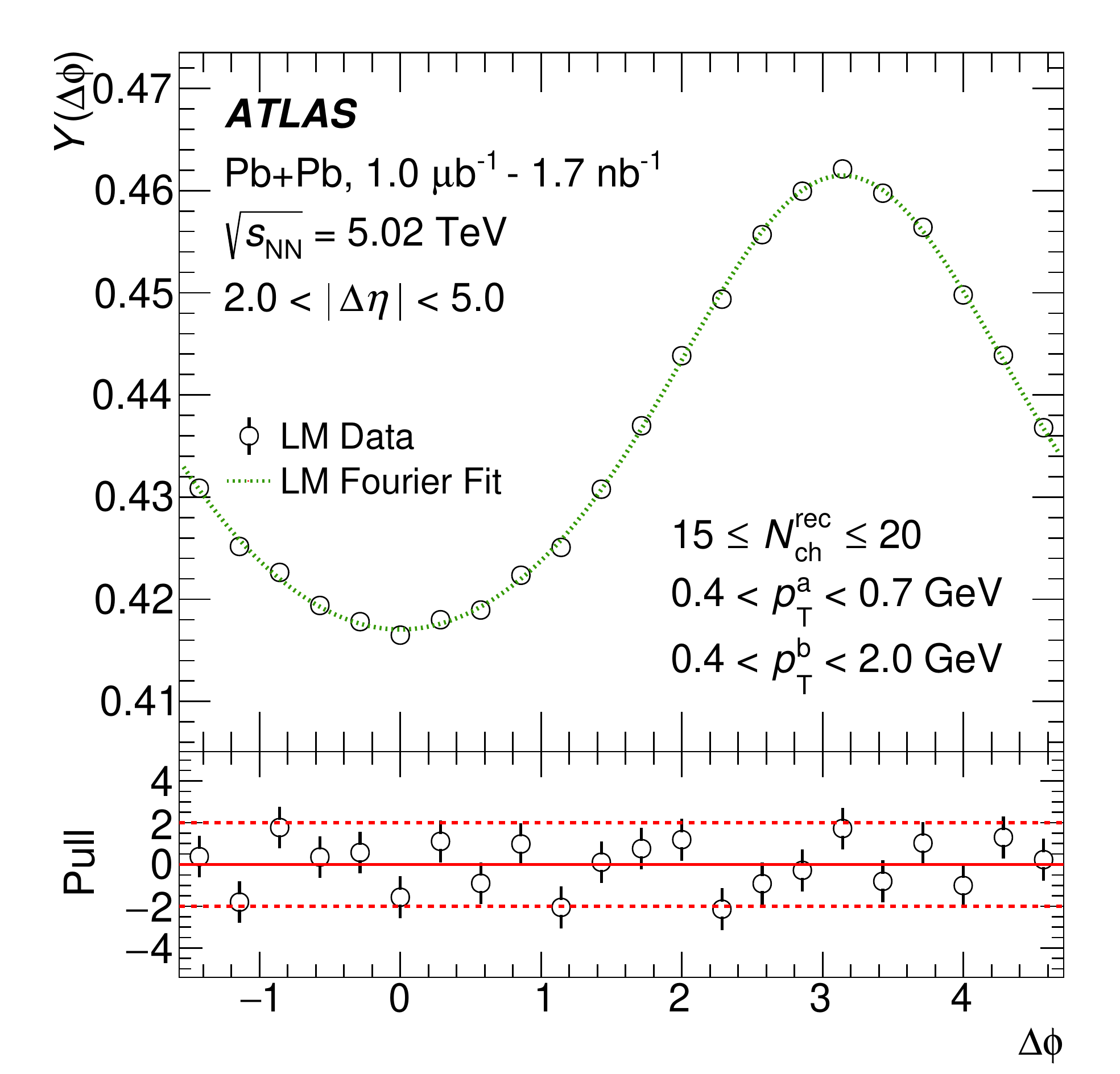}
\includegraphics[width=0.49\textwidth]{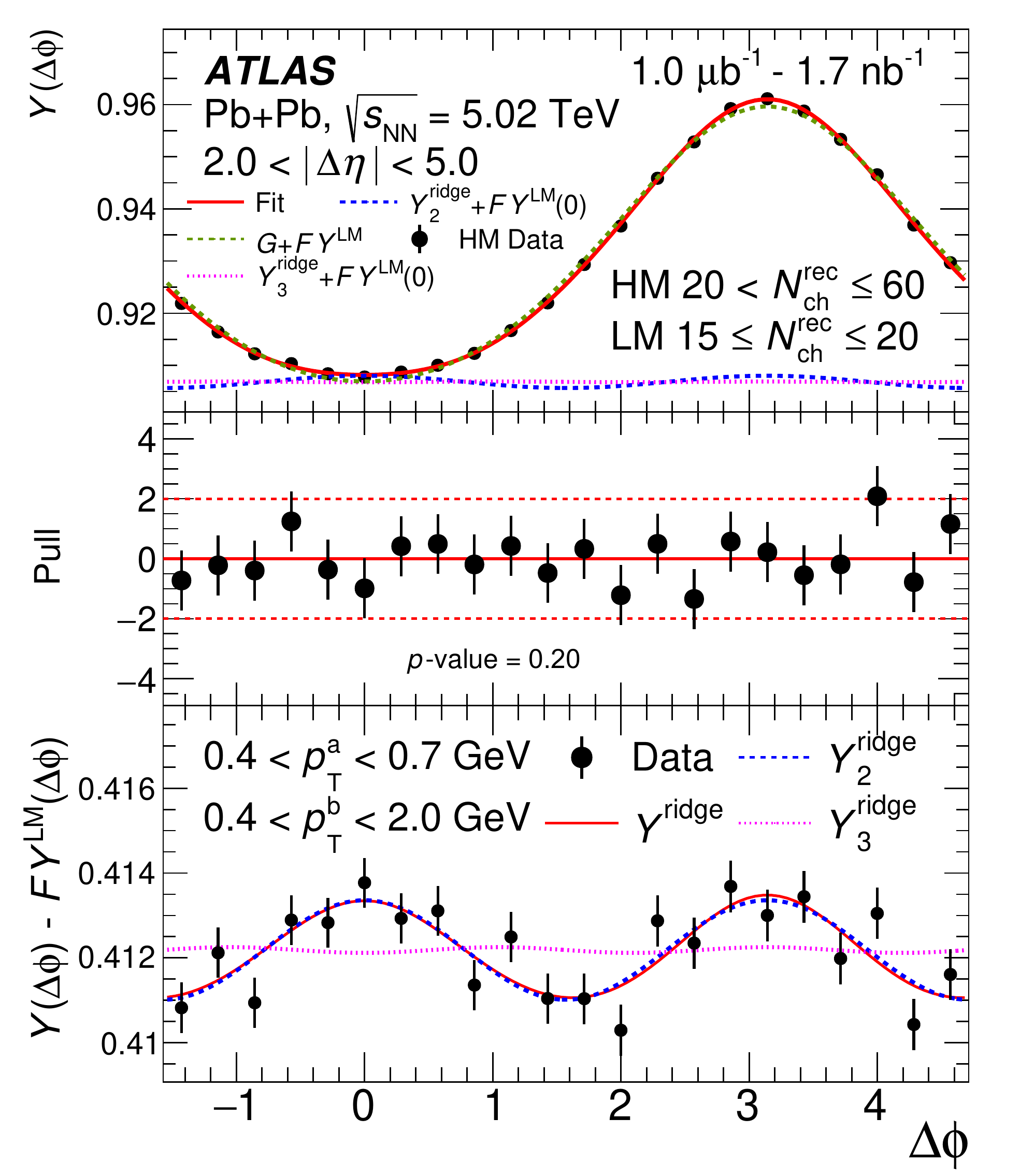}
\caption{An example of the template-fitting procedure for a selected \pt range.  The left plot displays the LM data with open markers and the simultaneous fit in the green dotted line.  The lower panel displays the pull distribution.
In the top panel of the right plot, the solid red line shows the total fit to the HM data in black markers. The dashed green line shows the scaled LM plus pedestal, while the dashed blue and dotted magenta lines indicate the two flow contributions to the fit, $Y_2^\mathrm{ridge} = G [1 +  2v_{2,2}\cos(2\Delta\phi)]$ and $Y_3^\mathrm{ridge} = G [1 + 2v_{3,3}\cos(3\Delta\phi)]$, shifted upwards by $FY^\mathrm{LM}(0)$ for visibility.  The middle-right panel shows the pull distribution for the template fit in the top panel.
The bottom-right panel shows the same set of data and fit components, where the scaled LM distribution has been subtracted to better isolate the modulation.}
\label{fig:templateFit_example}
\end{figure}

\begin{figure}[p]
\begin{center}
\includegraphics[width=0.45\textwidth]{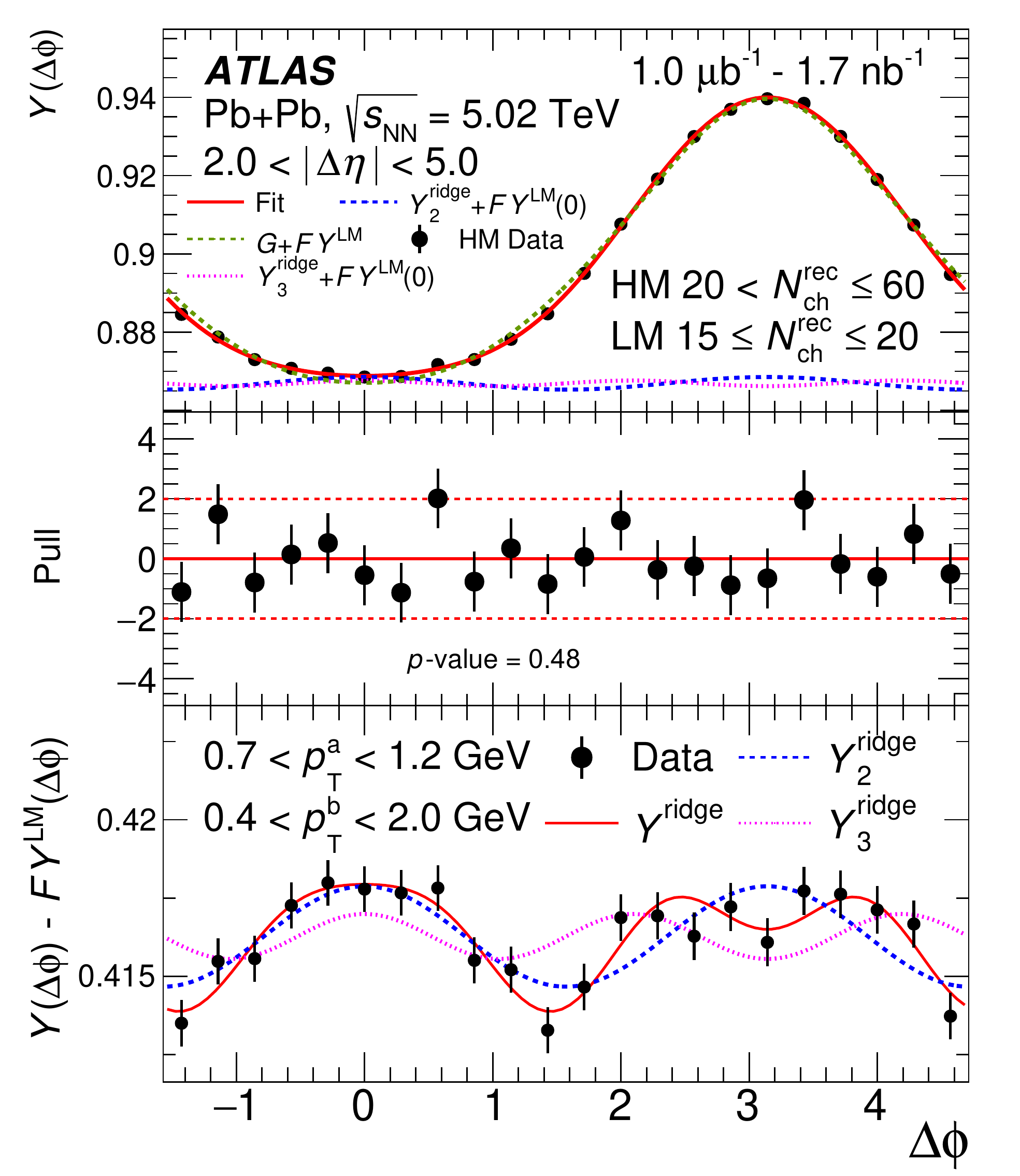}
\includegraphics[width=0.45\textwidth]{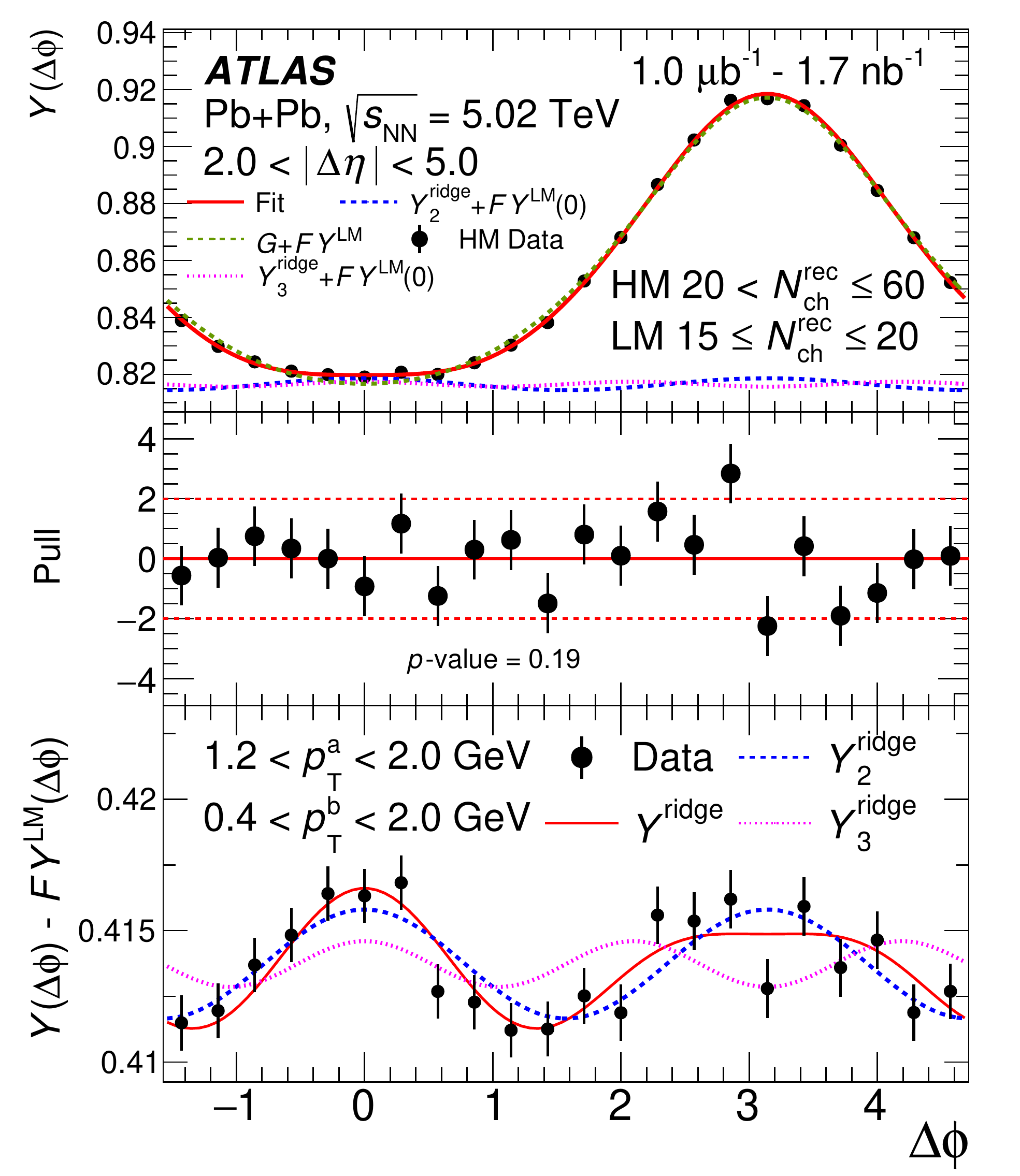} 		\includegraphics[width=0.45\textwidth]{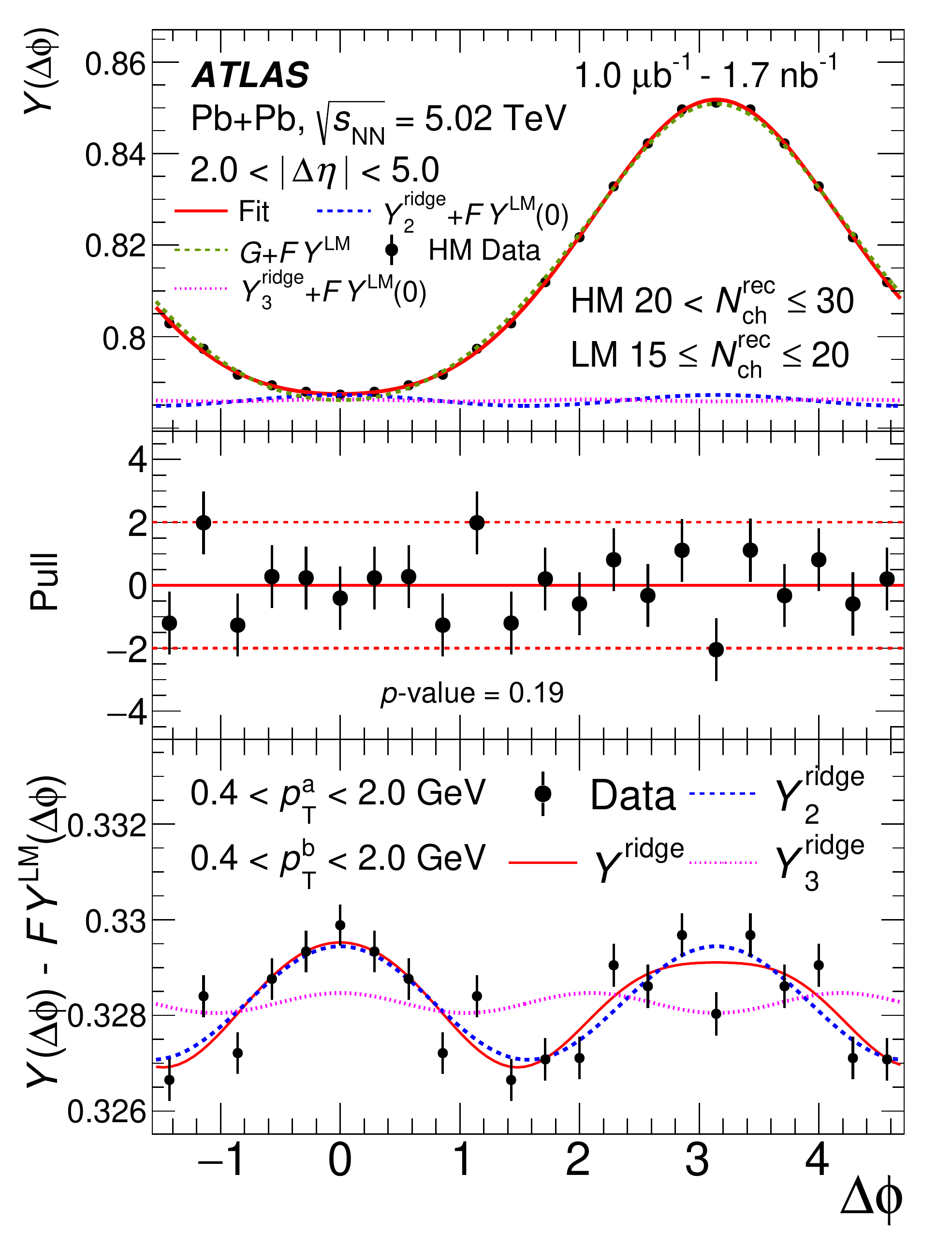}
\includegraphics[width=0.45\textwidth]{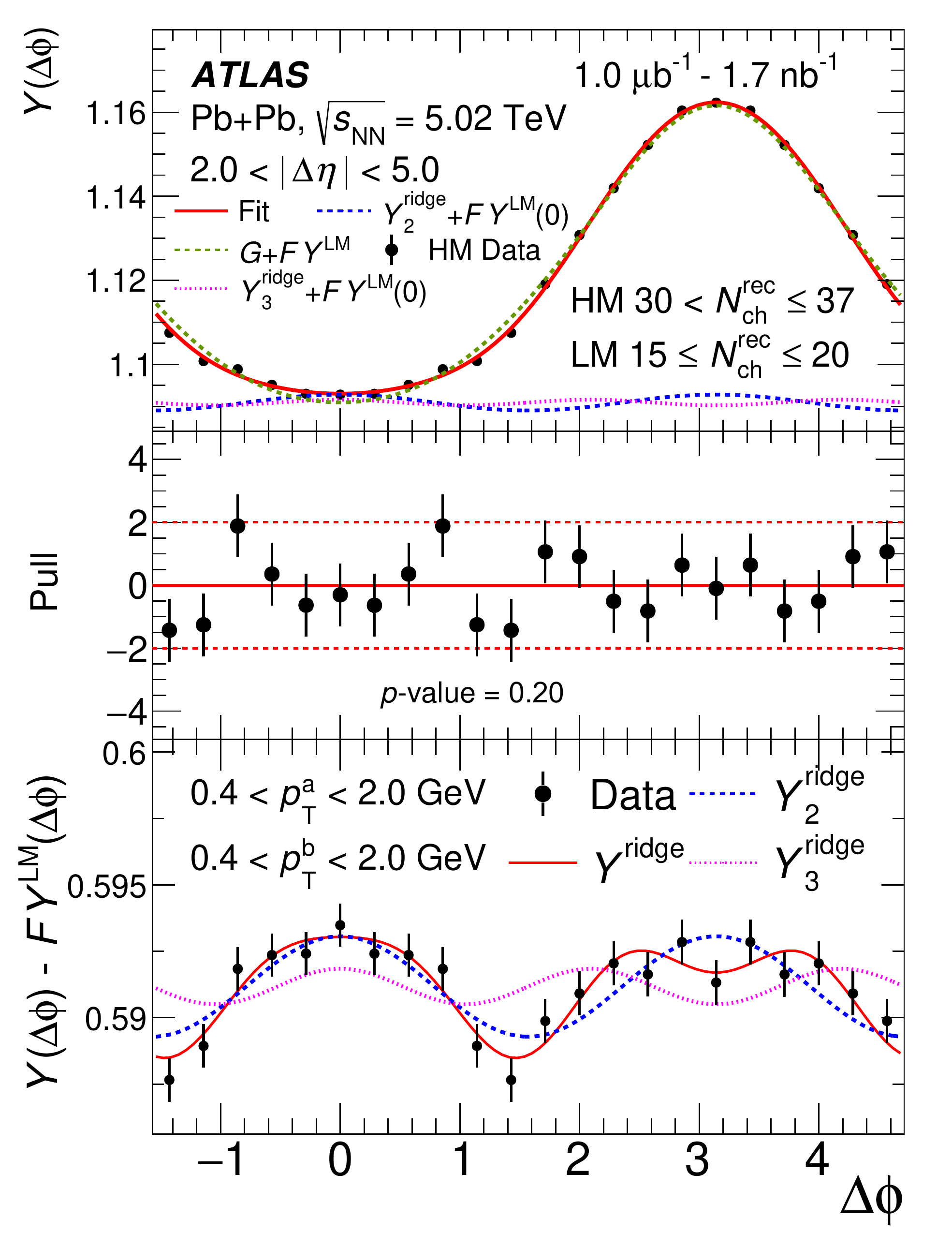}
\caption{Selected template fit results for two $p_\mathrm{T}^{a}$ intervals (top row) and two $N_\mathrm{ch}^\mathrm{rec}$ intervals (bottom row).
In the top panel of the right plot, the solid red line shows the total fit to the HM data in black markers. The dashed green line shows the scaled LM plus pedestal, while the dashed blue and dotted magenta lines
indicate the two flow contributions to the fit, $Y_2^\mathrm{ridge} = G [1 +  2v_{2,2}\cos(2\Delta\phi)]$ and $Y_3^\mathrm{ridge} = G [1 + 2v_{3,3}\cos(3\Delta\phi)]$, shifted upwards by $FY^\mathrm{LM}(0)$ for visibility.
The middle panels show the pull distribution for the template fits in the top panel.
The $p$-values from the bootstrapping experiment are also shown in the middle panels.
The bottom panels show the same set of data and fit components, where the scaled LM distribution has been subtracted from each to better isolate the modulation.
\label{fig:templateFit_pt_fine1} }
\end{center}
\end{figure}

Above, $Y^\mathrm{LM}(\Delta\phi)$ is the correlation function in LM events, which is parameterized as a truncated Fourier series up to the fourth order (with coefficients $c_n$ for $n=0,\ldots{},4$).
The values of $F$ and $G$, the three $v_{n,n}$ values, and the five parameters $c_n$, describing the LM reference, are free parameters in the fit,
but $F$ and $G$ are constrained such that the integrals of both sides of Eq.~(\ref{eq:tempfit_corr}) are the same.
Modulation terms up to fourth order ($v_{2,2}$, $v_{3,3}$, and $v_{4,4}$) are considered in the fit, in order to best describe the HM data.
By fitting $Y^\mathrm{LM}(\Delta\phi)$ and $Y^\mathrm{HM}(\Delta\phi)$ simultaneously, the extracted uncertainty in $F$, $G$, and $v_{n,n}$ correctly accounts for the statistical uncertainty of both the LM and HM samples. An example of the simultaneous fit of the HM selection and LM reference is shown in Figure~\ref{fig:templateFit_example}. Examples of the template fit in additional HM and $p_\mathrm{T}^{a}$ selections are shown in Figure~\ref{fig:templateFit_pt_fine1}. In the bottom panels of Figs.~\ref{fig:templateFit_example} and~\ref{fig:templateFit_pt_fine1}, the $p$-values are defined following the procedure described below.
 
The template fit is performed by minimizing the standard $\chi^2$ between the data points and the functional form.
However, the data points within the correlation functions contain nontrivial point-to-point correlations, since a single particle $b$ may be used in combination with multiple particles of type $a$.
The minimum of the $\chi^2$ statistic, when calculated in the traditional way, is found at the appropriate values of the fit parameters. However, the $p$-value and the uncertainty in the parameter values, if also determined in the standard way, would be inaccurate.
In order to properly account for these correlations and determine the parameter value uncertainties, a bootstrapping procedure was applied.
Pseudo-experiments were generated by giving a random Poisson weight (with a mean of one) to individual events.
These pseudo-experiments are independent sets of events drawn from the observed set of events with the same statistics and correlations as the real data.
The parameters values extracted in the pseudo-experiments agreed with those in data and followed Gaussian distributions.
The standard deviations of the parameters values extracted from the pseudo-experiments were used to set the statistical uncertainties in the final results, which are 10--30\% larger than if they were estimated in the na\"ive way without taking the point-to-point correlations into account.
The $p$-values from the bootstrapping experiment are presented in the template fit plots that follow.
 
\begin{figure}[p]
\begin{center}
\includegraphics[width=0.45\textwidth]{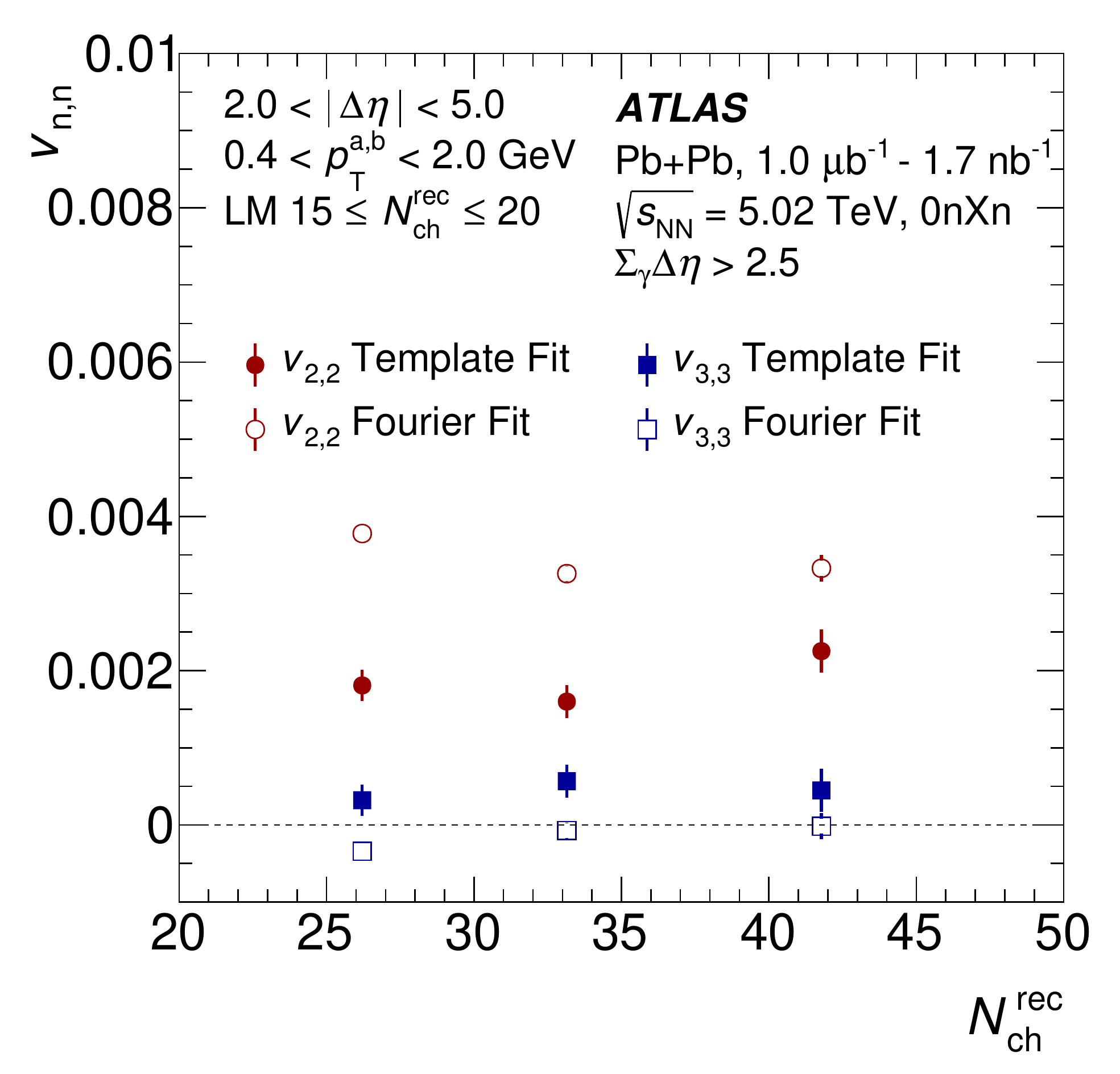}
\includegraphics[width=0.45\textwidth]{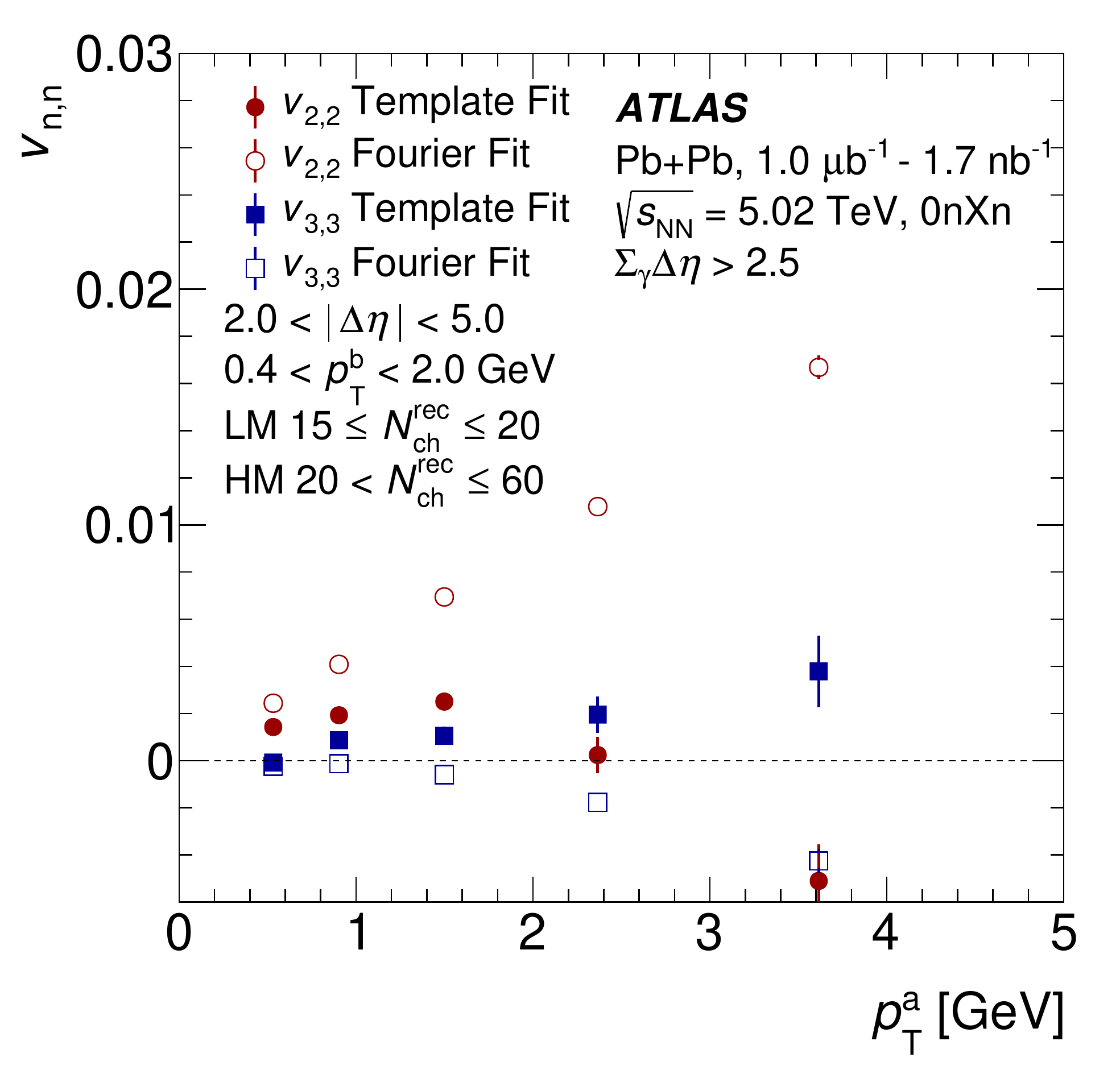}
\caption{Summary of $v_{2,2}$ and $v_{3,3}$ results as obtained from a direct Fourier fit (open markers) and those obtained after the template-fitting subtraction (filled markers). Results are shown differential in $N_\mathrm{ch}^\mathrm{rec}$ for fixed $p_\mathrm{T}^{a}$ (left) and differential in $p_\mathrm{T}^{a}$ for fixed $N_\mathrm{ch}^\mathrm{rec}$ (right). }
\label{fig:rawCoef}
\includegraphics[width=0.45\textwidth]{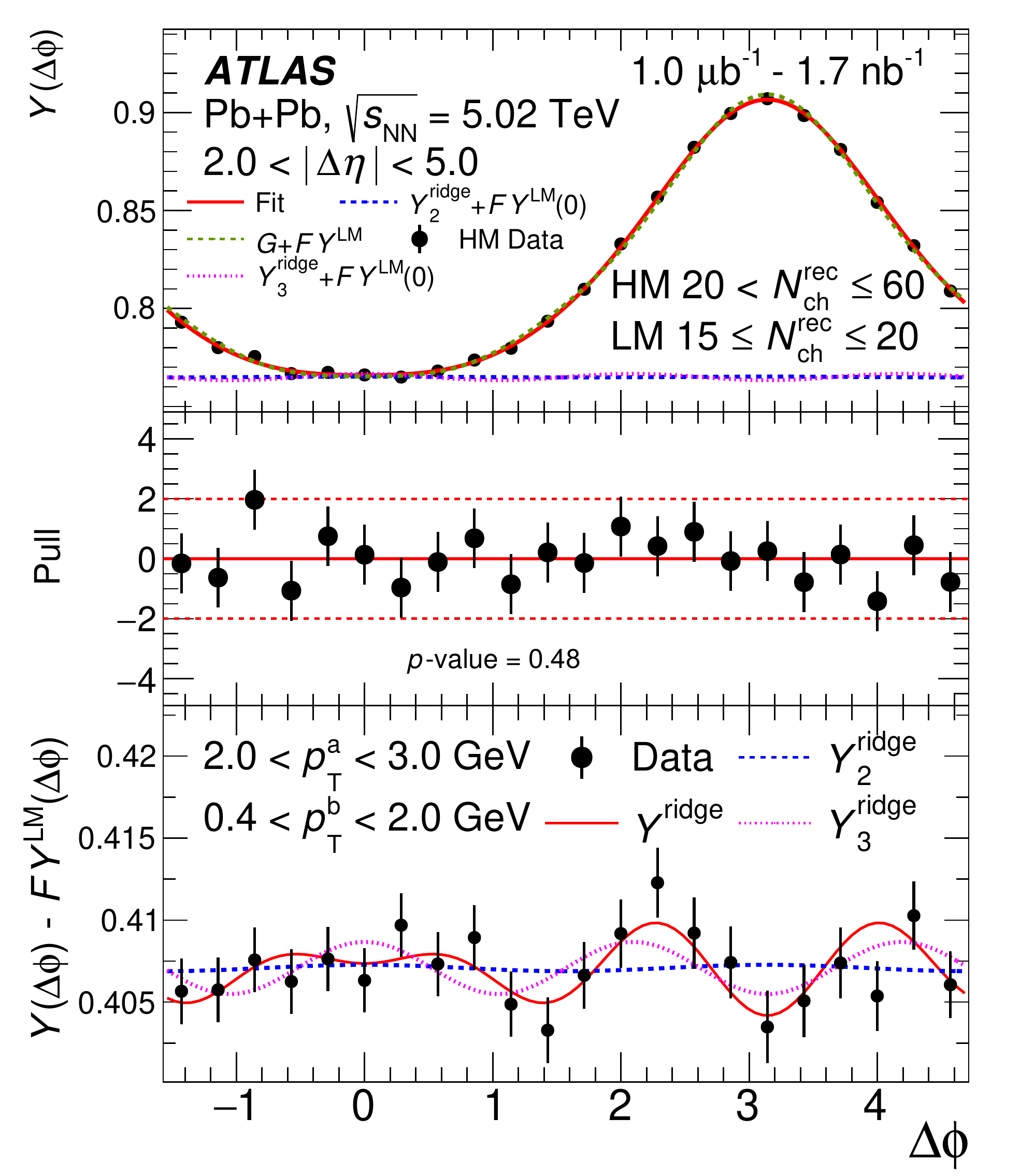}
\includegraphics[width=0.45\textwidth]{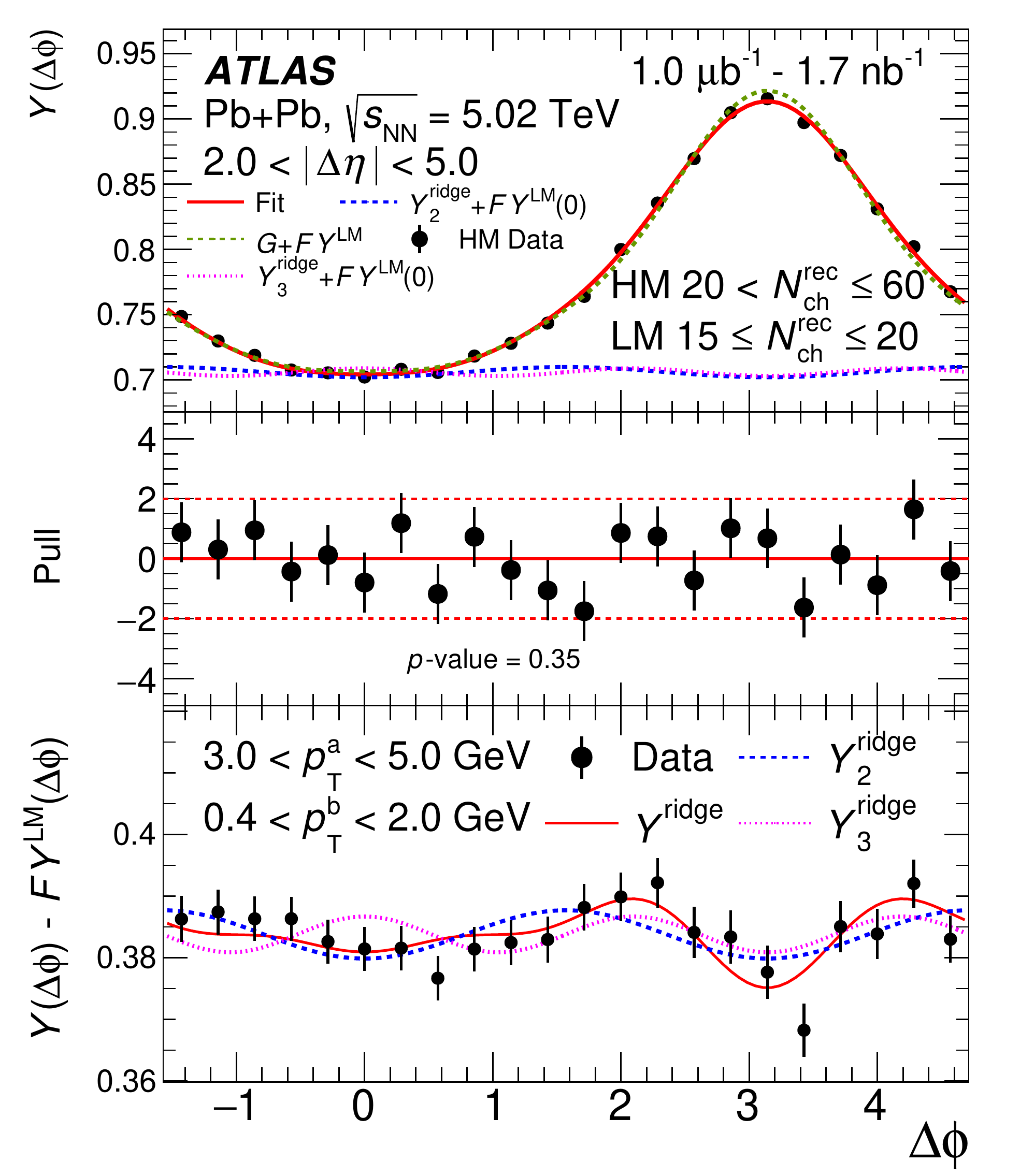}
\caption{Selected template fit results for the highest $p_\mathrm{T}^{a}$ intervals.
In the top panel of the right plot, the solid red line shows the total fit to the HM data in black markers. The dashed green line shows the scaled LM plus pedestal, while the dashed blue and dotted magenta lines indicate the two flow contributions to the fit, $Y_2^\mathrm{ridge} = G [1 +  2v_{2,2}\cos(2\Delta\phi)]$ and $Y_3^\mathrm{ridge} = G [1 + 2v_{3,3}\cos(3\Delta\phi)]$, shifted upwards by $FY^\mathrm{LM}(0)$ for visibility.
The middle panels show the pull distribution for the template fits in the top panel.
The $p$-values from the bootstrapping experiment are also shown in the middle panels.
The bottom panels show the same set of data and fit components, where the scaled LM distribution has been subtracted from each to better isolate the modulation.
\label{fig:templateFit_high_pt} }
\end{center}
\end{figure}
 
The full set of $v_{2,2}$ and $v_{3,3}$ values are
shown in Figure~\ref{fig:rawCoef}.
The filled points show the non-flow subtracted results according to the template fit procedure in Eq.~(\ref{eq:tempfit_corr}).
The open points show the $v_{n,n}$ values that would be obtained if instead a direct Fourier decomposition of $Y^\mathrm{HM}$ were performed, without any attempt to account for the non-flow contribution.
For most of the selections considered here, the non-flow subtraction has a significant effect on the extracted $v_{2,2}$ and $v_{3,3}$ values. The resulting $v_{n,n}$ values are positive in all selections, with one exception: in the $p_\mathrm{T}^{a}$-dependent results with a single HM selection, the $v_{2,2}$ value for $3 < p_\mathrm{T} < 5$~\GeV\ is negative. Additionally, the $v_{2,2}$ value for $2 < p_\mathrm{T} < 3$~\GeV\ is significantly lower than that for $1.2 < p_\mathrm{T} < 2$~\GeV. In these selections, the $v_{3,3}$ values also rise significantly. The template fits to these selections are shown in Figure~\ref{fig:templateFit_high_pt}, and are discussed further below.
 
\subsection{Factorization test}
\label{ssec:factorization}
\begin{figure}[t]
\begin{center}
\includegraphics[width=0.4\textwidth]{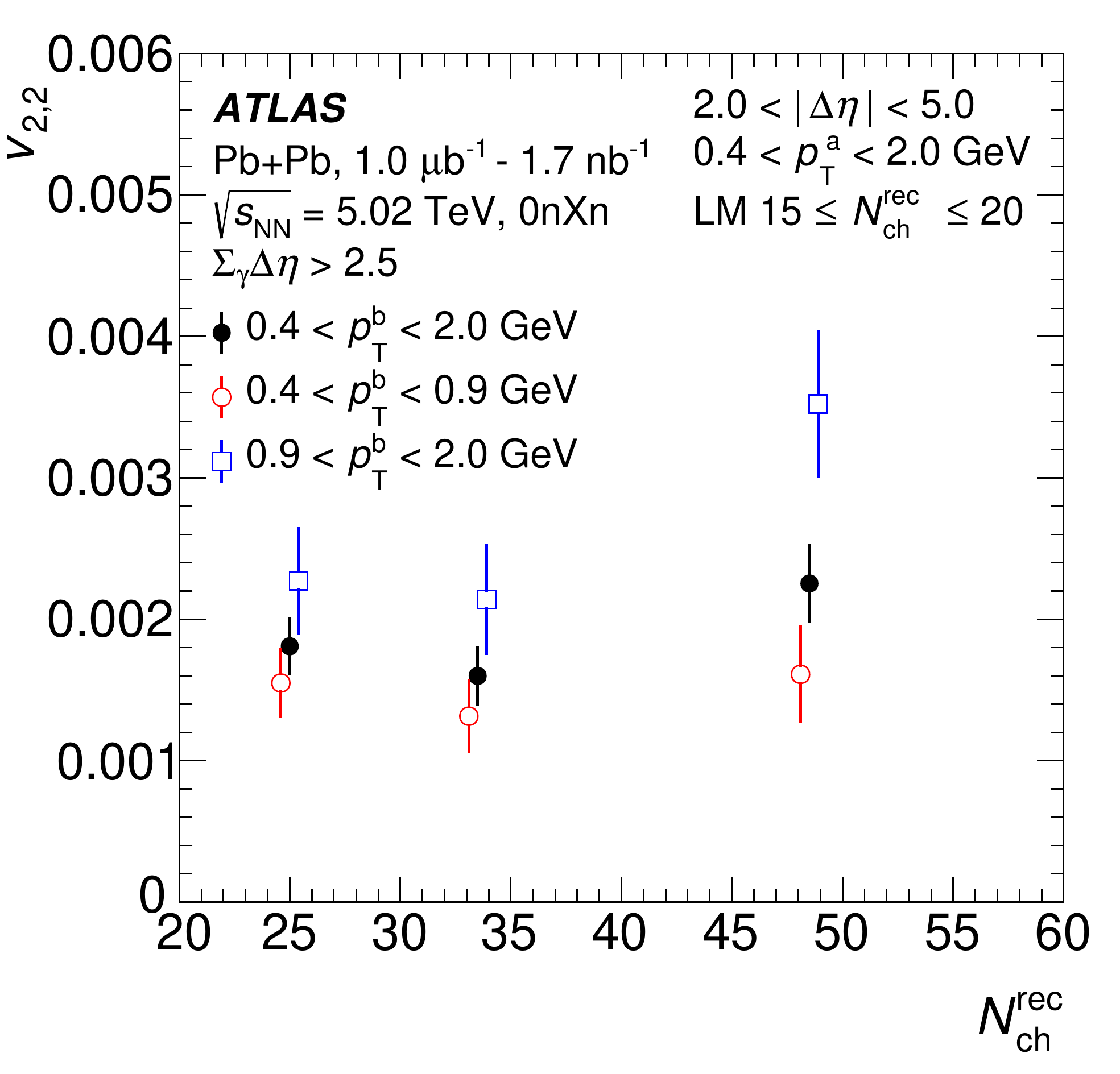}
\includegraphics[width=0.4\textwidth]{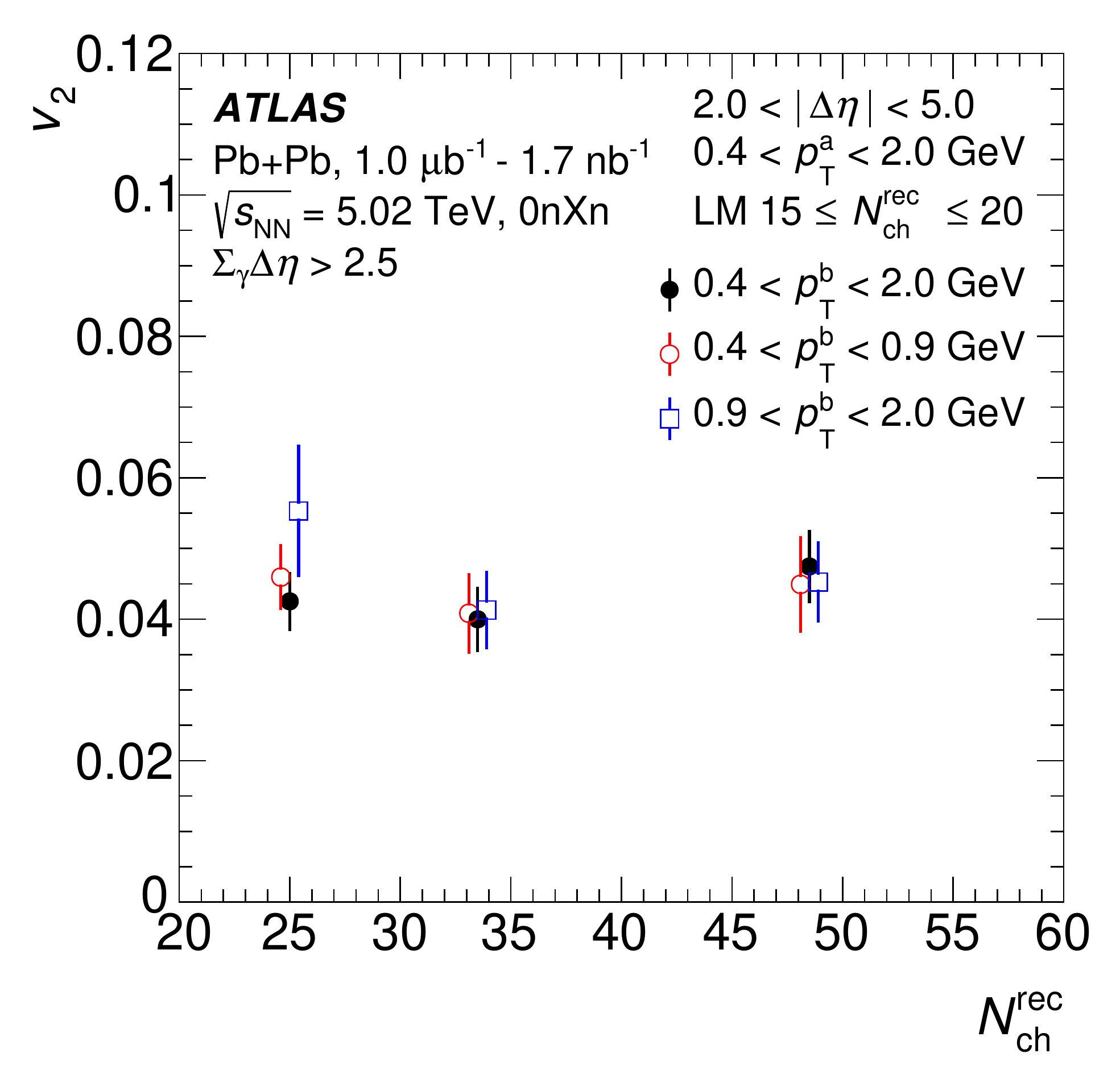}
\caption{Summary of the factorization test for $v_2$ values as a function of $N_\mathrm{ch}^\mathrm{rec}$. Comparison of $v_{2,2}$ values (left) and derived $v_{2}$ values (right) using the same particle $a$ selection, but different selections on particle $b$. Only statistical uncertainties are shown. The points are displaced horizontally for visibility.}
\label{fig:factorization_check}
\includegraphics[width=0.4\textwidth]{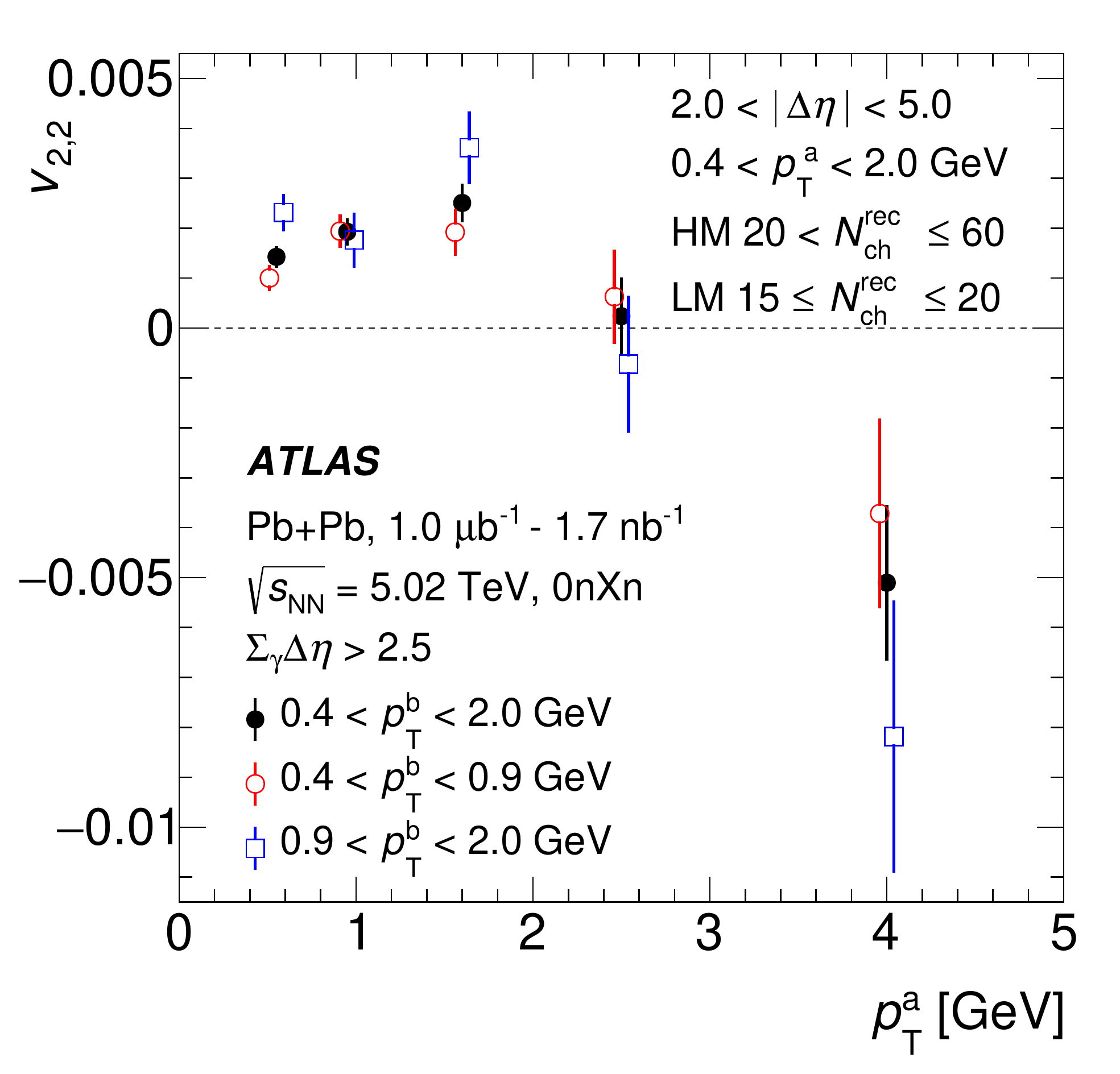}
\includegraphics[width=0.4\textwidth]{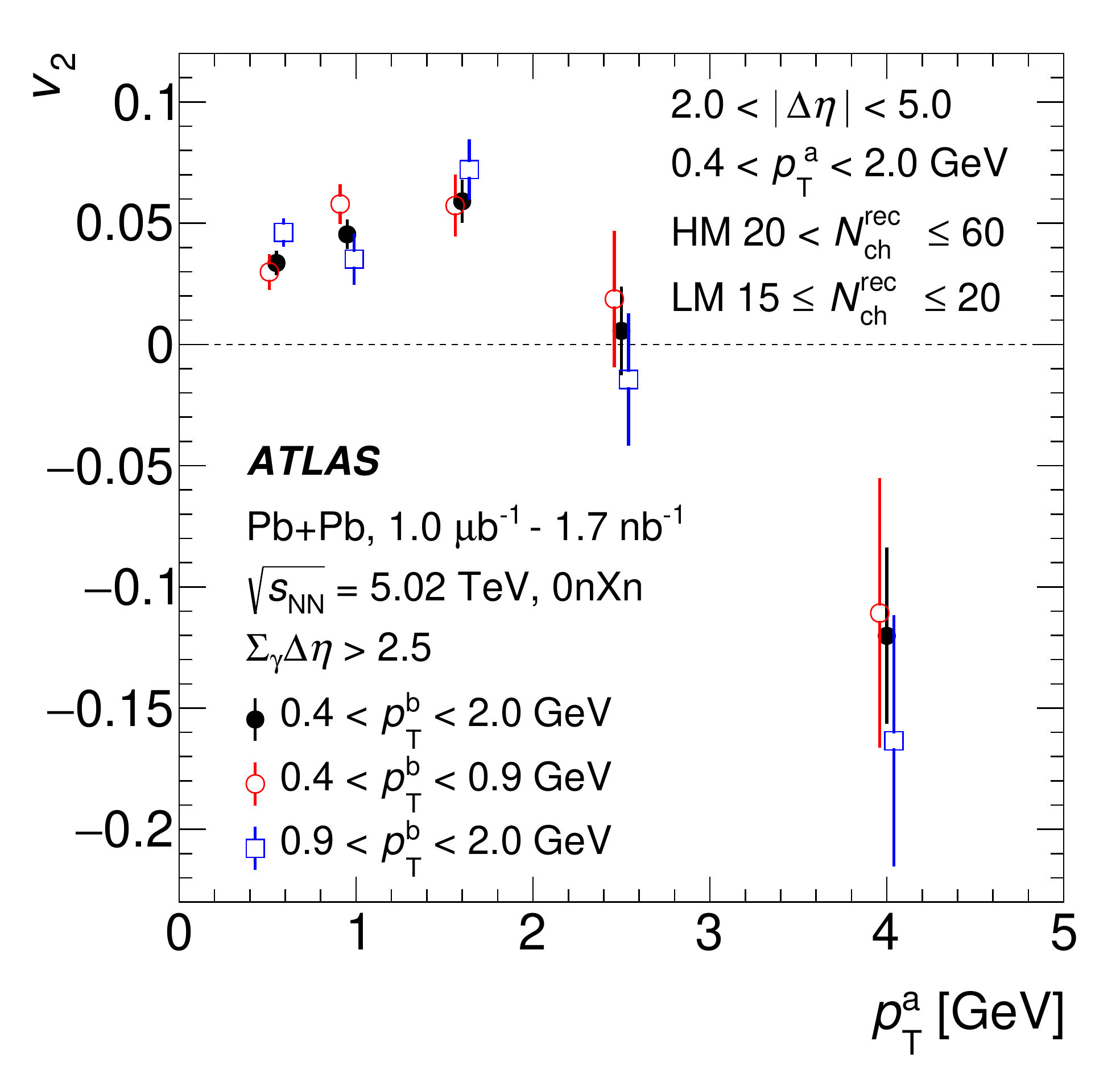}
\caption{Summary of the factorization test for $v_2$ values as a function of $p_\mathrm{T}^{a}$. Comparison of $v_{2,2}$ (left) and derived $v_{2}$ values (right) using the same particle $a$ selection, but with different selections on particle $b$. Only statistical uncertainties are shown. The points are displaced horizontally for visibility.}
\label{fig:factorization_pT}
\end{center}
\end{figure}

In the flow paradigm, a two-particle azimuthal modulation characterized by a $v_{n,n}$ value arises from the product of nonzero azimuthal anisotropies, $v_{n}$, for each particle. These are related via $v_{n,n}(\pt^{a},\pt^{b})  = v_{n}(\pt^{a}) v_{n}(\pt^{b})$, or $v_{n,n}(\pt^{a},\pt^{b})  = v_{n}(\pt)^2 $ if $a$ and $b$ are selected from the identical particle \pt range.
Thus, a single-particle flow coefficient $v_n(\pt^{a})$ may be determined from two-particle $v_{n,n}$ values through $v_n(\pt^{a}) = v_{n,n}(\pt^{a},\pt^{b}) / v_n(\pt^{b}) = v_{n,n}(\pt^{a},\pt^{b}) / \sqrt{v_{n,n}(\pt^{b},\pt^{b})}$ for a given selection on reference particle $b$.
To test whether the $v_{n,n}$ values in data are compatible with this picture, a factorization test can be performed in which $v_n$ values for particle $a$ are compared for different particle $b$ selections. The results of this test for the $v_2$ values as a function of $N_\mathrm{ch}^\mathrm{rec}$ are shown in Figure~\ref{fig:factorization_check}. The test demonstrates that while the $v_{2,2}$ values for different $p_\mathrm{T}^{b}$ selections may be different, the $v_{2}$ values obtained for particle $a$ as a function of $N_\mathrm{ch}^\mathrm{rec}$ are independent of the selection for particle $b$. This is a necessary condition of the paradigm that the observed $v_{2,2}$ values arise from the product of event-wide, single-particle $v_{2}$ azimuthal modulations.
 
\Fig{\ref{fig:factorization_pT}} shows the results of a factorization test as a function of particle $a$ $p_\mathrm{T}$. Although this test is more statistically limited than that for the $N_\mathrm{ch}^\mathrm{rec}$-dependent test above, the results for $p_\mathrm{T}^{a} < 2$~\GeV\ are compatible with the factorization assumption. Since non-flow effects are not necessarily expected to factorize, this validates an important assumption of the template procedure. For the selection with $p_\mathrm{T}^{a} > 3$~\GeV, the $v_{2,2}$ values obtained in Figure~\ref{fig:factorization_pT} are significantly negative, and are accompanied by large $v_{3,3}$ values. A negative $v_{2,2}$ value violates the expected factorization for particles with independent overall modulation, and thus cannot be interpreted as arising from hydrodynamic flow.
 
As can be seen from the growing Fourier $v_{2,2}$ values in Figure~\ref{fig:rawCoef} and the large away-side peaks in Figure~\ref{fig:templateFit_high_pt}, the non-flow contribution in these high-$p_\mathrm{T}^{a}$ selections is significantly larger than the magnitude of any possible modulation. Thus, the extraction of any genuine $\cos(n\Delta\phi)$ modulation is highly sensitive to assumptions in the template method about the scaling of the non-flow contribution between multiplicity classes. If violated, this could manifest as a reduced $v_{2,2}$ and increased $v_{3,3}$. This assumption is difficult to test given the limited data sample size for LM variations and factorization tests.
Therefore, within the present uncertainties and techniques, the $v_2$ cannot be reliably extracted at high $p_\mathrm{T}$.   Finally, the statistical precision of the data does not allow similar tests of the $v_{3}$ values. In this case, their presented values assume that factorization holds.
 
In Section~\ref{sec:result} the $v_{2}$ and $v_{3}$ values are presented for all $p_\mathrm{T}$ for completeness, although it should be noted that factorization has been demonstrated only for $v_2$ in the range $0.4 \le p_\mathrm{T} \le 2.0$~\GeV.
 
\subsection{Physics backgrounds}
\label{ssec:purity}
 
The measured $v_n$ values are potentially biased by the presence of background events, primarily from peripheral hadronic Pb+Pb events, in the selected event sample. The purity of photonuclear events is estimated by fitting templates derived from simulated photonuclear and peripheral Pb+Pb events to distributions measured in data, and determining the fraction of selected events which are compatible with photonuclear interactions.
 
The purity was first estimated by fitting the \gapDef distribution in data to the combination of distributions from \textsc{DPMJET}-III $\gamma$+Pb events and peripheral \textsc{Hijing} Pb+Pb events over the range $0.5 < \gapDef < 4.9$, with the relative contributions from each as free parameters. An example of these two contributions and the distribution in data is shown in the right panel of Figure~\ref{fig:DnDeta_Ntrk}. This fit was performed separately in each of the $N_\mathrm{ch}^\mathrm{rec}$ selections used in the measurement. The purity for the $\gapDef > 2.5$ selection is estimated to be greater than 98\% in all $N_\mathrm{ch}^\mathrm{rec}$ selections used in the analysis. Thus, for the nominal results, no correction for a possible impurity in the selected events is performed.
 
\section{Systematic uncertainties}
\label{sec:syst}
 
The primary sources of systematic uncertainty are grouped by category and discussed below.
The systematic uncertainties are evaluated by varying the procedure applied to the data, and are not directly derived from the simulation samples described in Section~\ref{ssec:event_prop}. Thus, the uncertainties do not depend on an accurate modeling of the photonuclear processes in simulation.
The uncertainties from each source are ultimately combined in quadrature to obtain the total systematic uncertainty.
For some individual sources below where only one variation of the analysis procedure was performed, the resulting uncertainty was symmetrized by adding an uncertainty of equal magnitude in the other direction. However, in general the final systematic uncertainties are not symmetric. The typical magnitudes of the systematic uncertainties are summarized in Table~\ref{tab:syst}.
 
\begin{table}[t]
\begin{center}
\caption{Summary of sources of systematic uncertainty, and the typical values of their absolute magnitude.  The last row shows the typical total uncertainties from the quadrature sum of all uncertainties.}
\vspace{0.25cm}
\begin{tabular}{ l|r|r|r|r}
Source & $v_{2}(N_\mathrm{ch}^\mathrm{rec})$~~~~ &  $v_{2}(p_\mathrm{T})$~~~~ & $v_{3}(N_\mathrm{ch}^\mathrm{rec})$~~~~ &  $v_{3}(p_\mathrm{T})$~~~~\\
\hline
$\Delta\eta$ pair cut   & 0--0.002     & 0.01--0.02   & 0--0.015     & 0--0.03      \\
LM reference            & 0.001        & 0--0.02      & 0--0.005     & 0.03--0.06   \\
$\phi$ binning          & 0.001        & 0--0.01      & 0.001        & 0--0.01      \\
Mixed-event correction       & 0.001        & 0--0.01      & 0--0.005     & 0--0.01      \\
Track \gapDef exclusion & 0.001        & 0--0.02      & 0.002        & 0--0.01      \\
Purity                  & 0--0.002    & 0--0.02      & 0.005        & 0--0.02      \\ \hline
Total                   & 0.001--0.002 & 0.005--0.04  & 0.005--0.015 & 0.03--0.07
\end{tabular}
\label{tab:syst}
\end{center}
\end{table}
 
\emph{Pair-pseudorapidity difference.} The two-particle correlation function is measured for pairs with $2.0 < |\Delta\eta| < 5.0$ as a way to strongly reduce the non-flow contribution, which varies much more strongly with $\Delta\eta$ than the genuine long-range azimuthal modulation. To test the sensitivity of the analysis to a potential residual contribution, the lower bound on $\Delta\eta$ is varied to $1.8$ and to $2.2$, and the impact of the resulting variations on the data is included as an uncertainty.
 
\emph{Low-multiplicity reference.} Photonuclear events with $15 \leq N_\mathrm{ch}^\mathrm{rec} \leq 20$ are used as the LM reference in the template fit to the other $N_\mathrm{ch}^\mathrm{rec}$ selections. In the template method, the shape of the non-flow contribution is assumed to be the same in the LM and HM selections.
To test the possible variation of the non-flow contribution with multiplicity, the analysis may be repeated using an alternative reference for LM events.
Additionally, a different multiplicity range may select a different sample of photonuclear events (i.e.\ distribution of photon energies, or mix of resolved and direct processes~\cite{Bertulani:2005ru}), to which the results may be sensitive. An alternative selection, $10 \leq N_\mathrm{ch}^\mathrm{rec} < 15$, composed from the MB trigger data, is considered for the LM reference. The analysis is repeated and the difference in the results is included in the uncertainties.
 
\emph{$\Delta\phi$ binning.} The sensitivity to the particular choice of $\Delta\phi$ binning was evaluated by repeating the fitting procedure after halving the number of bins. The resulting difference is included in the total uncertainty.
 
\emph{Acceptance effects in correlation function.} For the nominal analysis, the $\Delta\phi$-dependent correlation functions are formed by projecting the 2-D correlation functions in the region $2 < |\Delta\eta| < 5$. To test the possible impact of acceptance effects that may result in a distortion of the $\Delta\phi$ distribution, the correlation functions are corrected by the projection of the mixed-event 2-D correlation function onto $\Delta\phi$, as described in Section~\ref{sec:2PC}.
The differences in the results are assigned as a systematic uncertainty.
 
\emph{Exclusion of tracks from \gapDef.}  The default analysis excludes reconstructed tracks
that are within the sum-of-gaps range from the two-particle correlations, i.e.\ those with $\eta > 4.9 - \gapDef$.
As a systematic variation, these tracks are included in constructing the correlation functions.   The resulting
differences are included as a systematic uncertainty.
 
\emph{Photonuclear event purity.} The estimate of the purity of photonuclear events in data relies on the description of $\gamma$+Pb event properties in simulation. To test the sensitivity to this description, the fit described in Section~\ref{ssec:purity} was repeated using \PYTHIA $\gamma$+$p$ events as the model for photonuclear events and, alternatively, by fitting to the $N_\mathrm{ch}^\mathrm{rec}$ distribution, separately in different slices of \gapDef. In the latter case, the purity decreases to approximately 90\% in events with $N_\mathrm{ch}^\mathrm{rec} = 15$--$20$ and to 80\% in those with $N_\mathrm{ch}^\mathrm{rec} = 10$--$15$, which are the ranges used to define the nominal and alternative LM selections for the two-particle correlation analysis.
 
In the default analysis, no correction is made for a possible presence of backgrounds, primarily peripheral Pb+Pb events, in the selected sample of high-multiplicity $\gamma$+Pb events. As a systematic variation, the $\Delta\phi$ distributions in data are corrected by subtracting the contribution from these background events, scaled to match the purity in each $N_\mathrm{ch}^\mathrm{rec}$ range.
Since the potential impact of the purity is much larger for events with low $N_\mathrm{ch}^\mathrm{rec}$ than those with high $N_\mathrm{ch}^\mathrm{rec}$, the effect on the analysis is evaluated by accounting for the possible presence of backgrounds only in the LM reference.
Additionally, since the purity was observed to be lower in the alternative LM reference events, $N_\mathrm{ch}^\mathrm{rec} = 10$--$15$, than in the nominal LM selection, the difference in the results using the alternative LM reference, with and without the purity correction, was taken as the uncertainty for this source.
The model $\Delta\phi$ distributions for the background events are taken from $pp$ collision data at $\sqrt{s} = 5$~\TeV~\cite{HION-2015-14}, with the same reconstruction, definition of $N_\mathrm{ch}^\mathrm{rec}$, and gap-based event selection, as this analysis.
The resulting
differences are included as a systematic uncertainty.
 
\emph{Other uncertainty sources and checks.}
Several other potential sources were investigated, and were found to have a negligible effect on the results and thus are not included in the total uncertainty. These include: comparing the results using only events where the photon is headed towards the A side of the detector with those where it is headed towards the C side; the number of Fourier terms used in parameterizing the LM reference distribution; the number of additional Fourier terms beyond $n=4$ used in the template fit; the impact of a possible misalignment of the tracker on the charged-particle measurement; the impact of the correction for the finite tracking efficiency; and the impact of correcting for the trigger efficiency and different luminosities of the triggered event samples as a function of $N_\mathrm{ch}^\mathrm{rec}$ on a per-event basis.
 
While the $\gapDef > 2.5$ requirement is treated as an operational event definition which is not a source of uncertainty, the results using alternative definitions $\gapDef > 2.0$ and $\gapDef > 3.0$ were not significantly different from the nominal results. This demonstrates that the  sample of photonuclear events in data is robust against small changes in their selection criteria and against detector effects in the measurement of \gapDef.
 
Finally, the template method carries an implicit assumption that the modulation in the HM selection is the same as that in the LM selection. If this is not the case, the extracted $v_n$ values in the HM selection will be biased to be systematically closer to those in the LM selection. In recent measurements of azimuthal anisotropies in small systems~\cite{HION-2017-07}, a `one-step' correction was employed using two adjacent LM reference bins to account for the possibly changing $v_n$ magnitudes with multiplicity. Since the measured $v_n$ values were observed to have no $N_\mathrm{ch}^\mathrm{rec}$ dependence within uncertainties, this alternative method was not utilized here.
 
As can be seen in Table~\ref{tab:syst}, the systematic uncertainty is not dominated by any single source, and typically those from the pair-pseudorapidity difference, the choice of LM reference, and the event purity are co-dominant. The systematic uncertainties tend to be larger for the \pt-dependent results than for the $N_\mathrm{ch}^\mathrm{rec}$-dependent results, and are typically larger for the $v_3$ results than for the $v_2$ results. For the $v_2$ and $v_3$ results as a function of $N_\mathrm{ch}^\mathrm{rec}$, the statistical uncertainties are typically larger than the systematic uncertainties. However, for the results as a function of $p_\mathrm{T}^\mathrm{ch}$, the systematic uncertainties are generally larger. This reflects, for example, the greater sensitivity to the choice of LM reference, which must be divided into different  $p_\mathrm{T}^\mathrm{ch}$ selections.
 
\section{Results and discussion}
\label{sec:result}
 
This section presents the measurement of the second- and third-order flow coefficients in photonuclear collisions, in both cases as a function of $N_\mathrm{ch}^\mathrm{rec}$ and charged-particle $p_\mathrm{T}$. These results are compared with previous measurements of $v_2$ and $v_3$ for inclusive charged hadrons by ATLAS in inelastic, minimum-bias $13$~\TeV\ $pp$ collisions and in 5.02 TeV $p$+Pb collisions~\cite{HION-2016-01}.
 
\begin{figure}[t]
\begin{center}
\includegraphics[width=0.60\textwidth]{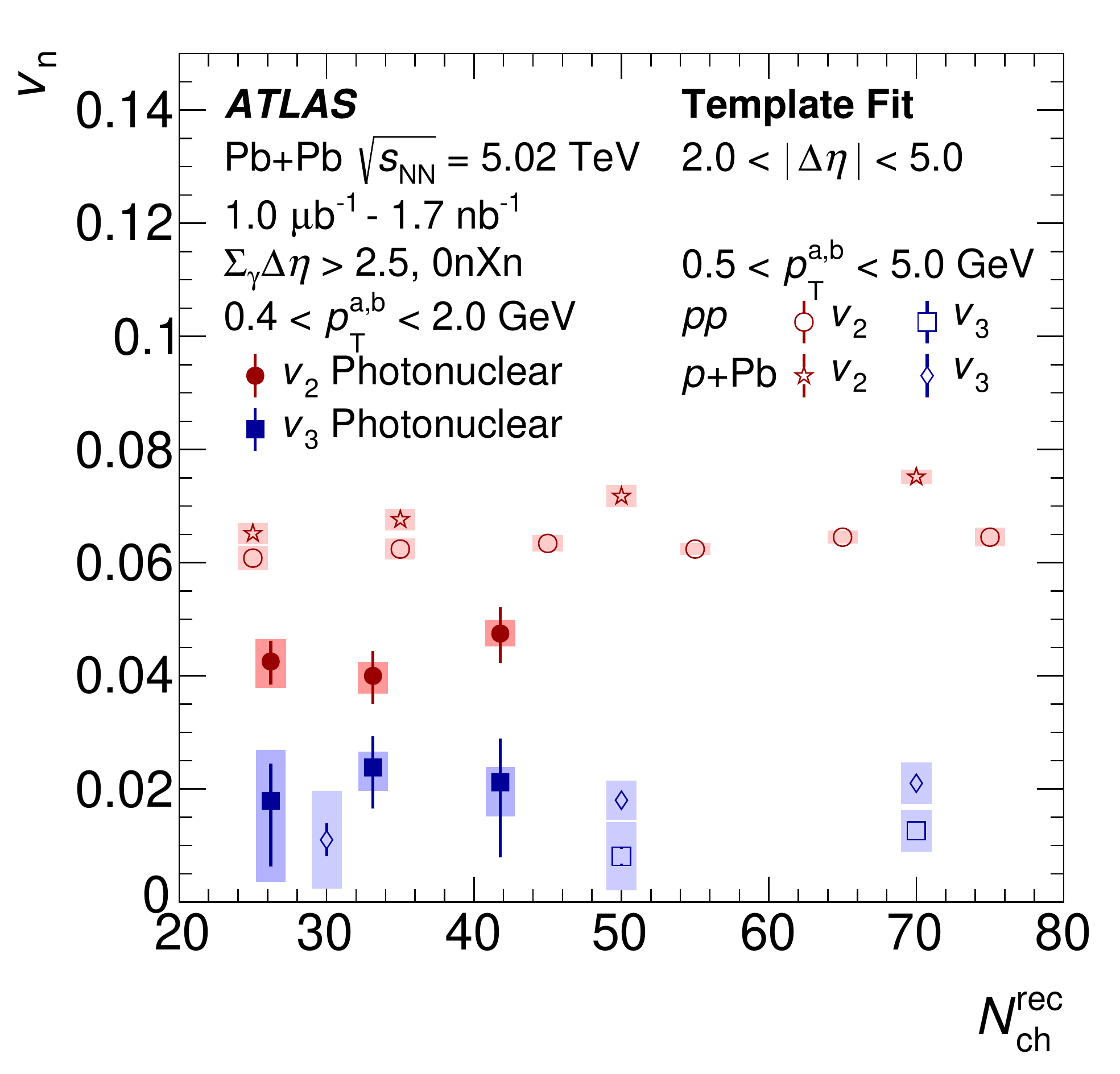}
\caption{Flow coefficients $v_2$ and $v_3$ for charged particles with $0.5 < \pt < 2.0~\GeV$ in photonuclear events, reported as a function of charged-particle multiplicity $N_\mathrm{ch}^\mathrm{rec}$.  The vertical error bars and colored boxes represent the statistical and total systematic uncertainties, respectively. The photonuclear data points are positioned at the average $N_\mathrm{ch}^\mathrm{rec}$ value in each interval.
The measurements in photonuclear events (solid symbols) are compared with those in \pp collisions at $13~\TeV$  and \pPb collisions at $5.02$~\TeV~\cite{HION-2016-01} (open symbols),
integrated over $0.5 < \pt < 5.0~\GeV$.
\label{fig:vn_dNch}}
\end{center}
\end{figure}

Figure~\ref{fig:vn_dNch} presents the $v_2$ and $v_3$ values as a function of the event charged-particle multiplicity $N_\mathrm{ch}^\mathrm{rec}$.
Significant, nonzero values for $v_2$ and $v_{3}$ are observed and they are compatible with no $N_\mathrm{ch}^\mathrm{rec}$ dependence within uncertainties.
The results are compared with previous measurements in $pp$ and $p$+Pb collisions, presented here as a function of $N_\mathrm{ch}^\mathrm{rec}$ for a slightly different $p_\mathrm{T}$ selection, $0.5 < p_\mathrm{T} < 5$~\GeV.
Given the shape of the $p_\mathrm{T}$ distribution in high-multiplicity $pp$ collisions~\cite{Acharya:2019mzb} and the $p_\mathrm{T}$-dependence of the $v_2$ values, the $v_2$ in $pp$ collisions for $0.4 < p_\mathrm{T} < 2$~\GeV\ particles (the selection used in this measurement) may be at most 10\% lower than the values presented for $0.5 < p_\mathrm{T} < 5$~\GeV.
In photonuclear events, the $v_2$ values as a function of $N_\mathrm{ch}^\mathrm{rec}$ are systematically below those in $pp$ and $p$+Pb events, even after accounting for the different $p_\mathrm{T}$ ranges in the datasets. The $v_3$ values are compatible with those in $pp$ and $p$+Pb events within significantly larger statistical and systematic uncertainties.
 
\begin{figure}[p]
\begin{center}
\includegraphics[width=0.47\textwidth]{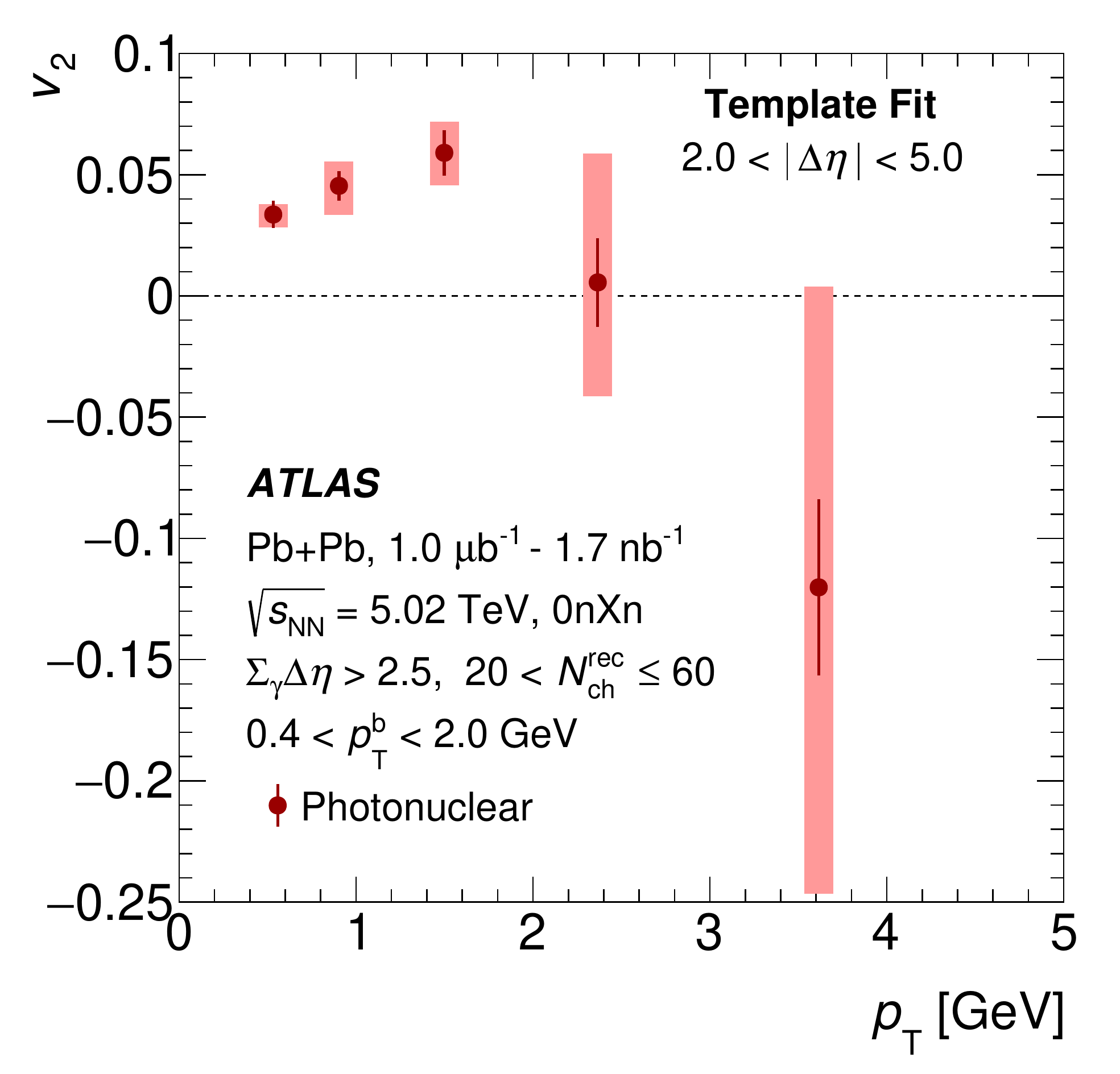}
\includegraphics[width=0.47\textwidth]{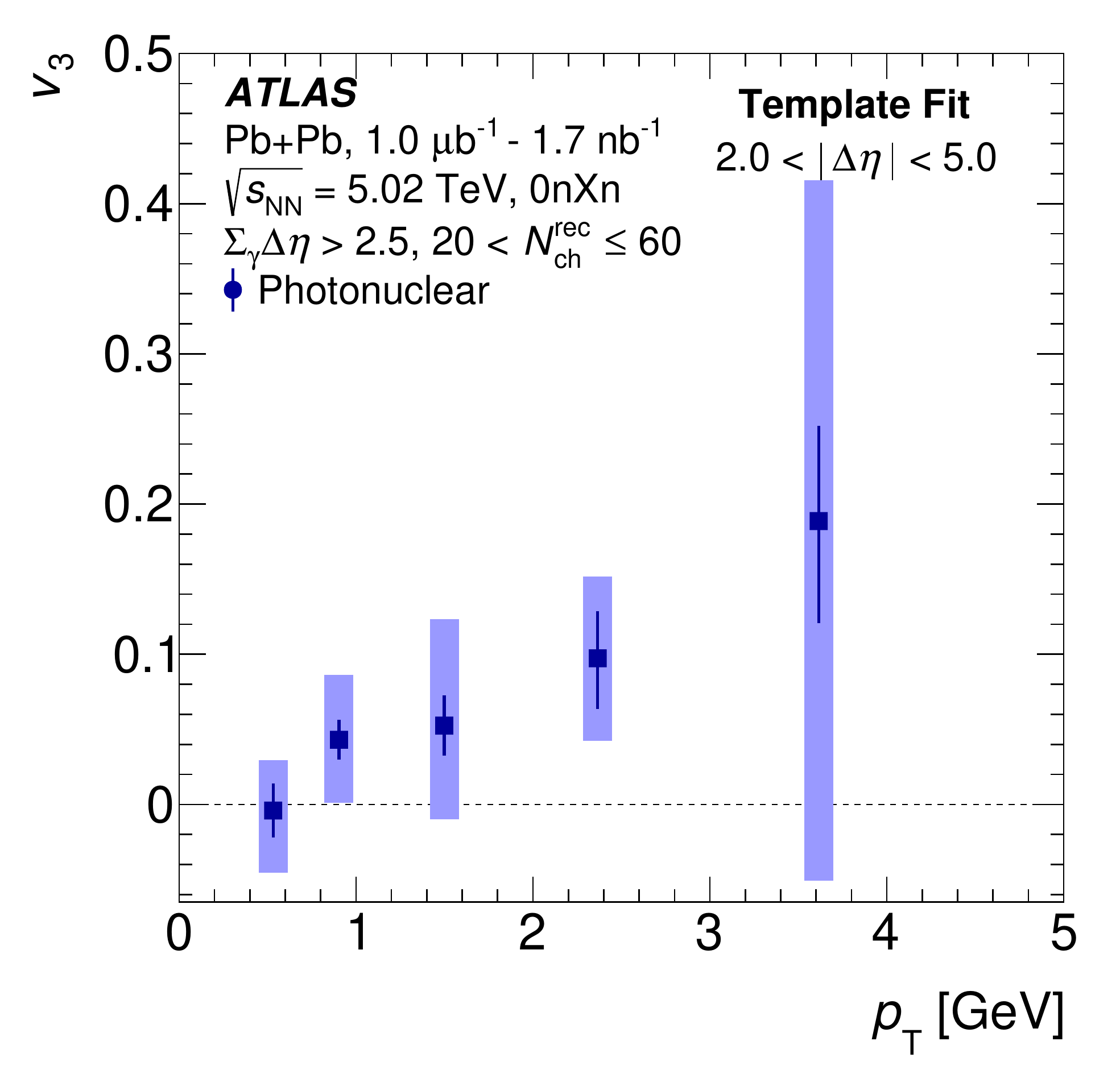}
\caption{Charged-particle flow coefficients $v_2$ (left) and $v_3$ (right) in photonuclear events with $20 < N_\mathrm{ch}^\mathrm{rec} \leq 60$, reported as a function of particle $p_\mathrm{T}$.  The vertical error bars and colored boxes represent the statistical and total systematic uncertainties, respectively.
The photonuclear data points are positioned at the average $p_\mathrm{T}$ value in each interval.
\label{fig:vn_dpt2}}
\end{center}
\begin{center}
\includegraphics[width=0.47\textwidth]{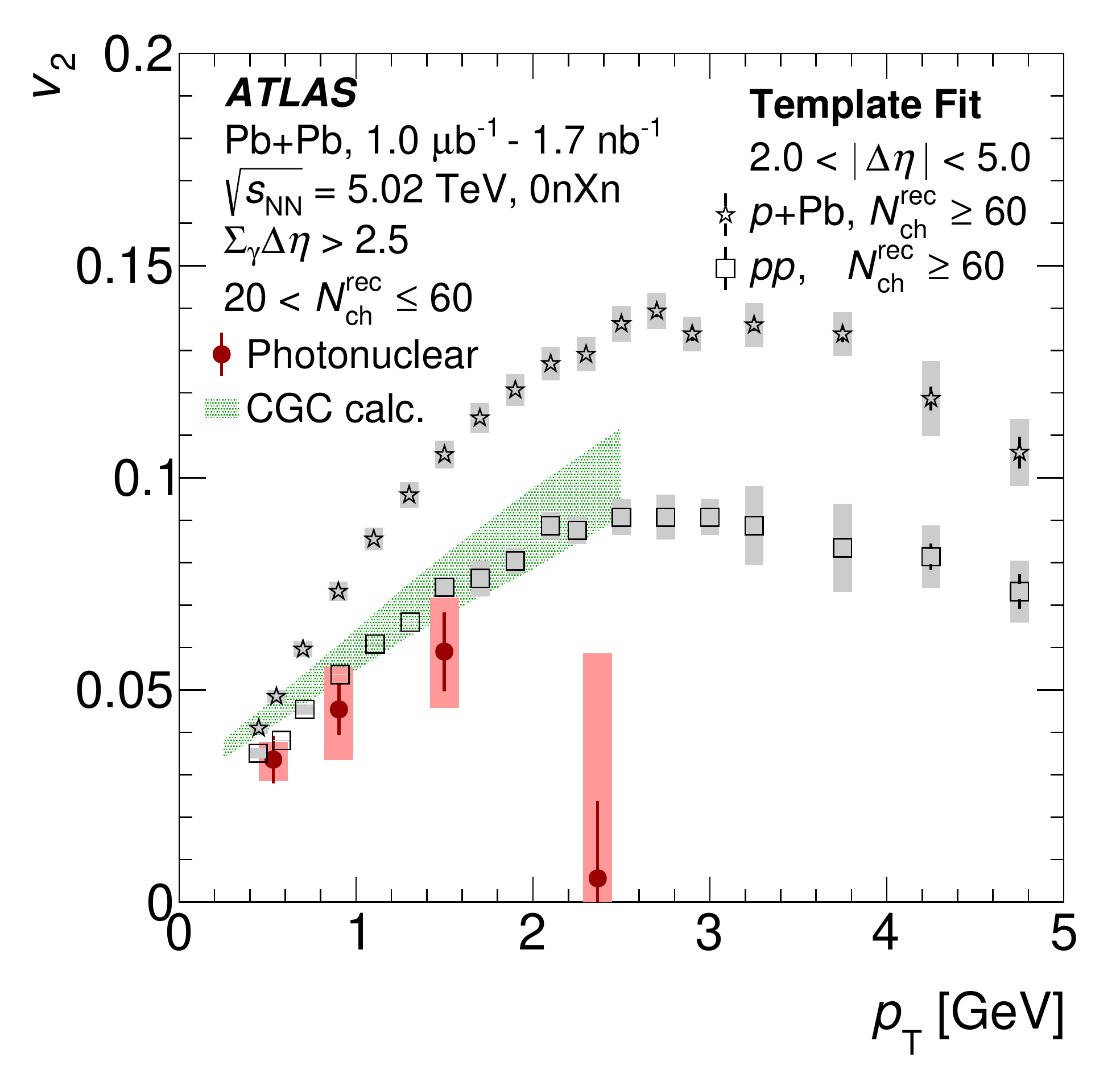}
\includegraphics[width=0.47\textwidth]{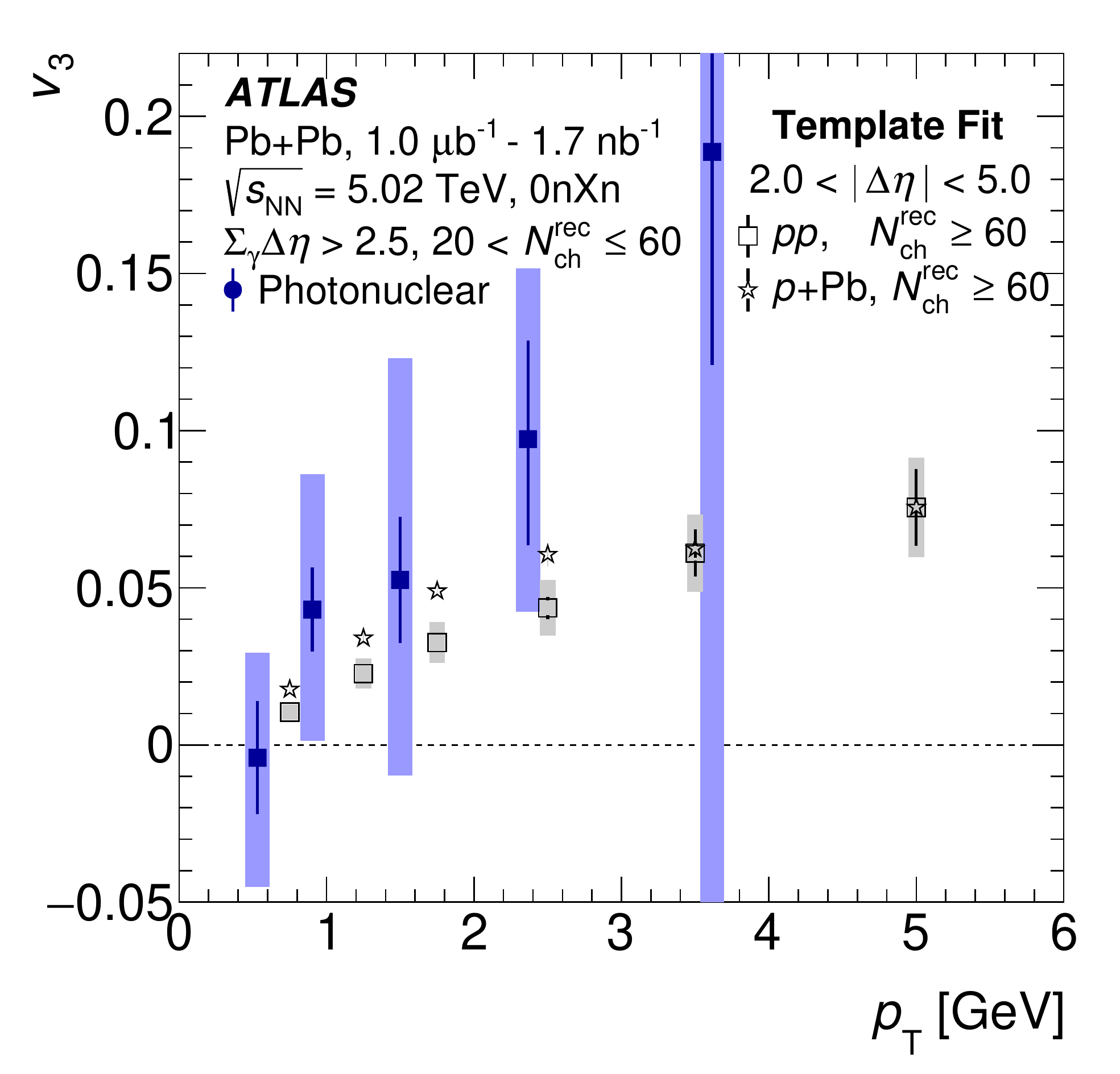}\\
\caption{Charged-particle flow coefficients $v_2$ (left) and $v_3$ (right) in photonuclear events with $20 < N_\mathrm{ch}^\mathrm{rec} \leq 60$, reported as a function of particle $p_\mathrm{T}$.  The vertical error bars and colored boxes represent the statistical and total systematic uncertainties, respectively.
The photonuclear data points are positioned at the average $p_\mathrm{T}$ value in each interval.
The data are compared with the analogous measurements in $pp$ collisions at 13~\TeV\ and $p$+Pb collisions at 5.02~\TeV\ for $N_\mathrm{ch}^\mathrm{rec} \geq 60$~\cite{HION-2016-01}.   The $v_{2}$ data are also compared with a CGC-based theory calculation from Ref.~\cite{Shi:2020djm}. These photonuclear data are the same as in Figure~\ref{fig:vn_dpt2} but with different y-axes ranges to allow comparison with additional data and theoretical predictions.
\label{fig:vn_dpt}}
\end{center}
\end{figure}

The $v_2$ and $v_3$ results as a function of charged-particle $p_\mathrm{T}$ are presented in Figure~\ref{fig:vn_dpt2} for events with $20 < N_\mathrm{ch}^\mathrm{rec} \leq 60$. The $v_2$ results have central values similar to those plotted as a function of $N_\mathrm{ch}^\mathrm{rec}$ in Figure~\ref{fig:vn_dNch} but have larger systematic and statistical uncertainties, due to the more limited size of the available data sample for the $p_\mathrm{T}$-dependent LM references for these measurements.  For $v_{2}$, the results trend towards negative values for $p_\mathrm{T} \ge 2.0$~\GeV, although with very large systematic uncertainties.   It is notable that there is a statistically significant confirmation of factorization, as detailed in Section~\ref{ssec:factorization}, only for $v_2$ with the selection $0.4 \le p_\mathrm{T} \le 2.0$~\GeV.
Hence the values of $v_2$ and $v_3$ for $p_\mathrm{T} \ge 2$~\GeV\ have an additional caveat, arising from the fact that the magnitude of the flow signal relative to that of the non-flow contribution is significantly smaller than at lower $p_\mathrm{T}$. In particular, the trend towards negative $v_2$ values and rising $v_3$ values suggests that the factorization assumption could be violated.
 
\Fig{\ref{fig:vn_dpt}} shows the same data as \Fig{\ref{fig:vn_dpt2}}, but zoomed in on the vertical axis to allow a better comparison with the analogous $p_\mathrm{T}$-dependent values in the $pp$ and $p$+Pb measurements described above, with the selection $N_\mathrm{ch}^\mathrm{rec} \geq 60$.
In the region $0.4 < p_\mathrm{T} < 2$~\GeV\, the central values of the $v_2$ are smaller than those in $pp$ and $p$+Pb collisions, similar to that observed in the $N_\mathrm{ch}^\mathrm{rec}$-dependent results in Figure~\ref{fig:vn_dNch}. However, due to the larger uncertainties in the $p_\mathrm{T}$-dependent case, the $v_\mathrm{2}$ values for photonuclear and $pp$ collisions are compatible within the uncertainties of the former in the range $p_\mathrm{T} < 2$~\GeV.   The $v_{3}$ values are compatible between systems within large uncertainties.
 
There are currently no published theoretical predictions for flow coefficients in photonuclear collisions within a hydrodynamic or parton transport framework.   In such frameworks, the elliptic and triangular flow coefficients scale with the initial geometry eccentricities, $\varepsilon_{2}$ and $\varepsilon_{3}$ respectively, and the charged-particle multiplicity $dN_\mathrm{ch}/d\eta$.
In the vector-meson dominance picture, photon--hadron interactions arise through fluctuations of the photon into hadronic states with the same quantum numbers as vector mesons, which have a nontrivial initial transverse geometry.
This geometry is determined by the spectrum of these fluctuations, and while models of this spectrum exist~\cite{Alvioli:2016gfo}, they have not yet been adapted to provide quantitative models.
In the absence of a complete model, the magnitude of the eccentricities can be estimated by noting that fluctuations of the photon into light vector-meson states such as the $\rho$ give the largest contribution to the cross section.
The initial geometries for $\rho$+Pb collisions can be computed with a Monte Carlo Glauber calculation~\cite{Loizides:2016djv} which treats the $\rho$ meson as having two constituent quarks.
The resulting mean values of the second- and third-order spatial eccentricities, $\varepsilon_{2}$ and $\varepsilon_{3}$, are nearly identical to those in the \pPb case.
Also, when comparing \pPb and photonuclear events with the same $N_\mathrm{ch}^\mathrm{rec}$, in fact the relevant $dN_\mathrm{ch}/d\eta$ is larger in the photonuclear events since the particles are distributed over a smaller pseudorapidity region.    Thus, one might na\"ively expect the flow coefficients to be similar in photonuclear events and \pPb collisions.   However, in order to compare any such calculation with data, a full modeling of the photon fluctuations in the selected $\gamma$+Pb events needs to be carried out. In addition, correctly accounting for the boosted kinematics and limited acceptance using a fully three-dimensional simulation may be important.
 
An alternative interpretation of two-particle correlations in small collision systems involves interactions at the earliest time between gluon fields in the Color Glass Condensate (CGC) framework~\cite{Lappi:2015vta}.
Recently such calculations have described heavy-flavor hadron and quarkonia azimuthal anisotropies in \pPb collisions~\cite{Zhang:2020ayy,Zhang:2019dth}, although calculations in the CGC framework fail to describe other aspects of the data, such as the charged-hadron flow coefficients in \pPb at the LHC and small-systems collisions at RHIC~\cite{Mace:2018vwq,Mace:2018yvl}.  The authors have extended these calculations to consider a color dipole interacting with a Pb nucleus either at a future Electron Ion Collider or in photonuclear collisions at the LHC~\cite{Shi:2020djm}.   The CGC calculation for photonuclear collisions is shown in Figure~\ref{fig:vn_dpt} and is in reasonable agreement with the $v_2$ data within uncertainties.  In these calculations, the Pb nucleus is described with a saturation scale $Q_{s}^2 = 5~\GeV^{2}$ and typical parton transverse momentum $\Delta = 0.5~\GeV$, as used in calculations of $v_2$ for heavy-flavor mesons and quarkonia~\cite{Zhang:2020ayy,Zhang:2019dth}.
However, in the calculation for the photonuclear case, the parameter $B_{p} = 25~\GeV^{-2}$, which controls the transverse area of the interaction and thus the number of color domains from the Pb nucleus taking part in the interaction, is significantly larger than in the previous calculations~\cite{Zhang:2020ayy,Zhang:2019dth}. A larger transverse area generally contains more color domains and results in a smaller value of $v_2$.
 
\begin{figure}[t]
\centering
\includegraphics[width=0.47\textwidth]{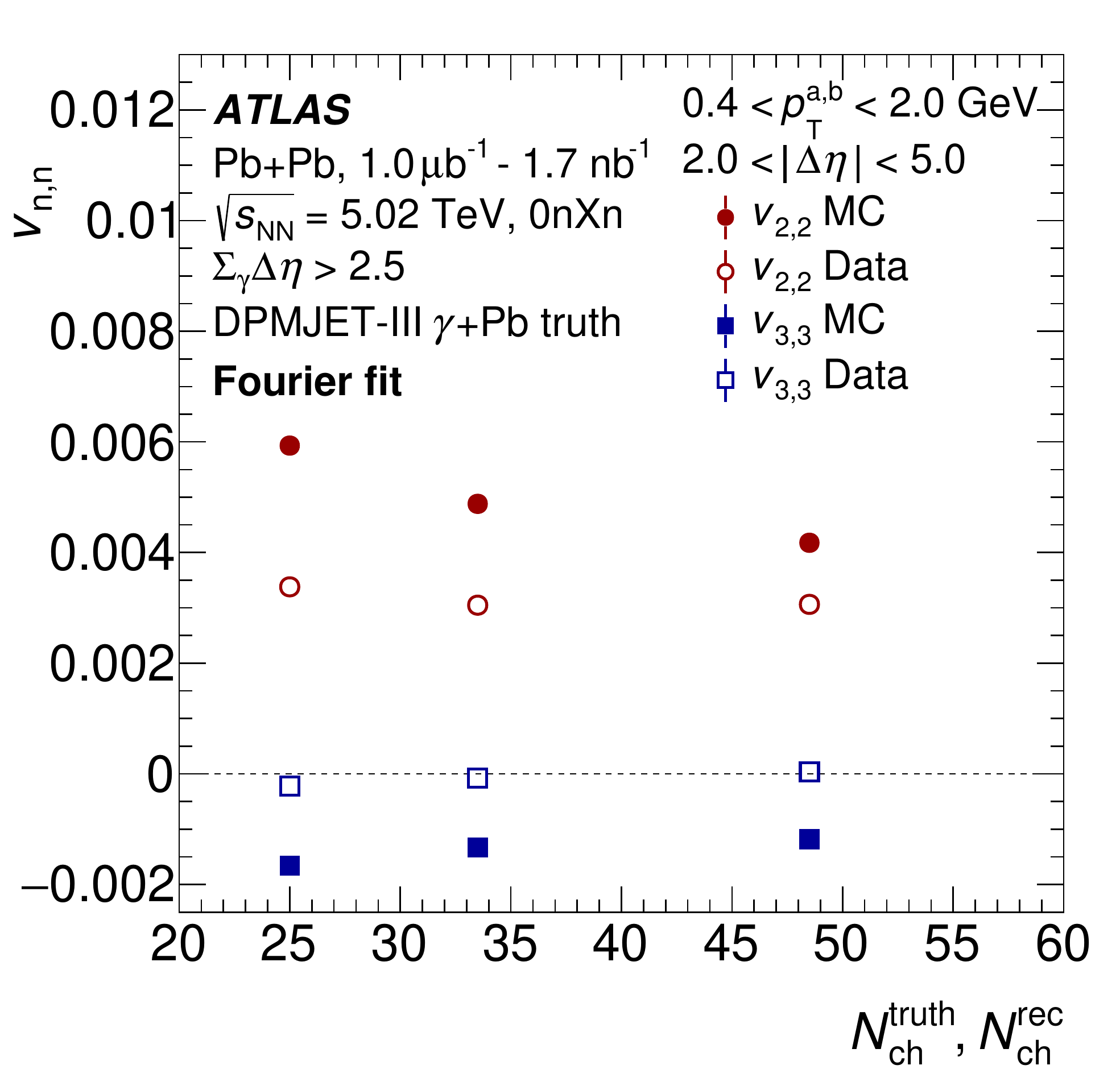}
\includegraphics[width=0.47\textwidth]{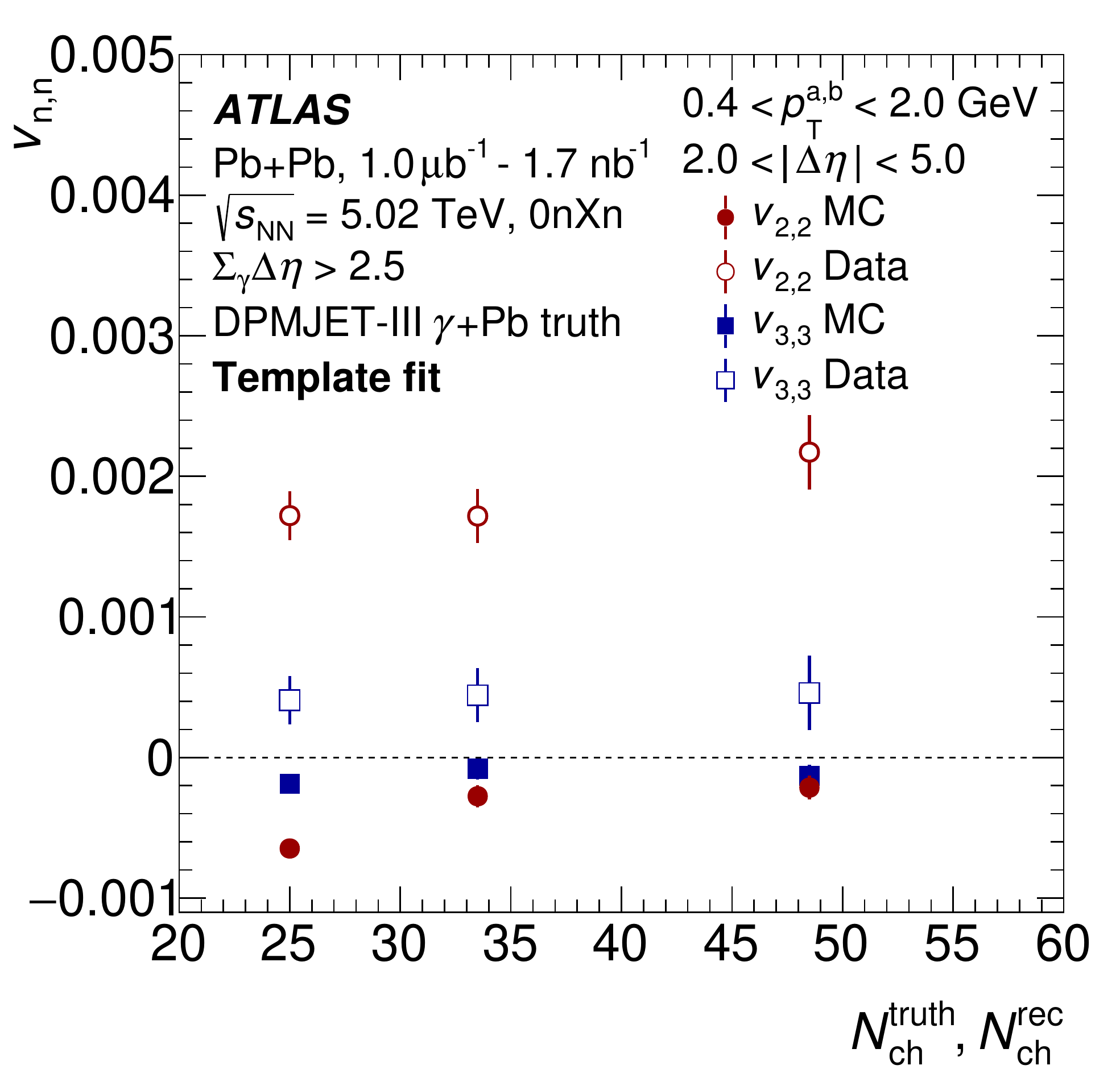}
\caption{Comparison of results for raw Fourier coefficients  $v_{2,2}$ and $v_{3,3}$ (left, without non-flow subtraction) and for non-flow subtracted coefficients $v_{2,2}$ and $v_{3,3}$ (right, with non-flow subtraction using the template method), shown in data (open points) and in \textsc{DPMJET}-III (filled points). The results in data and \textsc{DPMJET}-III are presented as a function of $N_\mathrm{ch}^\mathrm{rec}$ and $N_\mathrm{ch}^\mathrm{truth}$, respectively.}
\label{fig:dpmjet_vnn}
\end{figure}
 
Finally, it is interesting to compare the data with results from \textsc{DPMJET}-III, which has neither CGC-related correlations nor hydrodynamic or parton scattering effects.
This comparison sets a baseline and provides a check of the possible impact of issues such as the LM template choice (which accounts for non-flow) selecting events with photon energies that differ from the HM sample.    The photon energy is expected to be a function of $N_\mathrm{ch}^\mathrm{rec}$, which is true within \textsc{DPMJET}-III as shown in Figure~\ref{fig:photonE_in_MC}.    If the non-flow contribution changes with photon energy, it may be that the non-flow correlation shape would not be the same in the LM and HM selections, which is an assumption of the non-flow subtraction procedure.    In order to test if this effect might mimic the $v_2$ and $v_3$
signals observed, the \textsc{DPMJET}-III output has been run through the full analysis procedure, including the LM and HM selection and non-flow template fit.   The resulting $v_{n,n}$ values are shown in Figure~\ref{fig:dpmjet_vnn} for both data and \textsc{DPMJET}-III. The left panel shows the values that would result from a Fourier fit to the HM correlation function, without accounting for non-flow, while the right panel shows the template-subtracted results. In \textsc{DPMJET}-III, the template-subtracted $v_{n,n}$ values are both negative, in contradistinction to the opposite sign for these values observed in data.   Since they give negative values, which are impossible in a flow factorization interpretation, one cannot extract $v_{2}$ and $v_{3}$ values to compare with the data.   Nevertheless, this demonstrates that, at least within \textsc{DPMJET}-III, there is no trivial effect in the non-flow subtraction issue which could result in a false signal.

\FloatBarrier
 
\section{Conclusion}
 
This paper reports a measurement of long-range two-particle correlations in high-energy photonuclear collisions. Events are selected from 1.7~nb$^{-1}$ of Pb+Pb collision data at 5.02~\TeV\ recorded by the ATLAS detector at the LHC in 2018 using a combination of criteria for the energy deposited exclusively in one side of the zero-degree calorimeter, and requirements on the sum-of-gaps in the photon-going direction ($\gapDef > 2.5$). Azimuthal correlation functions are reported for pairs of charged particles separated by $2 < |\Delta\eta| < 5$ units of pseudorapidity. A template-fitting method is used to subtract the non-flow contribution.  The single-particle flow coefficients $v_2$ and $v_3$ are reported as a function of charged-particle multiplicity and single-particle $p_\mathrm{T}$.
Significant, nonzero $v_2$ and $v_{3}$ values are observed in photonuclear events, indicating that particles produced in these events participate in azimuthally dependent, collective motion. The $v_2$ values are smaller than those reported in \pp and \pPb collisions at similar particle multiplicities.  The results presented here provide new information about long-range collective behavior in an exotic collision system with different initial conditions than those in ordinary hadronic collisions. As such, they provide a new testing ground for different pictures of the physical origin of these correlations, and a starting point for understanding the systems which will be studied at the future Electron Ion Collider.

\section*{Acknowledgments}

We thank CERN for the very successful operation of the LHC, as well as the
support staff from our institutions without whom ATLAS could not be
operated efficiently.
 
We acknowledge the support of ANPCyT, Argentina; YerPhI, Armenia; ARC, Australia; BMWFW and FWF, Austria; ANAS, Azerbaijan; SSTC, Belarus; CNPq and FAPESP, Brazil; NSERC, NRC and CFI, Canada; CERN; ANID, Chile; CAS, MOST and NSFC, China; COLCIENCIAS, Colombia; MSMT CR, MPO CR and VSC CR, Czech Republic; DNRF and DNSRC, Denmark; IN2P3-CNRS and CEA-DRF/IRFU, France; SRNSFG, Georgia; BMBF, HGF and MPG, Germany; GSRT, Greece; RGC and Hong Kong SAR, China; ISF and Benoziyo Center, Israel; INFN, Italy; MEXT and JSPS, Japan; CNRST, Morocco; NWO, Netherlands; RCN, Norway; MNiSW and NCN, Poland; FCT, Portugal; MNE/IFA, Romania; JINR; MES of Russia and NRC KI, Russian Federation; MESTD, Serbia; MSSR, Slovakia; ARRS and MIZ\v{S}, Slovenia; DST/NRF, South Africa; MICINN, Spain; SRC and Wallenberg Foundation, Sweden; SERI, SNSF and Cantons of Bern and Geneva, Switzerland; MOST, Taiwan; TAEK, Turkey; STFC, United Kingdom; DOE and NSF, United States of America. In addition, individual groups and members have received support from BCKDF, CANARIE, Compute Canada, CRC and IVADO, Canada; Beijing Municipal Science \& Technology Commission, China; COST, ERC, ERDF, Horizon 2020 and Marie Sk{\l}odowska-Curie Actions, European Union; Investissements d'Avenir Labex, Investissements d'Avenir Idex and ANR, France; DFG and AvH Foundation, Germany; Herakleitos, Thales and Aristeia programmes co-financed by EU-ESF and the Greek NSRF, Greece; BSF-NSF and GIF, Israel; La Caixa Banking Foundation, CERCA Programme Generalitat de Catalunya and PROMETEO and GenT Programmes Generalitat Valenciana, Spain; G\"{o}ran Gustafssons Stiftelse, Sweden; The Royal Society and Leverhulme Trust, United Kingdom.
 
The crucial computing support from all WLCG partners is acknowledged gratefully, in particular from CERN, the ATLAS Tier-1 facilities at TRIUMF (Canada), NDGF (Denmark, Norway, Sweden), CC-IN2P3 (France), KIT/GridKA (Germany), INFN-CNAF (Italy), NL-T1 (Netherlands), PIC (Spain), ASGC (Taiwan), RAL (UK) and BNL (USA), the Tier-2 facilities worldwide and large non-WLCG resource providers. Major contributors of computing resources are listed in Ref.~\cite{ATL-SOFT-PUB-2020-001}.
 
 
\FloatBarrier
 
\clearpage

\printbibliography
 
\clearpage
 
\begin{flushleft}
\hypersetup{urlcolor=black}
{\Large The ATLAS Collaboration}

\bigskip

\AtlasOrcid[0000-0002-6665-4934]{G.~Aad}$^\textrm{\scriptsize 101}$,    
\AtlasOrcid[0000-0002-5888-2734]{B.~Abbott}$^\textrm{\scriptsize 127}$,    
\AtlasOrcid[0000-0002-7248-3203]{D.C.~Abbott}$^\textrm{\scriptsize 102}$,    
\AtlasOrcid[0000-0002-2788-3822]{A.~Abed~Abud}$^\textrm{\scriptsize 36}$,    
\AtlasOrcid[0000-0002-1002-1652]{K.~Abeling}$^\textrm{\scriptsize 53}$,    
\AtlasOrcid[0000-0002-2987-4006]{D.K.~Abhayasinghe}$^\textrm{\scriptsize 93}$,    
\AtlasOrcid[0000-0002-8496-9294]{S.H.~Abidi}$^\textrm{\scriptsize 29}$,    
\AtlasOrcid[0000-0002-8279-9324]{O.S.~AbouZeid}$^\textrm{\scriptsize 40}$,    
\AtlasOrcid{N.L.~Abraham}$^\textrm{\scriptsize 155}$,    
\AtlasOrcid[0000-0001-5329-6640]{H.~Abramowicz}$^\textrm{\scriptsize 160}$
% ...
\bigskip
\\

% ...
$^{29}$Physics Department, Brookhaven National Laboratory, Upton NY; United States of America.\\
% ...
$^{36}$CERN, Geneva; Switzerland.\\
% ...
$^{40}$Niels Bohr Institute, University of Copenhagen, Copenhagen; Denmark.\\
% ...
$^{53}$II. Physikalisches Institut, Georg-August-Universit\"{a}t G\"ottingen, G\"ottingen; Germany.\\
% ...
$^{93}$Department of Physics, Royal Holloway University of London, Egham; United Kingdom.\\
% ...
$^{101}$CPPM, Aix-Marseille Universit\'e, CNRS/IN2P3, Marseille; France.\\
$^{102}$Department of Physics, University of Massachusetts, Amherst MA; United States of America.\\
% ...
$^{127}$Homer L. Dodge Department of Physics and Astronomy, University of Oklahoma, Norman OK; United States of America.\\
% ...
$^{155}$Department of Physics and Astronomy, University of Sussex, Brighton; United Kingdom.\\
% ...
$^{160}$Raymond and Beverly Sackler School of Physics and Astronomy, Tel Aviv University, Tel Aviv; Israel.\\
% ...

\end{flushleft}


\end{document}